\documentclass[a4paper,11pt]{article}
\pdfoutput=1 
\usepackage{jheppub} 

\usepackage[english]{babel}
\usepackage{microtype,xspace}
\usepackage[utf8]{inputenc}	

\usepackage{amsfonts,amsmath,amssymb,bm,mathrsfs,mathtools,dsfont} 	
\allowdisplaybreaks 	
\usepackage[dvipsnames]{xcolor}
\usepackage{slashed}
\usepackage{fontawesome}

\usepackage{siunitx}
\usepackage{graphicx}
\graphicspath{{./Figures/}}
\usepackage{multirow}
\renewcommand{\arraystretch}{1.2}
\usepackage{tabularx}
\newcolumntype{Y}{>{\centering\arraybackslash}X}

\usepackage[shortlabels]{enumitem}
\setlist{itemsep=.1em,topsep=.5em}
\SetEnumerateShortLabel{i}{\textit{\roman*}}

\definecolor{red}{rgb}{0.6,.0706,.1373}
\definecolor{blue}{rgb}{0,0.396,0.741}
\colorlet{blueRef}{blue!80!black}
\colorlet{blueLink}{blue!80!black}
\hypersetup{
	colorlinks, 
	bookmarksopen, 
	bookmarksnumbered,
	citecolor=blueRef, 		
	linkcolor=blueLink,		
	urlcolor=blueLink,		
}

\DeclareMathVersion{sans}
\SetSymbolFont{operators}{sans}{OT1}{cmbr}{m}{n}
\SetSymbolFont{letters}{sans}{OML}{cmbrm}{m}{it}
\SetSymbolFont{symbols}{sans}{OMS}{cmbrs}{m}{n}
\SetMathAlphabet{\mathit}{sans}{OT1}{cmbr}{m}{sl}
\SetMathAlphabet{\mathbf}{sans}{OT1}{cmbr}{bx}{n}
\SetMathAlphabet{\mathtt}{sans}{OT1}{cmtl}{m}{n}
\SetSymbolFont{largesymbols}{sans}{OMX}{iwona}{m}{n}

\DeclareMathVersion{boldsans}
\SetSymbolFont{operators}{boldsans}{OT1}{cmbr}{b}{n}
\SetSymbolFont{letters}{boldsans}{OML}{cmbrm}{b}{it}
\SetSymbolFont{symbols}{boldsans}{OMS}{cmbrs}{b}{n}
\SetMathAlphabet{\mathit}{boldsans}{OT1}{cmbr}{b}{sl}
\SetMathAlphabet{\mathbf}{boldsans}{OT1}{cmbr}{bx}{n}
\SetMathAlphabet{\mathtt}{boldsans}{OT1}{cmtl}{b}{n}
\SetSymbolFont{largesymbols}{boldsans}{OMX}{iwona}{bx}{n}

\usepackage{needspace}
\usepackage[noindentafter
]{titlesec} 
\titleformat{\section}{\needspace{4\baselineskip} \large \bfseries \sffamily \mathversion{boldsans} \color{blue!70!black} }{\thesection}{10pt}{}{}
\titlespacing{\section}{0pt}{10pt}{5pt}
\titleformat{\subsection}{\needspace{3\baselineskip} \sffamily \mathversion{sans} \itshape \color{blue!80!black} }{\thesubsection}{10pt}{}{}
\titlespacing{\subsection}{0pt}{8pt}{3pt}
\titleformat{\subsubsection}{\needspace{3\baselineskip} \normalsize \sffamily \itshape \mathversion{sans} \color{blue!80!black} }{\thesubsubsection}{10pt}{}{}
\titlespacing{\subsubsection}{0pt}{8pt}{2pt}

\usepackage{braket} 
\usepackage{cancel}
\usepackage{multicol}
\usepackage{multirow}
\usepackage{vcell}
\usepackage{pdflscape}
\usepackage{titletoc}

\newcommand{\lzm}{\left(}
\newcommand{\dzm}{\right)}
\newcommand{\lzs}{\left[}
\newcommand{\dzs}{\right]}

\newcommand{\cL}{\mathcal{L}}
\newcommand{\cO}{\mathcal{O}}

\newcommand{\hermc}{\text{H.c.}}

\newcommand{\eminus}{\vcenter{\hbox{\scalebox{0.6}[1]{$ - $}}}}	
\newcommand{\ord}[1]{\mathcal{O}\!\left( #1 \right)}

\newcommand{\hc}{\; + \; \mathrm{H.c.} \;}

\newcommand{\ud}[2]{\phantom{}^{#1}\phantom{}_{#2}}
\newcommand{\du}[2]{\phantom{}_{#1}\phantom{}^{#2}} 
\newcommand{\diag}{\mathop{\mathrm{diag}}}

\newcommand{\rep}[1]{\mathbf{#1}}
\newcommand{\repbar}[1]{\overline{\mathbf{#1}}}

\newcommand{\sscript}[1]{{\scriptscriptstyle \mathrm{#1}}}

\newcommand{\LL}{\mathrm{L}}
\newcommand{\RR}{\mathrm{R}}
\newcommand{\U}{\mathrm{U}}
\newcommand{\SU}{\mathrm{SU}}

\newcommand{\opemph}[1]{{\color{blue!80!black} #1}}

\newcommand{\coloredbox}[3]{\colorbox{#1}{\texttt{{\color{#2}#3}}}}

\begin{document}

\author{Admir Greljo,}
\author{Ajdin Palavrić}
\author{and Anders Eller Thomsen}

\affiliation{Albert Einstein Center for Fundamental Physics, Institute for Theoretical Physics, University of Bern, CH-3012 Bern, Switzerland}

\emailAdd{admir.greljo@unibe.ch}
\emailAdd{ajdin.palavric@unibe.ch}
\emailAdd{thomsen@itp.unibe.ch}

\title{\mathversion{boldsans}
Adding Flavor to the SMEFT
}

\abstract{
We study the flavor structure of the lepton and baryon number--conserving dimension-6 operators in the Standard Model effective field theory (SMEFT). Building on the work of~\cite{Faroughy:2020ina}, we define several well-motivated flavor symmetries and symmetry-breaking patterns that serve as competing hypotheses about the ultraviolet (UV) dynamics beyond the SM, not far above the TeV scale. In particular, we consider four different structures in the quark sector and seven in the charged lepton sector. The set of flavor-breaking spurions is (almost) always taken to be the minimal one needed to reproduce the observed charged fermion masses and mixings.
For each case, we explicitly construct and count the operators to the first few orders in the spurion expansion, providing ready-for-use setups for phenomenological studies and global fits.
We provide a \texttt{Mathematica} package \texttt{SMEFTflavor} \href{https://github.com/aethomsen/SMEFTflavor}{\faicon{github}} to facilitate similar analyses for flavor symmetries not covered in this work.
}

\maketitle
\newpage

\section{Introduction}
\label{sec:intro}

Our current understanding of particle physics is condensed into the Standard Model (SM). It is a quantum field theory with Poincaré spacetime symmetry and $\SU(3)_c\times \SU(2)_\LL \times \U(1)_Y $ gauge invariance.  The field content includes five different gauge representations of Weyl fermions $q_i$, $\ell_i$, $u_i$, $d_i$, $e_i$ each coming in three flavors (copies, $i=1,2,3$) and a single scalar field that condenses at the electroweak scale, breaking the gauge symmetry down to $\SU(3)_c \times \U(1)_\sscript{EM} $. The Lagrangian is constructed from all local operators consistent with the gauge symmetries up to the mass dimension four, known as the renormalizable operators. In the initial days of the SM, renormalizability was considered to be an important consistency condition for valid theories. With better understanding, it became clear that this was only a low-energy property of an effective field theory~\cite{Weinberg:1968de,Wilson:1969zs,Wilson:1970ag,Georgi:1974yf,Weinberg:1978kz,Weinberg:1979sa,Weinberg:1980wa,Georgi:1994qn,Manohar:2018aog}. With more precision and/or higher energy, we will eventually discover the leading higher-dimensional operators in the infinite series organized in the inverse powers of the new physics (NP) cutoff $\Lambda$---an almost inevitable consequence of the known shortcomings of the SM in explaining the observed universe.

The SM effective field theory (SMEFT) is a framework which has gained popularity in recent years~\cite{Buchmuller:1985jz,Giudice:2007fh,Grzadkowski:2010es,Alonso:2013hga,Jenkins:2013wua,Jenkins:2013zja,Henning:2014wua,Fuentes-Martin:2016uol,Brivio:2017vri,Celis:2017hod,Wells:2015uba,Englert:2019zmt,deBlas:2017xtg,Fuentes-Martin:2020zaz,Fuentes-Martin:2020udw,Cohen:2020qvb,Carmona:2021xtq}. The Large Hadron Collider (LHC) discovered the Higgs boson $h$~\cite{ATLAS:2012yve,CMS:2012qbp} completing the list of the light degrees of freedom in the SMEFT. The Higgs mechanism with $h$ embedded in an $\SU(2)_\LL$ doublet (the linear realization) turned out to be a decent leading-order description of the electroweak symmetry breaking. At the same time, direct searches failed to discover any new particles despite tremendous efforts, suggesting a possible mass gap between the NP and SM states: $\Lambda > G_F^{\eminus 1/2}$. With this situation, the SMEFT analyses have become standard across different sectors of high-$p_T$ phenomenology, including top quark~\cite{Buckley:2015lku,Englert:2016aei,Hartland:2019bjb,Brivio:2019ius,Durieux:2018tev,vanBeek:2019evb,Bissmann:2020mfi,Bruggisser:2021duo,Ethier:2021bye,Ellis:2020unq}, electroweak vector bosons~\cite{Ethier:2021bye,Falkowski:2015jaa,Falkowski:2016cxu,Baglio:2017bfe,Ellis:2020unq,Panico:2017frx,Grojean:2018dqj,Gomez-Ambrosio:2018pnl,Dedes:2020xmo,Efrati:2015eaa}, Higgs~\cite{Pomarol:2013zra,deBlas:2016ojx,Ethier:2021bye,deBlas:2017wmn,Ellis:2020unq,Falkowski:2015jaa,Falkowski:2014tna}, jets~\cite{Krauss:2016ely,Alte:2017pme,Hirschi:2018etq,Goldouzian:2020wdq}, and high-mass Drell-Yan~\cite{Cirigliano:2012ab, deBlas:2013qqa, Gonzalez-Alonso:2016etj, Faroughy:2016osc, Greljo:2017vvb, Cirigliano:2018dyk, Greljo:2018tzh,Bansal:2018eha,Angelescu:2020uug, Farina:2016rws, Alioli:2017nzr, Raj:2016aky, Schmaltz:2018nls, Brooijmans:2020yij,Ricci:2020xre,Fuentes-Martin:2020lea,Alioli:2017ces,Alioli:2017jdo,Alioli:2018ljm,Alioli:2020kez,Panico:2021vav,Sirunyan:2021khd,ATLAS:2021pvh,Marzocca:2020ueu,Afik:2019htr,Alves:2018krf, Greljo:2021kvv}, aiming at finding smooth and correlated deviations from the SM predictions. To this end, a global SMEFT fit would optimally summarize the knowledge about ultraviolet physics (UV) accessible to us at low energies.

The largest obstacle to such an analysis is the proliferation in the number of independent operators in the SMEFT. For instance, there are 2499 independent baryon and lepton number--conserving SMEFT operators that arise at the leading order (dimension 6)~\cite{Alonso:2013hga}.\footnote{These many new operators also results in a large number of new CP-invariants~\cite{Bonnefoy:2021tbt,Yu:2022nxj,Yu:2022ttm}.} If the field content had included only a single generation instead of three, this number would have been 59~\cite{Grzadkowski:2010es}. 
This is not surprising, as most of the parameters are flavorful already in the SM at dimension 4. While the gauge sector and the scalar potential have a total of 4 parameters (modulo the strong CP-violating term), the Yukawa sector introduces an additional 13 parameters to accommodate the charged fermion masses and the CKM mixing matrix.

In this paper, we address the flavor structures of the baryon and lepton number--conserving dimension-6 operators in the SMEFT suitable for global fits, including high-$p_T$ data. For an anarchic flavor structure, with all real and imaginary coefficients of order one, charged lepton flavor violation, neutral meson oscillations, and electric dipole moments set a lower bound on the NP scale many decades above the TeV scale~\cite{Isidori:2010kg,EuropeanStrategyforParticlePhysicsPreparatoryGroup:2019qin,Aebischer:2020dsw,Silvestrini:2018dos,Pruna:2014asa,Feruglio:2015yua}. Only a few operators contributing to those rare transitions would be relevant for the phenomenology, leaving no hope of seeing new effects in the high-$p_T$ collider physics. This would also be in conflict with resolving the Higgs mass hierarchy problem at the TeV scale or, for instance, the $B$-anomalies~\cite{Hiller:2003js,LHCb:2017avl,LHCb:2021trn,LHCb:2020lmf,LHCb:2020gog,LHCb:2020zud,LHCb:2021awg,LHCb:2021vsc,LHCb:2014cxe,LHCb:2015wdu,LHCb:2016ykl,LHCb:2021zwz,Isidori:2021vtc} and $(g-2)_\mu$ anomaly~\cite{Muong-2:2006rrc,Muong-2:2021ojo} should those prove to originate in new physics.

There is, however, no reason to expect flavor anarchy in the dimension-6 operators, given the peculiar form of the dimension-4 Yukawa interactions. Rather, we observe hierarchical masses for different generations of charged fermions with Yukawa couplings spanning six orders of magnitude and an almost diagonal CKM mixing matrix.\footnote{In this work, we consider neutrinos to be massless. They obtain a mass from a different set of operators from those considered here, such as the lepton number--violating, dimension-5 Weinberg operator. We only consider symmetry-breaking spurions that always preserve the total lepton number, thus giving no contributions to the Weinberg operator. Neutrino masses can easily be accounted for by introducing additional spurions fully contracting $L^i L^j$ terms. Due to the smallness of the neutrino masses, the contribution from these spurions to dimension-6 operators is negligible.} 
The elusive why of these curious observations has long been sought after in the form of a theory of flavor. Almost regardless of what the solution is, an explanation will endow a flavor structure on the SMEFT above the EW scale.  

The renormalizable SM without the Yukawa interactions has a large global $\U(3)^5$ flavor symmetry. The observed parameters break the symmetry in a particular way. The largest breaking is due to the top quark Yukawa down to $\U(2)^2\times \U(1) \times \U(3)^3$ subgroup. This is a good approximate symmetry of the SM, which is further broken in steps by the smaller Yukawas. Eventually, the exact (classical) accidental symmetry of the dimension-4 Lagrangian, $\U(1)_B$ in the quark sector times $\U(1)^3$ for the leptons, is recovered. 
Higher-dimensional operators typically contribute to the breaking of the aforementioned (exact or approximate) accidental flavor symmetries. 

Postulating a flavor symmetry and its breaking pattern (via a set of spurions) means making a hypothesis about UV physics. A flavor spurion can be viewed as a non-dynamical (spurious) field that transforms under a nontrivial representation of the flavor group and whose background value breaks the flavor symmetry. In other words, we imagine that a UV theory will leave imprints on the flavor structure in the low-energy effective theory. The selection rules implied by the flavor symmetry have the advantage of reducing the number of important SMEFT operators (free parameters in the global fits) by truncating the flavor-spurion expansion to a given order. Global flavor symmetries and their breaking patterns, thus, provide a good organizing principle for the SMEFT, mapping the space of theories beyond the SM into universality classes. The induced model dependence should not be viewed as a drawback but rather as an opportunity to systematically learn about the UV from the data. 

The prototypical example is the minimal flavor violation (MFV) hypothesis~\cite{DAmbrosio:2002vsn}, which is a flavor structure based on the $\U(3)^5$ flavor symmetry broken by the SM Yukawa couplings $Y_u$, $Y_d$, and $Y_e$ promoted to spurions. Numerous analyses of the low-energy flavor physics data have shown that MFV allows for the NP cutoff as low as the TeV scale~\cite{Buras:2003td,Cirigliano:2005ck,Blanke:2006ig,UTfit:2005lis,Csaki:2011ge,Fitzpatrick:2007sa,Davidson:2006bd,Buras:2010wr,Isidori:2012ts,Hurth:2008jc}. Such a flavor structure naturally arises as the low-energy limit of large classes of models, including supersymmetric models with the anomaly or gauge mediation~\cite{Giudice:1998xp,Dine:1994vc}. The drawback of MFV is the ambiguity in the power counting, given that $y_t$ is a large parameter. A prominent competitor to MFV is the $\U(2)$~\cite{Barbieri:2011ci} (or general MFV~\cite{Kagan:2009bn}) flavor structures, which also provide sufficient protection against dangerous flavor violation~\cite{Barbieri:2011fc,Blankenburg:2012nx,Barbieri:2012uh,Barbieri:2012bh,Kley:2021yhn} and allow for a low NP scale. $\U(2)^5$ is an excellent approximate symmetry of the SM broken only at the level of $\mathcal{O}(10^{-2})$. The breaking spurions are already small parameters. Thus, the power counting is more transparent. Unlike MFV, it is also a good starting point for the combined explanation of $B$-anomalies in charged and neutral currents, where the third family plays a special role~\cite{Greljo:2015mma,Barbieri:2015yvd,Buttazzo:2017ixm,Cornella:2021sby,Fuentes-Martin:2019mun,Marzocca:2021miv,Bordone:2017anc}. Other flavor structures are also studied in the literature, e.g., Refs.~\cite{Bordone:2019uzc,Kobayashi:2021pav}.

\begin{table}[t]
\centering
\scalebox{0.685}{
\begin{tabular}{|cc|cccccccccccccccc|}
\hline
\multicolumn{2}{|c|}{\multirow{2}{*}{\begin{tabular}[c]{@{}c@{}} SMEFT $\cO(1)$ terms\\ {\footnotesize(dim-6, $\Delta B  = 0$)} \end{tabular}}}   & \multicolumn{16}{c|}{Lepton sector}    \\ \cline{3-18} 
\multicolumn{2}{|c|}{}                                                               & \multicolumn{2}{c|}{$\text{MFV}_L$} & \multicolumn{2}{c|}{$\U(3)_{V}$} & \multicolumn{2}{c|}{$\U(2)^2\!\times\! \U(1)^2$} & \multicolumn{2}{c|}{$\U(2)^2$} & \multicolumn{2}{c|}{$\U(2)_{V}$} & \multicolumn{2}{c|}{$\U(1)^6$} & \multicolumn{2}{c|}{$\U(1)^3$} & \multicolumn{2}{c|}{No symm.} \\ \hline
\multicolumn{1}{|c|}{\multirow{5}{*}{\begin{tabular}[c]{@{}c@{}}Quark\\ sector\end{tabular}}} & $\text{MFV}_Q$                 & 41     & \multicolumn{1}{c|}{6}     & 45        & \multicolumn{1}{c|}{9}         & 59        & \multicolumn{1}{c|}{6}         & 62  & \multicolumn{1}{c|}{9}   & 67        & \multicolumn{1}{c|}{13}        & 81  & \multicolumn{1}{c|}{6}   & 93  & \multicolumn{1}{c|}{18}  & 207             & 132            \\ \cline{2-18} 
\multicolumn{1}{|c|}{}                              & $\U(2)^2\times \U(3)_d$  & 72     & \multicolumn{1}{c|}{10}    & 78        & \multicolumn{1}{c|}{15}        & 95        & \multicolumn{1}{c|}{10}        & 100 & \multicolumn{1}{c|}{15}  & 107       & \multicolumn{1}{c|}{21}        & 122 & \multicolumn{1}{c|}{10}  & 140 & \multicolumn{1}{c|}{28}  & 281             & 169            \\ \cline{2-18} 
\multicolumn{1}{|c|}{}                              & $\U(2)^3 \times \U(1)_{d_3}$ & 86     & \multicolumn{1}{c|}{10}    & 92        & \multicolumn{1}{c|}{15}        & 111       & \multicolumn{1}{c|}{10}        & 116 & \multicolumn{1}{c|}{12}  & 123       & \multicolumn{1}{c|}{21}        & 140 & \multicolumn{1}{c|}{10}  & 158 & \multicolumn{1}{c|}{28}  & 305             & 175            \\ \cline{2-18} 
\multicolumn{1}{|c|}{}                              & $\U(2)^3$              & 93     & \multicolumn{1}{c|}{17}    & 100       & \multicolumn{1}{c|}{23}        & 118       & \multicolumn{1}{c|}{17}        & 124 & \multicolumn{1}{c|}{23}  & 132       & \multicolumn{1}{c|}{30}        & 147 & \multicolumn{1}{c|}{17}  & 168 & \multicolumn{1}{c|}{38}  & 321             & 191            \\ \cline{2-18} 
\multicolumn{1}{|c|}{}                              & No symmetry                    & 703    & \multicolumn{1}{c|}{570}   & 734       & \multicolumn{1}{c|}{600}       & 756       & \multicolumn{1}{c|}{591}       & 786 & \multicolumn{1}{c|}{621} & 818       & \multicolumn{1}{c|}{652}       & 813 & \multicolumn{1}{c|}{612} & 906 & \multicolumn{1}{c|}{705} & 1350            & 1149           \\ \hline
\end{tabular}}
\caption{Overview of the number of independent $\cO(1)$ terms for the dimension-6 SMEFT operators ($\Delta B = 0$) for different assignments of quark and lepton symmetries considered in the paper. The left (right) entry in each column gives the number of CP even (odd) coefficients for each symmetry combination. \label{tab:intro}}
\end{table}

Building on the work of Ref.~\cite{Faroughy:2020ina}, we identify several symmetry hypotheses capable, in particular, of allowing new physics to be at the TeV scale. 
Apart from theoretical motivations such as the  Higgs hierarchy problem, we are aiming at having an interesting interplay between low and high-$p_T$ physics, making the case for global fits. 
Our goal is to explore a broad spectrum of flavor structures beyond MFV and $\U(2)^5$. We extend the previous study~\cite{Faroughy:2020ina} by including four structures in the quark sector and seven in the lepton sector, for a total of 28 different flavor structures. 
The set of flavor-breaking spurions is taken to be the minimal one needed to reproduce the observed charged fermion masses and mixings (in some cases, other interesting spurions have been included for completeness). We benchmark the lepton sector densely compared to the quark sector. One motivation to do so is to pave the path for exploration of a broader class of models addressing only the neutral current $B$-anomalies and/or $(g-2)_\mu$ anomaly (see e.g.~\cite{Greljo:2021xmg,Greljo:2021npi,Davighi:2022qgb,Isidori:2021gqe}). Unlike the CKM matrix of the quark sector, the PMNS matrix has only been probed by neutrino oscillation data and could come from physics at a very high scale completely decoupled from the low-energy flavor structure of the charged leptons. Without any firm clues as to the underlying flavor structure, we are open to many possibilities. 
For each of the flavor structures, we determine and provide the full basis of dimension-6 SMEFT operators beyond NLO in the spurion expansion. 

A highlight of this work is Table~\ref{tab:intro} summarizing the number of independent $\Delta B = 0$ operators at dimension-6 in the SMEFT for different flavor symmetry assumptions in the absence of spurions (i.e., $\mathcal{O}(1)$ terms). 
A general expectation is that low-energy flavor physics will not be very sensitive to these $\mathcal{O}(1)$ operators thanks to the flavor symmetry protection leaving room for high-$p_T$ phenomenology~\cite{Aoude:2020dwv} (see, however, Ref.~\cite{Brod:2014hsa,Bobeth:2015zqa}). The other operators, which come with the flavor-spurion suppression, will be constrained chiefly by low-energy flavor physics. 
For any of the symmetries in Table~\ref{tab:intro}, if an operator containing fermions turns out to be Hermitian, it is counted in the left column since it introduces a single real parameter. Otherwise, it appears both in the left and right columns for the real and imaginary parameters, respectively (some care has to be taken when connecting the imaginary parameters with CP violation~\cite{Bonnefoy:2021tbt}). Additionally, there are always 9 CP-even (added to the left column) and 6 CP-odd (added to the right column) parameters for the pure bosonic dimension-6 operators, which do not carry any flavor. Table~\ref{tab:intro} illustrates how the number of independent parameters gradually increases with smaller (less restrictive) symmetries. Different flavor symmetries are considered to finely chart the space of SMEFT operators.
For reference, these are compared with the no-symmetry case shown in the last row and column for quarks and leptons, respectively. All the examples have considerably fewer $\mathcal{O}(1)$ parameters than the anarchic case, making global fits more feasible. During  this project, we developed a {\tt Mathematica} package \texttt{SMEFTflavor} to automatically construct the SMEFT operators given a flavor group (for details, see Appendix~\ref{sec:tool}). Should the user have a different symmetry  group or breaking spurions in mind, the package can be downloaded at the \texttt{github} page \href{https://github.com/aethomsen/SMEFTflavor}{\faicon{github}}.

The paper is organized as follows: In Section~\ref{sec:quark}, we first discuss the decomposition and the counting of the pure quark operators, while in Section~\ref{sec:leptonic}, we consider pure lepton operators. Along the way, we always present an explicit parametrization of the spurions and Wilson coefficients ready to be employed in phenomenological studies. The Warsaw basis used throughout this work is summarized in Appendix~\ref{app:Warsaw}. The {\tt Mathematica} package is documented in Appendix~\ref{sec:tool}. Appendix~\ref{sec:mixed} lists the counting of the mixed quark-lepton operators for all 28 flavor structures,  while Appendix~\ref{app:d} is reserved for useful group identities. We conclude in Section~\ref{sec:conc}.

\section{Quark Sector}
\label{sec:quark}

The kinetic Lagrangian is invariant under flavor rotations of the matter, that is, the unitary transformation between fields in the same gauge and Lorentz representations. In the SM quark sector, these transformations make up the group $ G_Q = \U(3)_q \times \U(3)_u \times \U(3)_d  $. Assuming/imposing a flavor structure of NP then comes down to proposing a finite set of spurions transforming in a subgroup of $ G\subset G_Q $, such that the whole SMEFT Lagrangian is invariant under $ G $. These assumptions can severely limit the number of operators, which can occur to a given order in the spurion expansion and will, generically, impose correlation between various operators.

In the SM, the $ G_Q $ quark flavor symmetry is broken classically by the two quark Yukawa coupling matrices $ Y_{u,d} $:
\begin{equation}\label{eq:YukLag}
    \cL\supset - \bar{q}_\LL Y_d d_\RR H - \bar{q}_\LL Y_u u_\RR \tilde{H} \hc
\end{equation}
Formally, these couplings can be considered spurions transforming as $ Y_u \sim (\rep{3},  \repbar{3}, \rep{1}) $ and $ Y_d \sim (\rep{3}, \rep{1},  \repbar{3}) $, such as to leave the SM Lagrangian invariant under flavor rotations.\footnote{Here we write only the $\SU(3)$ representation but assume that the global $\U(1)$ charges are also defined by Eq.~\eqref{eq:YukLag}. There is no difference between $\U(3)$ and its $\SU(3)$ subgroup for dimension-6 baryon number conserving operators. The global $\U(1)$ charges will however play an important role for $\U(2)$ versus $\SU(2)$ symmetries. }
MFV~\cite{DAmbrosio:2002vsn} is a popular framework that assumes that the SM contains the full flavor information of the underlying theory; no additional flavored spurions are introduced in the UV, and all other couplings are singlet under $ G_Q $. Accordingly, all flavor structure is contained in $ Y_u $ and $ Y_d $. 
In this framework, flavor violation in operators, whether fundamental or effective, can only occur through some combination of the $ Y_u, Y_d $ spurions.
This strongly constrains flavor transitions due to NP and presents a mechanism to avoid the strong flavor bounds from FCNC processes. 

If we wish to classify the SMEFT operators consistent with MFV, it is important to consider an organizing principle. One challenge is that, e.g., $ (Y_u^\dagger Y_u)^{n\geq 0} $ all transform in a similar manner, thus operators can always be dressed with higher powers of $ Y_{u}^\dagger Y_{u} $. However, not all of these are independent. In fact, three of these are enough to span the space, and higher powers can be absorbed into the coefficients of the operators with lower powers: a finite set is sufficient to capture all physics. 
A proper organizing principle exists when the spurions are small (e.g. if $ Y_u $ always comes with a small parameter $ \varepsilon_u\ll 1 $), and the MFV operators can be organized by powers of the spurions.    
This naive expansion in powers of $ Y_{u,d} $ is not necessarily possible since $ y_t \sim 1 $, and in 2HDM type models, even $ y_b $ can be order 1. The authors of Ref.~\cite{Kagan:2009bn} were able to show that non-linearly realized MFV, where a power expansion is impossible, can be effectively captured as a special case of the later, much acclaimed $ \U(2)^3 $ flavor symmetry~\cite{Barbieri:2011ci}. 

Here we consider a spectrum of viable flavor symmetries: 
	\begin{enumerate}[i)]
	\item $ G = \U(2)^3 $ decouples the third-generation quarks entirely, yet it gives decent protection against FCNCs.
	\item $ G = \U(2)^3 \times \U(1)_b $ decouples only the third generation of down-quarks and keeps $ y_b $, a spurion of $ \U(1)_b $, perturbatively small.
	\item $ G= \U(2)^2 \times \U(3)_d $ for when there is no suppression of $ y_t\simeq 1 $ in the SMEFT operators. The enhanced symmetry allows for a spurion expansion of all but the top quark.  
	\item $ G= \U(3)^3 $ linearly realized MFV, provides strong constraints on NP and effectively protects against NP contributions to rare SM processes.  
	\end{enumerate}

In this section, we explore these 4 different flavor structures for the quark sector. In each case, we will assume that a perturbative expansion in spurion insertions is possible. For each symmetry, we provide a parametrization of the spurions, list all flavor contractions that can occur up to dimension 6 in the SMEFT, and finally provide a counting of the quark operators at dimension 6.

\subsection{$ \U(2)^3 $ symmetry} \label{sec:quark_U2^3}
We assume that the NP posses a symmetry $ G = \U(2)_q \times \U(2)_u \times \U(2)_d \subset G_Q $, under which the SM quarks decompose as
\begin{align}
	 q &=\begin{bmatrix} q^a\sim(\rep{2},\rep{1},\rep{1}) \\
	 q_3 \sim(\rep{1},\rep{1},\rep{1}) \end{bmatrix}, &
	 u &=\begin{bmatrix}u^a\sim(\rep{1},\rep{2},\rep{1}) \\
	 u_3 \sim(\rep{1},\rep{1},\rep{1}) \end{bmatrix}, &
	 d &=\begin{bmatrix}d^a\sim(\rep{1},\rep{1},\rep{2})\\
	 d_3\sim(\rep{1},\rep{1},\rep{1}) \end{bmatrix}.
\end{align}
The minimal set of spurions needed to reproduce the SM masses and CKM matrix is  
	\begin{equation}
	V_q \sim (\rep{2},\rep{1},\rep{1})~, \qquad \Delta_u \sim (\rep{2}, \repbar{2},\rep{1})~, \qquad \Delta_d \sim (\rep{2},\rep{1}, \repbar{2})~.
	\end{equation}
These spurions generally allow for a slew of Yukawa operators, which contributes to the Yukawa coupling matrices as 
	\begin{equation}
	Y_{u,d} = \begin{bmatrix}
	a^{u,d}_{1} \Delta_{u,d} + a^{u,d}_{2} \Delta_{u} \Delta_{u}^\dagger \Delta_{u,d} + \ldots &
	b^{u,d}_{1} V_q + b^{u,d}_{2} \Delta_{u} \Delta_{u}^\dagger V_q + \ldots \\
	c^{u,d}_{1} V_q^\dagger \Delta_{u,d} + \ldots &
	d^{u,d}_{1} + d^{u,d}_{2} V_q^\dagger V_q + \ldots
	\end{bmatrix}
	\end{equation}
for $ \mathcal{O}(1) $ parameters $ a^{u,d}_{n}, \ldots d^{u,d}_{n}$, parametrizing all covariant combinations of the spurions at each entry in the coupling matrix. 		

We now wish to point out a redundancy, which to our knowledge, has been overlooked in previous literature on the $ \U(2)^3 $ flavor symmetry: despite the assumption of the $ G $ symmetry, the quark kinetic terms are, nevertheless, invariant under all $ G_Q $ transformations.
In particular, rotations from the $ G_Q/G $ coset space have not been utilized because generic field transformations of this kind would ruin the explicit $ G $ invariance of the model. The spurions, however, allow for the construction of covariant $ G_Q/G $ rotation matrices
	\begin{equation}
	\begin{split}
	U_q &= \exp \! \begin{bmatrix}
	0 & 
	\lambda^q_1 V_q + \lambda^q_2 \Delta_{u} \Delta_{u}^\dagger V_q + \ldots \\
	-(\lambda^{q}_1)^\ast V_q^\dagger - (\lambda^{q}_2)^\ast V_q^\dagger \Delta_{u} \Delta_{u}^\dagger - \ldots 
	& 0
	\end{bmatrix}, \\
	U_{u,d} &= \exp \! \begin{bmatrix}
	0 &
	\lambda^{u,d}_1 \Delta_{u,d}^\dagger V_q + \ldots \\
	-(\lambda^{u,d}_1)^\ast V_q^\dagger \Delta_{u,d} - \ldots & 0 
	\end{bmatrix}.
	\end{split}
	\end{equation}
For a suitable choice of $ \lambda^{f}_{n} $, a field transformation will bring the Yukawa couplings on the form\footnote{This can be seen order by order in the perturbative expansion.}
	\begin{equation}
	\begin{split}
	Y'_{u} &= \begin{bmatrix}
	a^{u\prime}_{1} \Delta_{u} + a^{u\prime}_{2} \Delta_{u} \Delta_{u}^\dagger \Delta_{u} + \ldots &
	b^{u\prime}_{1} V_q + b^{u\prime}_{2} \Delta_{u} \Delta_{u}^\dagger V_q + \ldots \\
	0&
	d^{u\prime}_{1} + d^{u\prime}_{2} V_q^\dagger V_q + \ldots
	\end{bmatrix},\\
	Y'_{d} &= \begin{bmatrix}
	a^{d\prime}_{1} \Delta_{d} + a^{d\prime}_{2} \Delta_{u} \Delta_{u}^\dagger \Delta_{d} + \ldots &
	0 \\
	0&
	d^{u\prime}_{1} + d^{u\prime}_{2} V_q^\dagger V_q + \ldots
	\end{bmatrix}.
	\end{split}
	\end{equation}	
Finally, we can do a redefinition of the spurions (in the event that $ a_{1}^{u \prime }, a_{1}^{d \prime }, b_{1}^{u \prime } $ are not $ \mathcal{O}(1) $, it can be necessary to maintain explicit real coefficients multiplying the new spurions to maintain the perturbative spurion expansion) to write 
	\begin{equation} \label{eq:yuk_3U(2)}
	Y'_{u} = \begin{bmatrix}
	\Delta_{u}  & V_q\\
	0& y_t
	\end{bmatrix}, \qquad 
	Y'_{d} = \begin{bmatrix}
	\Delta_{d} & 0 \\
	0& y_b
	\end{bmatrix}.
	\end{equation}
This spurion redefinition shuffles the coefficients of the higher-dimension SMEFT operators but otherwise does not change the spurions being generic matrices. $ G_Q/G $ contains a final phase transformation of the third generation quarks, which allows for setting $ y_t, y_b $ to be real parameters.   
The Yukawa matrices~\eqref{eq:yuk_3U(2)} are, thus, completely generic representations in the $ \U(2)^3 $ symmetry. This parametrization effectively breaks $ G_Q \to G \times \U(1)_B $, where the spurions transform under the $ G $ symmetry. 

We next consider how the spurions break the $ G $ symmetry and how the broken symmetry allows us to choose a minimal parametrization of the spurions. For simplicity, we consider the realistic case where the singular values of each of the spurions are finite and non-degenerate in agreement with current data.\footnote{Such degeneracy could cause an enhanced remnant symmetry (see, e.g., Ref.~\cite{Bonnefoy:2021tbt}).} 
First, fixing the $ \U(2)_q $ doublet gives\footnote{To see what parameters can be removed from the spurions, it is useful to write them in terms of their singular value decomposition such that the flavor rotations can directly remove phases and angles from the left and right rotation matrices.}
	\begin{align}
	V_q &\longrightarrow \begin{bmatrix} 0\\ \epsilon_q \end{bmatrix}:  
	&&\U(2)_q \longrightarrow \U(1)_{q_1}. \label{eq:Vq,U(2)^3}
\intertext{Next $ \Delta_u $, breaks}
	\Delta_u &\longrightarrow\begin{bmatrix}	c_u & -s_u \\ s_u & c_u	\end{bmatrix}
		\begin{bmatrix} \delta_u & 0\\ 0& \delta_u' \end{bmatrix}: &&
	\U(1)_{q_1} \times \U(2)_u \longrightarrow \emptyset,
\intertext{where we adopt the notation $ s_u, c_u$ for sine and cosine of the same angle, and}
	\Delta_d &\longrightarrow \begin{bmatrix}	c_d & -s_d e^{i\alpha} \\ s_d e^{-i\alpha} & c_d \end{bmatrix}
		\begin{bmatrix} \delta_d &0 \\ 0& \delta_d' \end{bmatrix}: 
	&&\U(2)_d \longrightarrow \emptyset. \label{eq:Dd,U(2)^3}
	\end{align} 
The complete breaking of $ G\to \emptyset $ by the spurions makes it possible to remove 12 unphysical parameters from the spurions, reducing the naive 10 complex parameters down to a total of 5 real positive parameters, 2 mixing angles, and 1 phase. At dimension 4, together with $y_b$ and $y_t$, these give the quark masses and the CKM mixing matrix. The breaking pattern and utilization of the full group of $ G_Q $ transformations at dimension 4 means that all coefficients of the baryon number--conserving SMEFT operators are physical.\footnote{Some of these coefficients are unphysical when using the redundant spurion parametrization of Ref.~\cite{Faroughy:2020ina}. With this in mind, the parametrization in Ref.~\cite{Faroughy:2020ina} can still be useful in model building.}

For convenience, we compute numerical values for the relevant parameters reproducing the observed CKM mixing matrix~\cite{ParticleDataGroup:2020ssz} and quark masses (taken from Ref.~\cite{Martin:2019lqd} at the renormalisation scale set to $m_t$) from the Yukawa terms in Eq.~\eqref{eq:yuk_3U(2)} at tree level:\footnote{A suitable parameter point can be found by going to the down aligned basis, which allows for relating $ \delta_d^{(\prime)} $ and $ y_b$ directly to the down-type Higgs Yukawas. After this, the remaining parameters can be determined from $ V_\sscript{CKM}^\dagger 
\hat{Y}^2_u V_\sscript{CKM} = \Lambda Y_u Y_u^\dagger \Lambda^\dagger$. Here the left-hand side consists of the observed up-type Yukawa couplings (diagonal) and the CKM matrix, while the right-hand side consists of our Yukawa matrix in the down-aligned basis and $ \Lambda = \diag(e^{i \theta}, \, e^{i \phi},\, e^{\eminus i (\theta+\phi)}) $ are additional unphysical phases. This provides a total of 9 constraints fixing all 9 parameters of the right-hand side.}
    \begin{align}\label{eq:U(2)numerical}
    \delta_d &= \num{1.46e-5}, & \delta_d' &= \num{2.91e-4}, & y_b &= 0.0155, \nonumber\\
    \delta_u &= \num{6.72e-6}, & \delta_u' &= \num{3.38e-3}, & y_t &= 0.934, \\
    \epsilon_q &= 0.0380, & \theta_d &= 0.210, & \theta_u &= 0.0888, & \alpha &= -1.57. \nonumber
    \end{align}
Although there are corrections from higher order terms and radiative corrections, this provides a decent estimate. As already anticipated, the largest breaking of the symmetry occurs at $\mathcal{O}(10^{-2})$ by $\epsilon_q$, while the symmetry-allowed parameter $y_t $ is $ \mathcal{O}(1)$. In other words, the approximate $\U(2)^3$ successfully explains these features. However, it fails to explain the smallness of $y_b$ and the hierarchy in $\Delta_{u,d}$. This flavor structure also gives a suppression in FCNCs due to the smallness of spurions and allows for a new physics scale not far above the TeV scale~\cite{Barbieri:2011fc,Blankenburg:2012nx,Barbieri:2012uh,Barbieri:2012bh,Kley:2021yhn}. Since this is the least restrictive symmetry in the quark sector we consider, similar (stronger) conclusions will hold in the subsequent three sections as well.

We determine the dimension-6 SMEFT operators allowed by the flavor structure up to several insertions of the spurions. We base this on the Warsaw basis for the SMEFT (cf. Appendix~\ref{app:Warsaw}), where we identify the unique fermion combinations that appear, the bilinear and quartic structures. We present the possible flavor contractions (including spurions) for these field combinations, and these, in turn, can be directly inserted back into the appropriate Warsaw basis operators to recover the full set of SMEFT operators compatible with the flavor structure.  
We present the decompositions of the bilinear structures in Eqs.~(\ref{eq212}--\ref{eq217}) and decompositions of the unique quartic structures in Eqs.~(\ref{eq218}--\ref{eq223}). This also includes the expanded set of structures available in case the flavor symmetry assumption is reduced to $ \SU(2)^3 $. The spurion counting of the pure quark SMEFT operators assuming $\U(2)^3$ ($\SU(2)^3$) symmetry in the quark sector is presented in Table \ref{tab:quark3U2} (\ref{tab:quark3SU2}).

\begin{table}[t]
\centering
\scalebox{0.95}{
\begin{tabular}{|cc|cc|cc|cc|cc|cc|cc|}
\hline
\multicolumn{2}{|c|}{$\U(2)_q \times \U(2)_u \times \U(2)_d$}                                 & \multicolumn{2}{c|}{$\cO(1)$} & \multicolumn{2}{c|}{$\cO(V)$} & \multicolumn{2}{c|}{$\cO(V^2)$} & \multicolumn{2}{c|}{$\cO(V^3)$} & \multicolumn{2}{c|}{$\cO(\Delta)$} & \multicolumn{2}{c|}{$\cO(\Delta V)$} \\ \hline
\multicolumn{1}{|c|}{\multirow{2}{*}{$\psi^2 H^3$}} & $Q_{uH}$                          & 1              & 1              & 1              & 1              &                 &               &                &                & 1                   & 1                  & 1                   & 1                  \\
\multicolumn{1}{|c|}{}                                  & $Q_{dH}$                          & 1              & 1              & 1              & 1              &                 &               &                &                & 1                   & 1                  & 1                   & 1                  \\ \hline
\multicolumn{1}{|c|}{\multirow{2}{*}{$\psi^2 XH$}}  & $Q_{u(G,W,B)}$        & 3              & 3              & 3              & 3              &                 &               &                &                & 3                   & 3                  & 3                   & 3                  \\
\multicolumn{1}{|c|}{}                                  & $Q_{d(G,W,B)}$        & 3              & 3              & 3              & 3              &                 &               &                &                & 3                   & 3                  & 3                   & 3                  \\ \hline
\multicolumn{1}{|c|}{\multirow{3}{*}{$\psi^2 H^2 D$}}   & $Q_{Hq}^{(1,3)}$     & 4              &                & 2              & 2              & 2               &               &                &                &                     &                    &                     &                    \\
\multicolumn{1}{|c|}{}                                  & $Q_{Hu}$,\,$Q_{Hd}$                 & 4              &                &                &                &                 &               &                &                &                     &                    & 2                   & 2                  \\
\multicolumn{1}{|c|}{}                                  & $Q_{Hud}$                     & 1              & 1              &                &                &                 &               &                &                &                     &                    & 2                   & 2                  \\ \hline
\multicolumn{1}{|c|}{$(LL)(LL)$}                        & $Q_{qq}^{(1,3)}$     & 10             &                & 6              & 6              & 10              & 2             & 2              & 2              &                     &                    &                     &                    \\ \hline
\multicolumn{1}{|c|}{\multirow{2}{*}{$(RR)(RR)$}}       & $Q_{uu}$,\,$Q_{dd}$                 & 10             &                &                &                &                 &               &                &                &                     &                    & 6                   & 6                  \\
\multicolumn{1}{|c|}{}                                  & $Q_{ud}^{(1,8)}$     & 8              &                &                &                &                 &               &                &                &                     &                    & 8                   & 8                  \\ \hline
\multicolumn{1}{|c|}{$(LL)(RR)$}                        & $Q_{qu}^{(1,8)}$,\,$Q_{qd}^{(1,8)}$ & 16             &                & 8              & 8              & 8               &               &                &                & 4                   & 4                  & 12                  & 12                 \\ \hline
\multicolumn{1}{|c|}{$(LR)(LR)$}                     & $Q_{quqd}^{(1,8)}$                & 2              & 2              & 4              & 4              & 2               & 2             &                &                & 8                   & 8                  & 12                  & 12                 \\ \hline\hline
\multicolumn{2}{|c|}{Total}                                                                 & 63             & 11             & 28             & 28             & 22              & 4             & 2              & 2              & 20                  & 20                 & 50                  & 50                 \\ \hline
\end{tabular}}
\caption{Counting of the pure quark SMEFT operators (see Appendix \ref{app:Warsaw}) assuming $\U(2)_q \times \U(2)_u \times \U(2)_d$ symmetry in the quark sector. The counting is performed taking up to three insertions of the $V_q$ spurion, one insertion of $\Delta_{u,d}$ and one insertion of the $\Delta_{u,d}V_q$ spurion product. The left (right) entry in each column gives the number of CP even (odd) coefficients at the given order in spurion counting.}
\label{tab:quark3U2}
\end{table}

\begin{table}[t]
\centering
\scalebox{0.925}{\begin{tabular}{|cc|cc|cc|cc|cc|cc|cc|}
\hline
\multicolumn{2}{|c|}{$\SU(2)_q \times \SU(2)_u \times \SU(2)_d$}                          & \multicolumn{2}{c|}{$\cO(1)$} & \multicolumn{2}{c|}{$\cO(V)$} & \multicolumn{2}{c|}{$\cO(V^2)$} & \multicolumn{2}{c|}{$\cO(V^3)$} & \multicolumn{2}{c|}{$\cO(\Delta)$} & \multicolumn{2}{c|}{$\cO(\Delta V)$} \\ \hline
\multicolumn{1}{|c|}{\multirow{2}{*}{$\psi^2 H^3$}}   & $Q_{uH}$                          & 1             & 1             & 2             & 2             &                &                &                &                & 2                & 2               & 4                 & 4                \\
\multicolumn{1}{|c|}{}                                & $Q_{dH}$                          & 1             & 1             & 2             & 2             &                &                &                &                & 2                & 2               & 4                 & 4                \\ \hline
\multicolumn{1}{|c|}{\multirow{2}{*}{$\psi^2 XH$}}    & $Q_{u(G,W,B)}$                    & 3             & 3             & 6             & 6             &                &                &                &                & 6                & 6               & 12                & 12               \\
\multicolumn{1}{|c|}{}                                & $Q_{d(G,W,B)}$                    & 3             & 3             & 6             & 6             &                &                &                &                & 6                & 6               & 12                & 12               \\ \hline
\multicolumn{1}{|c|}{\multirow{3}{*}{$\psi^2 H^2 D$}} & $Q_{Hq}^{(1,3)}$                  & 4             &               & 4             & 4             & 4              & 2              &                &                &                  &                 &                   &                  \\
\multicolumn{1}{|c|}{}                                & $Q_{Hu}$,$Q_{Hd}$                 & 4             &               &               &               &                &                &                &                &                  &                 & 8                 & 8                \\
\multicolumn{1}{|c|}{}                                & $Q_{Hud}$                     & 1             & 1             &               &               &                &                &                &                &                  &                 & 8                 & 8                \\ \hline
\multicolumn{1}{|c|}{$(LL)(LL)$}                      & $Q_{qq}^{(1,3)}$                  & 10            &               & 14            & 14            & 20             & 12             & 8              & 8              &                  &                 &                   &                  \\ \hline
\multicolumn{1}{|c|}{\multirow{2}{*}{$(RR)(RR)$}}     & $Q_{uu}$,$Q_{dd}$                 & 10            &               &               &               &                &                &                &                &                  &                 & 28                & 28               \\
\multicolumn{1}{|c|}{}                                & $Q_{ud}^{(1,8)}$                  & 8             &               &               &               &                &                &                &                &                  &                 & 32                & 32               \\ \hline
\multicolumn{1}{|c|}{$(LL)(RR)$}                      & $Q_{qu}^{(1,8)}$,$Q_{qd}^{(1,8)}$ & 16            &               & 16            & 16            & 16             & 8              &                &                & 16               & 16              & 56                & 56               \\ \hline
\multicolumn{1}{|c|}{$(LR)(LR)$}                      & $Q_{quqd}^{(1,8)}$                & 4             & 4             & 8             & 8             & 8              & 8              &                &                & 16               & 16              & 60                & 60               \\ \hline\hline
\multicolumn{2}{|c|}{Total}                                                               & 65            & 13            & 58            & 58            & 48             & 30             & 8              & 8              & 48               & 48              & 224               & 224              \\ \hline
\end{tabular}}
\caption{Counting of the pure quark SMEFT operators (see Appendix \ref{app:Warsaw}) assuming $\SU(2)_q \times \SU(2)_u \times \SU(2)_d$ symmetry in the quark sector. The counting is performed taking up to three insertions of the $V_q$ spurion, one insertion of $\Delta_{u,d}$ and one insertion of the $\Delta_{u,d}V_q$ spurion product. The left (right) entry in each column gives the number of CP even (odd) coefficients at the given order in spurion counting. Due to the presence of the additional $\SU(2)$ structures in the decompositions, the counting is different compared to Table \ref{tab:quark3U2}.}
\label{tab:quark3SU2}
\end{table}

\subsubsection*{Decomposition of bilinear structures}
In this section, we present the construction of bilinear structures invariant under the $\U(2)^3$ flavor symmetry, starting with the $\cO(1)$ structures. Since $q$, $u$, and $d$ all decompose as $\rep{2}_{q,u,d} \oplus \rep{1}$, respectively, under $\U(2)^3$ group, the $\cO(1)$ bilinears can be formed either by contracting two doublets or singlets of the same field or by contracting singlets of the different fields. By doing this, we end up with nine $\cO(1)$ bilinears: $(\bar{q}q)$, $(\bar{q}_3 q_3)$, $(\bar{u}u)$, $(\bar{u}_3 u_3)$, $(\bar{d}d)$, $(\bar{d}_3 d_3)$, $(\bar{u}_3 d_3)$, $(\bar{q}_3 u_3)$ and $(\bar{q}_3 d_3)$.

There are only three bilinears that can be formed with one insertion of the $V_q$ spurion. All of these include the singlet contraction with the quark doublet and one additional singlet from the field decomposition: $(\bar{q}V_q q_3)$, $(\bar{q}V_q u_3)$ and $(\bar{q}V_q d_3)$. However, in the case of $\SU(2)^3$, there are additional bilinears, such as $\opemph{ V_q^a \varepsilon_{ab} (\bar{q}_3 q^b ) }$.
Following analogous reasoning for the case of $\cO(V^2)$ bilinears, we obtain $(\bar{q}V_q V_q^\dag q)$ structure for the $ \U(2)^3 $ case and ${\color{blueRef}\varepsilon_{bc}( \bar{q}V_q V_q^c q^b )}$ for the $\SU(2)^3$.

Before we proceed with our discussion about $\cO(\Delta)$ and $\cO(\Delta V)$ bilinears, let us introduce the shorthand notation
	\begin{equation}
	(\widetilde{\Delta}_{u,d})\ud{a}{d} = \varepsilon^{ab} (\Delta^{\ast}_{u,d})\du{b}{c} \varepsilon_{cd} \sim (\rep{2}, \repbar{2}, \rep{1}),~ (\rep{2}, \rep{1}, \repbar{2}),
	\end{equation}
which proves to be useful in constructing $\SU(2)^3$ invariant structures.	
The $ \mathcal{O}(\Delta) $ bilinears are formed with the $(\bar{q}\Delta_{u,d})_a$ contractions, which can then be contracted to $u^a$ or $d^a$ yielding two bilinears $(\bar{q} \Delta_u u )$ and $(\bar{q} \Delta_d d )$. In case of $\SU(2)^3$ we get two additional ones given by $\opemph{(\bar{q} \widetilde{\Delta}_u u )}$ and $\opemph{(\bar{q} \widetilde{\Delta}_d d )}$.

For the remaining $ \mathcal{O}(\Delta V)$ bilinears, we observe that the contractions $(V_q^\dag \Delta_{u,d})_a$ transform in the anti-fundamental representation of $\U(2)_{u,d}$ meaning they can form singlets when contracted to $u^a$ or $d^a$.  
We find six such structures: $(\bar{u} \Delta_u^\dagger V_q u_3 )$, $(\bar{d} \Delta_d^\dagger V_q d_3 )$, $(\bar{u} \Delta_u^{\dag} V_q   d_3)$, $(\bar{u}_3 V_q^{\dag} \Delta_d  d)$, $(\bar{q}_3 V_q^\dagger \Delta_u u )$ and $(\bar{q}_3 V_q^\dagger \Delta_d d )$. The complete list of bilinears is presented below, and the new structures that appear in case of $\SU(2)^3$ symmetry are denoted in {\color{blueRef} blue}:\\\\
\noindent $\boxed{(\bar{q} q)}$
	\begin{equation}
	\label{eq212}
	\begin{alignedat}{4}
	& \ord{1}:~  (\bar{q} q)~, \quad (\bar{q}_3 q_3)~,\qquad \ord{V}:~ (\bar{q} V_q q_3 )~, \quad \opemph{ V_q^a \varepsilon_{ab} (\bar{q}_3 q^b ) }~, \quad \hermc~,\\ 
	&\ord{V^2}:~  (\bar{q}  V_q V_q^\dagger q )~,\quad \lzs {\color{blueRef}\varepsilon_{bc}( \bar{q}V_q V_q^c q^b )}~,\quad\hermc \dzs~.
	\end{alignedat}
	\end{equation}
$\boxed{(\bar{u} u)}$
	\begin{equation}
	\begin{alignedat}{4}
	& \ord{1}:\quad &&(\bar{u} u)~, \quad (\bar{u}_3 u_3)~, \\
	&\ord{\Delta V}:\quad   &&(\bar{u} \Delta_u^\dagger V_q u_3 )~,\quad \opemph{(\bar{u}_a u_3) \varepsilon^{ab} (V_q^\dagger \Delta_u)_b }~,\quad {\color{blueRef}\varepsilon^{ad}\varepsilon_{bc}[\bar{u}^a V_q^b (\Delta_u)\ud{c}{d} u_3]}~,\quad \hermc~,\\
	&&&{\color{blueRef}\varepsilon_{bc}[ \bar{u}_3 V_q^b (\Delta_u)\ud{c}{a} u^a ]}~,\quad     \hermc~.
	\end{alignedat}
	\end{equation}
$\boxed{(\bar{d} d)}$
	\begin{equation}
	\begin{alignedat}{4}
	& \ord{1}:\quad &&(\bar{d} d)~, \quad (\bar{d}_3 d_3)~,  \\
	&\ord{\Delta V}:\quad  &&(\bar{d} \Delta_d^\dagger V_q d_3 )~,\quad \opemph{(\bar{d}_a d_3) \varepsilon^{ab} (V_q^\dagger \Delta_d)_b}~,\quad {\color{blueRef}\varepsilon^{ad}\varepsilon_{bc}[ \bar{d}^a V_q^b (\Delta_d)\ud{c}{d}d_3 ]}~, \quad \hermc~,\\
	&&&{\color{blueRef}\varepsilon_{bc}[ \bar{d}_3 V_q^b (\Delta_d)\ud{c}{a} d^a ]}~,\quad \hermc~.
	\end{alignedat}
	\end{equation}
$\boxed{(\bar{u}d)}$\\
\begin{equation}
    \begin{alignedat}{2}
        &\mathcal{O}(1):\quad &&(\bar{u}_3 d_3)~, \\
        &\mathcal{O}(\Delta V):\quad && (\bar{u} \Delta_u^{\dag} V_q   d_3)~,\quad (\bar{u}_3 V_q^{\dag} \Delta_d  d)~,\quad \opemph{(\bar{u}_a d_3) \varepsilon^{ab} (V_q^\dagger \Delta_u)_b }~,\quad \opemph{(\bar{u}_3 d^a) \varepsilon_{ab} (V_q \Delta_d^\dag)^b }~,\quad \\&&&{\color{blueRef}\varepsilon^{ad}\varepsilon_{bc}[ \bar{u}_a V_q^b (\Delta_u)\ud{c}{d}d_3 ]}~,\quad {\color{blueRef}\varepsilon^{bc}\varepsilon_{ad}[ \bar{u}_3 (V_q^*)_b (\Delta_d^*)\du{c}{d} d^a ]}~,\quad{\color{blueRef}\varepsilon^{bc}[ \bar{u}_a (V_q^*)_b (\Delta_u^*)\du{c}{a} d_3 ]}~,\\
        &&&{\color{blueRef}\varepsilon_{bc}[ \bar{u}_3 V_q^b (\Delta_d)\ud{c}{a} d^a ]}~.
    \end{alignedat}
\end{equation}
$\boxed{(\bar{q} u)}$
	\begin{equation}
	\begin{alignedat}{4}
	 & \ord{1}:~ (\bar{q}_3 u_3)~,\qquad \ord{V}:~ (\bar{q} V_q u_3 )~, \quad \opemph{ (V^\ast_q)_a \varepsilon^{ab} (\bar{q}_b u_3 ) }~,\quad  \\
	 & \ord{\Delta}:~  (\bar{q} \Delta_u u )~, \quad \opemph{(\bar{q} \widetilde{\Delta}_u u )}~,\\
	 & \ord{\Delta V}:~ (\bar{q}_3 V_q^\dagger \Delta_u u )~, \quad \opemph{(\bar{q}_3 V_q^\dagger \widetilde{\Delta}_u u )}~,\quad {\color{blueRef}\varepsilon_{bc}[ \bar{q}_3 V_q^b (\Delta_u)\ud{c}{a}u^a ]}~,\quad {\color{blueRef}\varepsilon_{ac}[ \bar{q}_3 V_q^b (\Delta_u^*)\du{b}{c} u^a ]}~.
	\end{alignedat}
	\end{equation}
$\boxed{(\bar{q} d)}$
	\begin{equation}
	\label{eq217}
	\begin{alignedat}{4}
	 & \ord{1}:~   (\bar{q}_3 d_3)~,\qquad \ord{V}:~ (\bar{q} V_q d_3 )~, \quad \opemph{ (V^\ast_q)_a \varepsilon^{ab} (\bar{q}_b d_3 ) }~,\quad   \\
	 & \ord{\Delta}:~  (\bar{q} \Delta_d d )~, \quad \opemph{(\bar{q} \widetilde{\Delta}_d d )}~,\\ 
	 & \ord{\Delta V}:~  (\bar{q}_3 V_q^\dagger \Delta_d d )~, \quad \opemph{(\bar{q}_3 V_q^\dagger \widetilde{\Delta}_d d )}~,\quad {\color{blueRef}\varepsilon_{bc}[ \bar{q}_3 V_q^b (\Delta_d)\ud{c}{a} d^a ]}~,\quad {\color{blueRef}\varepsilon_{ac}[ \bar{q}_3 V_q^b (\Delta_d^*)\du{b}{c} d^a ]}~.
	\end{alignedat}
	\end{equation}

\subsubsection*{Decomposition of quartic structures}
Let us continue with the construction of the quartic structures. In what follows, we focus on the unique, non-factorizable\footnote{Epithets `unique' and `non-factorizable' are used interchangeably when dealing with the quartic structures. This nomenclature refers simply to the quartic structures that cannot be formed as a product of two factorizing bilinears fully invariant under the discussed flavor group. Needless to say, the final spurion counting of the SMEFT operators is performed taking the full set of quartic structures.} structures only. Starting with $\cO(1)$ structures, we follow a similar reasoning as in the case of bilinears, obtaining six structures: $(\bar{q}_a q^b) (\bar{q}_b q^a)$, $(\bar{q}_a q_3) (\bar{q}_3 q^a)$, $(\bar{u}_a u^b) (\bar{u}_b u^a)$, $(\bar{u}_a u_3) (\bar{u}_3 u^a)$, $(\bar{d}_a d^b) (\bar{d}_b d^a)$ and $(\bar{d}_a d_3) (\bar{d}_3 d^a)$. In the case of $\SU(2)^3$ symmetry, only one additional $\cO(1)$ structure appears: $\opemph{(\bar{q}_a u_3) \varepsilon^{ab} (\bar{q}_b d_3)}$.

At $\cO(V)$, the $\U(2)^3$ and $\SU(2)^3$ unique structures are $(\bar{q}_a q_3) (\bar{q} V_q q^a)$, ${\color{blueRef}(\bar{q}_3 q^a)(\bar{q}_a \varepsilon_{bc} V_q^c q^b)}$ and ${\color{blueRef}(\bar{q}_3 q^a)(\bar{q}V_q \varepsilon_{ac}q^c)}$, while at $\cO(V^2)$ there is only one structure of the form $(\bar{q}_a V_q^\dagger q) (\bar{q} V_q q^a)$.  
With one insertion of $\Delta$ spurion, there are four $\U(2)^3$ unique ones: $(\bar{q}_a V_q^\dagger q) (\Delta_u)\ud{a}{b} (\bar{u}_3 u^b)$, $(\bar{q}_a q_3) (\Delta_d)\ud{a}{b} (\bar{d}_3 d^b)$, $(\bar{q}_a u_3) (\Delta_d)\ud{a}{b} (\bar{q}_3 d^b)$ and $(\bar{q}_3 u^a) (\Delta_u)\ud{b}{a} (\bar{q}_b d_3)$.

With the insertion of both $\Delta_{u,d}$ and $V_q$ spurions, we obtain six $\cO(\Delta V)$ $\U(2)^3$ structures given by $(\bar{u}_a u_3) (\bar{u} \Delta_u^\dagger V_q u^a)$, $(\bar{d}_a d_3) (\bar{d} \Delta_d^\dagger V_q  d^a)$, $(\bar{q}_a V_q^\dagger q) (\Delta_u)\ud{a}{b} (\bar{u}_3 u^b)$, $(\bar{q}_a V_q^\dagger q) (\Delta_d)\ud{a}{b} (\bar{d}_3 d^b)$, $(\bar{q}_a u_3) (\Delta_d)\ud{a}{b} (\bar{q}  V_q  d^b)$ and $(\bar{q} V_q  u^a) (\Delta_u)\ud{b}{a} (\bar{q}_b d_3)$. There are, however, plenty of new $\SU(2)^3$ unique structures that emerge at both $\cO(\Delta)$ and $\cO(\Delta V) $. The complete list is presented below and the $\SU(2)^3$ structures are denoted in \opemph{blue}:\\

\noindent $\boxed{(\bar{q} q) (\bar{q} q)}$
	\begin{equation}
	\label{eq218}
	\begin{alignedat}{4}
	& \ord{1}:~   &&(\bar{q}_a q^b) (\bar{q}_b q^a)~, \quad (\bar{q}_a q_3) (\bar{q}_3 q^a)~,\\ 
	&\ord{V}:~   &&(\bar{q}_a q_3) (\bar{q} V_q q^a)~, \quad {\color{blueRef}(\bar{q}_3 q^a)(\bar{q}_a \varepsilon_{bc} V_q^c q^b)}~,\quad {\color{blueRef}(\bar{q}_3 q^a)(\bar{q}V_q \varepsilon_{ac}q^c)}~,\quad \hermc~,\\ 
	& \ord{V^2}: ~   &&(\bar{q}_a V_q^\dagger q) (\bar{q} V_q q^a)~.
	\end{alignedat}
	\end{equation}

\noindent $\boxed{(\bar{u} u) (\bar{u} u)}$
\begin{equation}
    \small
	\begin{alignedat}{4}
	& \ord{1}:~ && (\bar{u}_a u^b) (\bar{u}_b u^a)~, \quad (\bar{u}_a u_3) (\bar{u}_3 u^a)~,\\
	& \ord{\Delta V}:~ && (\bar{u}_a u_3) (\bar{u} \Delta_u^\dagger V_q u^a)~,\quad 
	{\color{blueRef}(\bar{u}_a u_3)\varepsilon^{ab}\varepsilon_{de}[ \bar{u}_b V_q^d (\Delta_u)\ud{e}{c} u^c ]}~,\quad 
	{\color{blueRef}\varepsilon^{be}\varepsilon_{cd}(\bar{u}_a u_3)[ \bar{u}_b V_q^c (\Delta_u)\ud{d}{e}u^a ]}~,\quad \hermc~,\\
	&&& {\color{blueRef}(\bar{u}_3 u^a)[ \bar{u}_a V_q^c \varepsilon_{cd}(\Delta_u)\ud{d}{b} u^b ]}~,\quad 
	{\color{blueRef}(\bar{u}_3 u^a)  [\bar{u}_a \varepsilon_{bd}V_q^c (\Delta_u^*)\du{c}{d} u^b]}~,\quad
	{\color{blueRef}\varepsilon_{ac}(\bar{u}_3 u^a)[ \bar{u}_b V_q^d (\Delta_u^*)\du{d}{b} u^c ]}~,\quad \hermc~.
	\end{alignedat}
	\end{equation}
\noindent $\boxed{(\bar{d} d) (\bar{d} d)}$
	\begin{equation}
{\small
        \begin{alignedat}{4}
	& \ord{1}:~  &&(\bar{d}_a d^b) (\bar{d}_b d^a)~, \quad (\bar{d}_a d_3) (\bar{d}_3 d^a)~,\\
	& \ord{\Delta V}:~  &&(\bar{d}_a d_3) (\bar{d} \Delta_d^\dagger V_q  d^a)~,\quad 
	{\color{blueRef}(\bar{d}_a d_3)\varepsilon^{ab}\varepsilon_{de}[ \bar{d}_b V_q^d (\Delta_d)\ud{e}{c} d^c ]}~,\quad 
	{\color{blueRef}\varepsilon^{be}\varepsilon_{cd}(\bar{d}_a d_3)[ \bar{d}_b V_q^c (\Delta_d)\ud{d}{e} d^a ]}~,\quad \hermc~,\\
	&&& {\color{blueRef}(\bar{d}_3 d^a)[ \bar{d}_a V_q^c \varepsilon_{cd}(\Delta_d)\ud{d}{b} d^b ]}~,\quad 
	{\color{blueRef}(\bar{d}_3 d^a)  [\bar{d}_a \varepsilon_{bd}V_q^c (\Delta_d^*)\du{c}{d} d^b]}~,\quad
	{\color{blueRef}\varepsilon_{ac}(\bar{d}_3 d^a)[ \bar{d}_b V_q^d (\Delta_d^*)\du{d}{b} d^c ]}~,\quad \hermc~.\\
	&&&
	\end{alignedat}}
	\end{equation}
\noindent $\boxed{(\bar{q} q) (\bar{u} u)}$
	\begin{equation}
    {\small
    \begin{alignedat}{4}
	& \ord{\Delta}:~  && (\bar{q}_a q_3) (\Delta_u)\ud{a}{b} (\bar{u}_3 u^b)~,\quad 
	{\color{blueRef}\varepsilon^{bc}(\bar{q}_a   q_3)(\Delta_u)\ud{a}{c}(\bar{u}_b u^3)}~,\quad 
	{\color{blueRef}\varepsilon^{bd}\varepsilon_{ac}(\bar{q}_3 q^a)(\Delta_u)\ud{c}{d}(\bar{u}_b u_3)}~,\quad \hermc~,\\
	&&&{\color{blueRef}\varepsilon_{ac}(\bar{q}_3 q^a)[ \bar{u}_3 (\Delta_u)\ud{c}{b} u^b ]}~,\quad \hermc~,\\ 
	& \ord{\Delta V}:~  && (\bar{q}_a V_q^\dagger q) (\Delta_u)\ud{a}{b} (\bar{u}_3 u^b)~,\quad 
	{\color{blueRef}\varepsilon^{ce}\varepsilon_{bd}(\bar{q}V_q q^b)(\Delta_u)\ud{d}{e}(\bar{u}_c u_3)}~,\quad 
	{\color{blueRef}\varepsilon^{ce}\varepsilon_{bd}(\bar{q}_aV_q^d q^b)(\Delta_u)\ud{a}{e}(\bar{u}_c u_3)}~,\quad\hermc~,\\
	&&&{\color{blueRef}\varepsilon_{bd}(\bar{q}V_q q^b)(\Delta_u)\ud{d}{c}(\bar{u}_3 u^c)}~,\quad 
	{\color{blueRef}\varepsilon_{bd}(\bar{q}_a V_q^d q^b)(\Delta_u)\ud{a}{c}(\bar{u}_3 u^c)}~,\quad 
	{\color{blueRef}\varepsilon_{cd}(\bar{q}V_q q^b)(\Delta_u^*)\du{b}{d}(\bar{u}_3 u^c)}~,\quad\hermc~.
	\end{alignedat}}
	\end{equation}
\noindent $\boxed{(\bar{q} q) (\bar{d} d)}$
	\begin{equation}
    {\small
    \begin{alignedat}{4}
	& \ord{\Delta}:~  && (\bar{q}_a q_3) (\Delta_d)\ud{a}{b} (\bar{d}_3 d^b)~,\quad 
	{\color{blueRef}\varepsilon^{bc}(\bar{q}_a   q_3)(\Delta_d)\ud{a}{c}(\bar{d}_b d^3)}~,\quad 
	{\color{blueRef}\varepsilon^{bd}\varepsilon_{ac}(\bar{q}_3 q^a)(\Delta_d)\ud{c}{d}(\bar{d}_b d_3)}~,\quad \hermc~,\\
	&&&{\color{blueRef}\varepsilon_{ac}(\bar{q}_3 q^a)[ \bar{d}_3 (\Delta_d)\ud{c}{b} d^b ]}~,\quad \hermc~,\\ 
	& \ord{\Delta V}:~  && (\bar{q}_a V_q^\dagger q) (\Delta_d)\ud{a}{b} (\bar{d}_3 d^b)~,\quad 
	{\color{blueRef}\varepsilon^{ce}\varepsilon_{bd}(\bar{q}V_q q^b)(\Delta_d)\ud{d}{e}(\bar{d}_c d_3)}~,\quad 
	{\color{blueRef}\varepsilon^{ce}\varepsilon_{bd}(\bar{q}_aV_q^d q^b)(\Delta_d)\ud{a}{e}(\bar{d}_c d_3)}~,\quad\hermc~,\\
	&&&{\color{blueRef}\varepsilon_{bd}(\bar{q}V_q q^b)(\Delta_d)\ud{d}{c}(\bar{d}_3 d^c)}~,\quad 
	{\color{blueRef}\varepsilon_{bd}(\bar{q}_a V_q^d q^b)(\Delta_d)\ud{a}{c}(\bar{d}_3 d^c)}~,\quad 
	{\color{blueRef}\varepsilon_{cd}(\bar{q}V_q q^b)(\Delta_d^*)\du{b}{d}(\bar{d}_3 d^c)}~,\quad\hermc~.
	\end{alignedat}}
	\end{equation}
\noindent $\boxed{(\bar{q} u) (\bar{q} d)}$
	\begin{equation}
	\label{eq223}
    {\small
    \begin{alignedat}{4}
	& \ord{1}:~  && \opemph{(\bar{q}_a u_3) \varepsilon^{ab} (\bar{q}_b d_3)}~,\quad  \\
	& \ord{\Delta}:~ && (\bar{q}_a u_3) (\Delta_d)\ud{a}{b} (\bar{q}_3 d^b)~, \quad 
		(\bar{q}_3 u^a) (\Delta_u)\ud{b}{a} (\bar{q}_b d_3)~,\quad\opemph{(\bar{q}_a u_3) (\widetilde{\Delta}_d)\ud{a}{b} (\bar{q}_3 d^b){\color{black}~,} \quad 
			(\bar{q}_3 u^a) (\widetilde{\Delta}_u)\ud{b}{a} (\bar{q}_b d_3) }~,\quad\\
	& \ord{\Delta V}:~ && (\bar{q}_a u_3) (\Delta_d)\ud{a}{b} (\bar{q}  V_q  d^b)~, \quad (\bar{q} V_q  u^a) (\Delta_u)\ud{b}{a} (\bar{q}_b d_3)~,\quad 
	{\color{blueRef}\varepsilon^{ac}\varepsilon_{be}(\bar{q}_a u^b) V_q^d (\Delta_u^*)\du{d}{e} (\bar{q}_c d_3)}~,\quad\\
	&&& {\color{blueRef}(\bar{q}V_q u^b)(\widetilde{\Delta}_u)\ud{c}{b}(\bar{q}_c d_3)}~,\quad 
	{\color{blueRef}\varepsilon^{ac}[\bar{q}_a (V_q^\dag \Delta_u)_b u^b](\bar{q}_c d_3)}~,\quad {\color{blueRef}\varepsilon^{ad}[\bar{q}_a (V_q^*)_d u^b](\Delta_u)\ud{c}{b}(\bar{q}_c d_3)}~,\quad\\
	&&&{\color{blueRef}\varepsilon^{ac}[ \bar{q}_a (V_q^\dag \widetilde{\Delta}_u)_b u^b ] (\bar{q}_c d^3)}~,\quad 
	{\color{blueRef}\varepsilon^{ad}[ \bar{q}_a (V_q^*)_d u^b ] (\widetilde{\Delta}_u)\ud{c}{b}(\bar{q}_c d^3)}~,\quad
	{\color{blueRef}\varepsilon^{ab}\varepsilon_{ce}(\bar{q}_a u_3)V_q^d (\Delta_d^*)\du{d}{e}(\bar{q}_b d^c)}~,\quad\\
	&&&{\color{blueRef}(\bar{q}_a u_3)(\widetilde{\Delta}_d)\ud{a}{c}(\bar{q}V_q d^c)}~,\quad 
	{\color{blueRef}\varepsilon^{ab}(\bar{q}_a u_3)[ \bar{q}_b (V_q^\dag \Delta_d)_c d^c ]}~,\quad 
	{\color{blueRef}\varepsilon^{bd}[\bar{q}_a (V_q^*)_d u_3](\Delta_d)\ud{a}{c}(\bar{q}_b d^c)}~,\quad\\
	&&&{\color{blueRef}\varepsilon^{ab}(\bar{q}_a u_3)[ \bar{q}_b (V_q^\dag \widetilde{\Delta}_d)_c d^c ]}~,\quad
	{\color{blueRef}\varepsilon^{bd}[ \bar{q}_a (V_q^*)_d u_3 ](\widetilde{\Delta}_d)\ud{a}{c}(\bar{q}_b d^c)}~.
	\end{alignedat}}
	\end{equation}


\subsection{$\U(2)^3 \times \U(1)_{d_3}$ symmetry}
If the NP possesses a $ G= \U(2)_q \times \U(2)_u \times \U(2)_d \times \U(1)_{d_3} \subset G_Q$, it is possible to introduce the $ b $ Yukawa as a spurion for the perturbative counting. Under this symmetry, the quark fields decompose as\footnote{The representations of the fields and spurions with respect to this particular group are labeled as $(\U(2)_q,\U(2)_u,\U(2)_d \lvert \U(1)_{d_3})$.}
\begin{equation}
\begin{alignedat}{2}
   &q= \begin{bmatrix}
    q^a\sim (\boldsymbol2,\boldsymbol1,\boldsymbol1 \lvert 0) \\ q_3\sim(\boldsymbol1,\boldsymbol1,\boldsymbol1\lvert 0)
    \end{bmatrix},\quad 
   u= \begin{bmatrix}
    u^a\sim (\boldsymbol1,\boldsymbol2,\boldsymbol1 \lvert 0) \\ u_3\sim(\boldsymbol1,\boldsymbol1,\boldsymbol1\lvert 0)
    \end{bmatrix},\quad
   d= \begin{bmatrix}
    d^a\sim (\boldsymbol1,\boldsymbol1,\boldsymbol2 \lvert 0) \\ d_3\sim(\boldsymbol1,\boldsymbol1,\boldsymbol1\lvert 1)
    \end{bmatrix},
\end{alignedat}
\end{equation}
and the minimal set of spurions required to write the Yukawa matrix is given as:
\begin{equation}
    \Delta_u\sim (\boldsymbol2,\boldsymbol{\bar{2}},\boldsymbol{1}\lvert 0),\quad \Delta_d\sim (\boldsymbol2,\boldsymbol1,\boldsymbol{\bar{2}}\lvert 0),\quad V_q\sim(\boldsymbol2,\boldsymbol1,\boldsymbol1\lvert 0),\quad X_b\sim(\boldsymbol1,\boldsymbol1,\boldsymbol1\lvert -1).
\end{equation}

Transforming the fields with $ G_Q/G$ rotations and redefining the spurions as in  Section~\ref{sec:quark_U2^3}, the Yukawa matrices can generically be written as:
\begin{equation}
    Y_u = \begin{bmatrix}
        \Delta_u & V_q \\
        0 & y_t
    \end{bmatrix},\quad
    Y_d = \begin{bmatrix}
        \Delta_d & 0 \\
         0&  X_b
    \end{bmatrix},
\end{equation}
for $ y_t > 0$.
The complete breaking of $ G $ by the spurions proceeds in the same way, leading to the parametrization of the $\Delta_{u,d}$ and $V_q$ spurions of Eqs.~(\ref{eq:Vq,U(2)^3}--\ref{eq:Dd,U(2)^3}). The only novelty compared to the $\U(2)^3$ case is the introduction of the $X_b$ spurion, which breaks the $\U(1)_{d_3} $ symmetry, allowing us to remove an extra phase from $X_b$ and make it real and positive. In total, we end up parametrizing spurions with 6 real and positive parameters, 2 mixing angles and 1 phase. The numerical values of the parameters are as in Eq.~\eqref{eq:U(2)numerical} but with the replacement $y_b = X_b$ explaining the smallness of the bottom Yukawa.

The spurion counting of the pure quark operators is presented in the Table \ref{tab:quark3U2xU1}, while the bilinear and unique quartic structures are listed in Eqs.~(\ref{eq229}--\ref{eq234}) and Eqs.~(\ref{eq235}--\ref{eq240}) respectively.

\begin{table}[t]
\centering
\scalebox{0.673}{\begin{tabular}{|cc|cc|cc|cc|cc|cc|cc|cc|cc|cc|cc|cc|}
\hline
\multicolumn{2}{|c|}{$\U(2)^3 \times \U(1)_{d_3}$}                           & \multicolumn{2}{c|}{$\cO(1)$} & \multicolumn{2}{c|}{$\cO(V)$} & \multicolumn{2}{c|}{$\cO(V^2)$} & \multicolumn{2}{c|}{$\cO(V^3)$} & \multicolumn{2}{c|}{$\cO(\Delta)$} & \multicolumn{2}{c|}{$\cO(\Delta V)$} & \multicolumn{2}{c|}{$\cO(X)$} & \multicolumn{2}{c|}{$\cO(VX)$} & \multicolumn{2}{c|}{$\cO(V^2 X)$} & \multicolumn{2}{c|}{$\cO(\Delta X)$} & \multicolumn{2}{c|}{$\cO(\Delta V X)$} \\ \hline
\multicolumn{1}{|c|}{\multirow{2}{*}{$\psi^2 H^3 $}} & $Q_{uH}$                      & 1              & 1            & 1             & 1             &                 &               &                &                & 1                & 1               & 1                 & 1                &               &               &                &               &                 &                 &                   &                  &                    &                   \\
\multicolumn{1}{|c|}{}                                  & $Q_{dH}$                      &                &              &               &               &                 &               &                &                & 1                & 1               & 1                 & 1                & 1             & 1             & 1              & 1             &                 &                 &                   &                  &                    &                   \\ \hline
\multicolumn{1}{|c|}{\multirow{2}{*}{$\psi^2 XH $}}  & $Q_{u(G,W,B)}$    & 3              & 3            & 3             & 3             &                 &               &                &                & 3                & 3               & 3                 & 3                &               &               &                &               &                 &                 &                   &                  &                    &                   \\
\multicolumn{1}{|c|}{}                                  & $Q_{d(G,W,B)}$    &                &              &               &               &                 &               &                &                & 3                & 3               & 3                 & 3                & 3             & 3             & 3              & 3             &                 &                 &                   &                  &                    &                   \\ \hline
\multicolumn{1}{|c|}{\multirow{4}{*}{$\psi^2 H^2 D$}}   & $Q_{Hq}^{(1,3)}$ & 4              &              & 2             & 2             & 2               &               &                &                &                  &                 &                   &                  &               &               &                &               &                 &                 &                   &                  &                    &                   \\
\multicolumn{1}{|c|}{}                                  & $Q_{Hu}$                      & 2              &              &               &               &                 &               &                &                &                  &                 & 1                 & 1                &               &               &                &               &                 &                 &                   &                  &                    &                   \\
\multicolumn{1}{|c|}{}                                  & $Q_{Hd}$                      & 2              &              &               &               &                 &               &                &                &                  &                 &                   &                  &               &               &                &               &                 &                 &                   &                  & 1                  & 1                 \\
\multicolumn{1}{|c|}{}                                  & $Q_{Hud}$                 &                &              &               &               &                 &               &                &                &                  &                 & 1                  & 1                 & 1             & 1             &                &               &                 &                 &                   &                  & 1                  & 1                 \\ \hline
\multicolumn{1}{|c|}{$(LL)(LL)$}                        & $Q_{qq}^{(1,3)}$ & 10             &              & 6             & 6             & 10              & 2             & 2              & 2              &                  &                 &                   &                  &               &               &                &               &                 &                 &                   &                  &                    &                   \\ \hline
\multicolumn{1}{|c|}{\multirow{3}{*}{$(RR)(RR)$}}       & $Q_{uu}$                      & 5              &              &               &               &                 &               &                &                &                  &                 & 3                 & 3                &               &               &                &               &                 &                 &                   &                  &                    &                   \\
\multicolumn{1}{|c|}{}                                  & $Q_{dd}$                      & 5              &              &               &               &                 &               &                &                &                  &                 &                   &                  &               &               &                &               &                 &                 &                   &                  & 3                  & 3                 \\
\multicolumn{1}{|c|}{}                                  & $Q_{ud}^{(1,8)}$ & 8              &              &               &               &                 &               &                &                &                  &                 & 4                 & 4                &               &               &                &               &                 &                 &                   &                  & 4                  & 4                 \\ \hline
\multicolumn{1}{|c|}{\multirow{2}{*}{$(LL)(RR)$}}       & $Q_{qu}^{(1,8)}$              & 8              &              & 4             & 4             & 4               &               &                &                & 2                & 2               & 6                 & 6                &               &               &                &               &                 &                 &                   &                  &                    &                   \\
\multicolumn{1}{|c|}{}                                  & $Q_{qd}^{(1,8)}$                & 8              &              & 4             & 4             & 4               &               &                &                &                  &                 &                   &                  &               &               &                &               &                 &                 & 2                 & 2                & 6                  & 6                 \\ \hline
\multicolumn{1}{|c|}{$(LR)(LR)$}                     & $Q_{quqd}^{(1,8)}$            &                &              &               &               &                 &               &                &                & 4                & 4               & 6                 & 6                & 2             & 2             & 4              & 4             & 2               & 2               & 4                 & 4                & 6                  & 6                 \\ \hline\hline
\multicolumn{2}{|c|}{Total}                                                             & 56             & 4            & 20            & 20            & 20              & 2             & 2              & 2              & 14               & 14              & 29                & 29               & 7             & 7             & 8              & 8             & 2               & 2               & 6                 & 6                & 21                 & 21                \\ \hline
\end{tabular}}
\caption{Counting of the pure quark SMEFT operators (see Appendix \ref{app:Warsaw}) assuming $\U(2)^3 \times \U(1)_{d_3}$ symmetry in the quark sector. The counting is done up to three insertions of the $V_q$ spurion and one insertion of $\Delta_{u,d}$ or $X_b$. For spurion products, the counting is presented for $\Delta_{u,d}V_q$, $V_q X_b$, and $\Delta_{u,d}X_b$ as well as for $V_q^2 X_b$ and $\Delta_{u,d}V_q X_b$ insertions. The left (right) entry in each column gives the number of CP even (odd) coefficients at the given order in spurion counting.}
\label{tab:quark3U2xU1}
\end{table}

\subsubsection*{Decomposition of bilinear structures}
We follow the analysis of Section~\ref{sec:quark_U2^3} to obtain all SMEFT structures in the presence of the $ \U(2)^3 \times \U(1)_{d_3} $ symmetry. The only difference is that when the $ \U(1)_{d_3} $ charge of a structure is nonzero, powers of the $ X_b $ spurion need to be included, giving further spurion suppression. The list of all invariant bilinears is presented below:\\

\noindent $\boxed{(\bar{q} q)}$
	\begin{equation}
	\label{eq229}
	\begin{alignedat}{4}
	& \ord{1}:~ (\bar{q} q)~, \quad (\bar{q}_3 q_3)~,\qquad \ord{V}:~  (\bar{q} V_q q_3 )~, \quad \hermc ~,\qquad \ord{V^2}:~ (\bar{q}  V_q V_q^\dagger q )~.
	\end{alignedat}
	\end{equation}
$\boxed{(\bar{u} u)}$
	\begin{equation}
	\begin{alignedat}{4}
	& \ord{1}:~ (\bar{u} u)~, \quad (\bar{u}_3 u_3)~,\qquad\ord{\Delta V}:~ (\bar{u} \Delta_u^\dagger V_q u_3 )~,\quad  \hermc~.
	\end{alignedat}
	\end{equation}
$\boxed{(\bar{d} d)}$
	\begin{equation}
	\begin{alignedat}{4}
	& \ord{1}:~   (\bar{d} d)~, \quad (\bar{d}_3 d_3)~,\qquad \ord{\Delta V X}:~  (\bar{d} \Delta_d^\dagger V_q X_b d_3 )~,\quad \hermc~.
	\end{alignedat}
	\end{equation}
$\boxed{(\bar{u}d)}$\\
\begin{equation}
    \begin{alignedat}{2}
        &\mathcal{O}(X):~(\bar{u}_3 X_b d_3)~,\qquad \cO(\Delta V):~ (\bar{u}_3 V_q^\dag \Delta_d d)~,\qquad\mathcal{O}(\Delta V X):~ (\bar{u} \Delta_u^{\dag} V_q   X_b d_3)~.
    \end{alignedat}
\end{equation}
$\boxed{(\bar{q} u)}$
	\begin{equation}
	\begin{alignedat}{4}
	 & \ord{1}:~  && (\bar{q}_3 u_3)~,\qquad\ord{V}:~ (\bar{q} V_q u_3 )~,\qquad\ord{\Delta}:~ (\bar{q} \Delta_u u )~,\qquad\ord{\Delta V}: \quad && (\bar{q}_3 V_q^\dagger \Delta_u u ) ~.
	\end{alignedat}
	\end{equation}
$\boxed{(\bar{q} d)}$
	\begin{equation}
	\label{eq234}
	\begin{alignedat}{4}
	 & \ord{X}:~ (\bar{q}_3 X_b d_3)~,\qquad\ord{VX}:~ (\bar{q} V_q X_b d_3 )~,\qquad\ord{\Delta}:~ (\bar{q} \Delta_d d )~,\\&\ord{\Delta V}: \quad (\bar{q}_3 V_q^\dagger \Delta_d d )~.
	\end{alignedat}
	\end{equation}
\subsubsection*{Decomposition of quartic structures}
Again following the construction of quartic SMEFT structures in Section~\ref{sec:quark_U2^3} and with suitable insertions of $ X_b$, the complete list of $\U(2)^3\times \U(1)_{d_3}$ unique quartic structures is presented below:\\

\noindent $\boxed{(\bar{q} q) (\bar{q} q)}$
	\begin{equation}
	\label{eq235}
	\begin{alignedat}{4}
	& \ord{1}:~ (\bar{q}_a q^b) (\bar{q}_b q^a)~, \quad (\bar{q}_a q_3) (\bar{q}_3 q^a)~,\qquad\ord{V}:~ (\bar{q}_a q_3) (\bar{q} V_q q^a)~,\quad \hermc~,\\ 
	& \ord{V^2}:~  (\bar{q}_a V_q^\dagger q) (\bar{q} V_q q^a)~.
	\end{alignedat}
	\end{equation}

\noindent $\boxed{(\bar{u} u) (\bar{u} u)}$
	\begin{equation}
	\begin{alignedat}{4}
	& \ord{1}:~ (\bar{u}_a u^b) (\bar{u}_b u^a)~, \quad (\bar{u}_a u_3) (\bar{u}_3 u^a)~,\qquad\ord{\Delta V}:~ (\bar{u}_a u_3) (\bar{u} \Delta_u^\dagger V_q u^a)~, \quad\hermc~.
	\end{alignedat}
	\end{equation}
\noindent $\boxed{(\bar{d} d) (\bar{d} d)}$
	\begin{equation}
	\begin{alignedat}{4}
	& \ord{1}:~ (\bar{d}_a d^b) (\bar{d}_b d^a)~, \quad (\bar{d}_a d_3) (\bar{d}_3 d^a)~,\qquad\ord{\Delta V X}:~ (\bar{d}_a X_b d_3) (\bar{d} \Delta_d^\dagger V_q  d^a)~, \quad\hermc~.
	\end{alignedat}
	\end{equation}
\noindent $\boxed{(\bar{q} q) (\bar{u} u)}$
	\begin{equation}
	\begin{alignedat}{4}
	& \ord{\Delta}:~ (\bar{q}_a q_3) (\Delta_u)\ud{a}{b} (\bar{u}_3 u^b)~, \quad \hermc~,\qquad\ord{\Delta V}:~ (\bar{q}_a V_q^\dagger q) (\Delta_u)\ud{a}{b} (\bar{u}_3 u^b)~,\quad  \hermc~.
	\end{alignedat}
	\end{equation}
\noindent $\boxed{(\bar{q} q) (\bar{d} d)}$
	\begin{equation}
	\begin{alignedat}{4}
	& \ord{\Delta X}:~ (\bar{q}_a q_3) (\Delta_d)\ud{a}{c} (\bar{d}_3 X_b^* d^c)~, \quad \hermc~,\qquad\ord{\Delta V X}:~ (\bar{q}_a V_q^\dagger q) (\Delta_d)\ud{a}{c} (\bar{d}_3 X_b^* d^c)~,\quad \hermc~.
	\end{alignedat}
	\end{equation}
\noindent $\boxed{(\bar{q} u) (\bar{q} d)}$
	\begin{equation}
	\label{eq240}
	\begin{alignedat}{4}
	& \ord{\Delta}:~ (\bar{q}_a u_3) (\Delta_d)\ud{a}{b} (\bar{q}_3 d^b)~,\qquad\cO(\Delta X):~ (\bar{q}_3 u^a) (\Delta_u)\ud{c}{a} (\bar{q}_c X_b d_3)~, \\
	& \ord{\Delta V}:~ (\bar{q}_a u_3) (\Delta_d)\ud{a}{b} (\bar{q}  V_q  d^b)~,\qquad\cO(\Delta V X):~ (\bar{q} V_q  u^a) (\Delta_u)\ud{c}{a} (\bar{q}_c X_b d_3)~.
	\end{alignedat}
	\end{equation}

\subsection{$\U(2)^2 \times \U(3)_d$ symmetry}
We consider the quark symmetry $ G= \U(2)_q \times \U(2)_u \times \U(3)_d \subset G_Q$, which is the flavor symmetry of the SM when only the top quark has a Yukawa coupling to the Higgs. This is also a good approximate symmetry of the realistic Yukawa sector. Here,
$q$ and $u$ decompose as
\begin{equation}
\begin{alignedat}{2}
    q &= \begin{bmatrix}q^a\sim (\rep{2},\rep{1},\rep{1}) \\ q_3\sim(\rep{1},\rep{1},\rep{1})\end{bmatrix},\quad
    u &= \begin{bmatrix}u^a\sim (\rep{1},\rep{2},\rep{1}) \\ u_3\sim (\rep{1},\rep{1},\rep{1})\end{bmatrix},
\end{alignedat}
\end{equation}
under $ G $, while $d$ transforms as $d^i \sim (\rep{1},\rep{1},\rep{3})$.
The minimal set of spurions needed to produce realistic Yukawa matrices for the quarks are 
\begin{equation}
    \Delta_u \sim (\boldsymbol2,\boldsymbol{\bar{2}},\boldsymbol1),\qquad \Sigma_d\sim (\boldsymbol2,\boldsymbol1,\boldsymbol{\bar{3}}),\qquad \Lambda_d\sim(\boldsymbol1,\boldsymbol1,\boldsymbol3),
\end{equation}
but we will also allow for $ V_q\sim (\boldsymbol2,\boldsymbol1,\boldsymbol1) $ to allow for further mixing between top quarks and light generations. In the minimal analysis, one can simply set $ V_q = 0 $. 
In either case, a combination of rotations in $ G_Q/G $ and redefinitions of the spurions, as detailed in Section~\ref{sec:quark_U2^3}, allows for the generic Yukawa matrices to be parametrized as
\begin{equation}
    Y_u = \begin{bmatrix} \Delta_u & 0\\0 & y_t \end{bmatrix}, \qquad  
    Y_d = \begin{bmatrix} \Sigma_d \\ \Lambda_d^\dagger \end{bmatrix},
\end{equation}
with a real coefficient $ y_t $.

To see how the spurions break $ G $, we begin with $ \Delta_u $, which, after a suitable rotation, can be parametrized as 
\begin{align}
	\Delta_u &\longrightarrow
	\begin{bmatrix} \delta_u &0 \\ 0& \delta_u' \end{bmatrix}: &&
	\U(2)_{q} \times \U(2)_u \longrightarrow \U(1)_{q+u}^2.
\intertext{Meanwhile, a $ \U(3)_d $ rotation can be used to align $ \Lambda_b $ to the 3rd generation:}
	\Lambda_d &\longrightarrow \begin{bmatrix} 0& 0& y_b\end{bmatrix}^\intercal: 
	&&\U(3)_d \longrightarrow \U(2)_d.
\end{align} 
The last part of the symmetry is used to parametrize
    \begin{equation}
    \Sigma_d \longrightarrow \begin{bmatrix} c_d & -s_d e^{i\alpha}\\s_d e^{-i\alpha} & c_d \end{bmatrix} \!
	\begin{bmatrix} \delta_d &0 &0\\ 0&\delta'_d& 0 \end{bmatrix} \!
	\begin{bmatrix}  
	c_{13}&0&-s_{13}\\
	-s_{13} s_{23} &c_{23}&-c_{13} s_{23}\\
	s_{13} c_{23} &s_{23}& c_{13} c_{23}
	\end{bmatrix}:\quad 
	\U(1)_{q+u}^2 \times \U(2)_d \longrightarrow \emptyset.
    \end{equation}
The singular values can always be taken to be $ \delta_u^{(\prime)}, \delta_d^{(\prime)}, y_b > 0 $. 
The complete breaking of $ G $ by the spurions makes it possible to remove 17 unphysical parameters from the spurions, reducing the naive 13 complex parameters down to a total of 5 real positive parameters, 3 mixing angles, and a phase.  
In case $ V_q $ is included, both of its complex parameters are physical.

Following a similar procedure to Eq.~\eqref{eq:U(2)numerical}, the observed CKM matrix and quark masses are reproduced for
    \begin{align}
    \delta_d &= \num{5.70e-5}, & \delta_d' &= \num{6.91e-4}, & y_b &= 0.0155, \nonumber\\
    \delta_u &= \num{6.72e-6}, & \delta_u' &= \num{3.38e-3}, & y_t &= 0.934,  \\
    \theta_d &= 0.096, & \theta_{13} &= 0.412, & \theta_{23} &= 1.453, & \alpha &= -1.19.  \nonumber
    \end{align}
The absolute value of all symmetry-breaking parameters (including $y_b$) is small compared with the symmetry-allowed $y_t$.

The spurion counting of the pure quark SMEFT operators assuming $\U(2)^2\times \U(3)_d$ symmetry is presented in Table \ref{tab:quark2U2xU3}. The decompositions of the bilinear structures are listed in (\ref{eq246})--(\ref{eq251}) and the unique quartic structures are given in (\ref{eq252})--(\ref{eq257}).

\begin{table}[t]
\centering
\scalebox{0.63}{\begin{tabular}{|cc|cc|cc|cc|cc|cc|cc|cc|cc|cc|cc|cc|cc|}
\hline
\multicolumn{2}{|c|}{$\U(2)_q \times \U(2)_u \times \U(3)_d$}                           & \multicolumn{2}{c|}{$\cO(1)$} & \multicolumn{2}{c|}{$\cO(V_q)$} & \multicolumn{2}{c|}{$\cO(V_q^2)$} & \multicolumn{2}{c|}{$\cO(V_q^3)$} & \multicolumn{2}{c|}{$\cO(\Delta)$} & \multicolumn{2}{c|}{$\cO(\Delta V_q)$} & \multicolumn{2}{c|}{$\cO(\Lambda_d)$} & \multicolumn{2}{c|}{$\cO(\Sigma_d)$} & \multicolumn{2}{c|}{$\cO(V_q\Lambda_d)$} & \multicolumn{2}{c|}{$\cO(\Lambda_d^2)$} & \multicolumn{2}{c|}{$\cO(V_q \Sigma_d)$} & \multicolumn{2}{c|}{$\cO(\Lambda_d \Sigma_d)$} \\ \hline
\multicolumn{1}{|c|}{\multirow{2}{*}{$\psi^2 H^3 $}} & $Q_{uH}$                      & 1              & 1            & 1              & 1              &                  &                &                 &                 & 1                & 1               & 1                  & 1                 &                &                &                &                &                  &                 &                  &                &                  &                  & 1                & 1                \\
\multicolumn{1}{|c|}{}                                  & $Q_{dH}$                      &                &              &                &                &                  &                &                 &                 &                  &                 &                    &                   & 1              & 1              & 1              & 1              & 1                & 1               &                  &                & 1                & 1                &                  &                  \\ \hline
\multicolumn{1}{|c|}{\multirow{2}{*}{$\psi^2 XH $}}  & $Q_{u(G,W,B)}$    & 3              & 3            & 3              & 3              &                  &                &                 &                 & 3                & 3               & 3                  & 3                 &                &                &                &                &                  &                 &                  &                &                  &                  & 3                & 3                \\
\multicolumn{1}{|c|}{}                                  & $Q_{d(G,W,B)}$    &                &              &                &                &                  &                &                 &                 &                  &                 &                    &                   & 3              & 3              & 3              & 3              & 3                & 3               &                  &                & 3                & 3                &                  &                  \\ \hline
\multicolumn{1}{|c|}{\multirow{4}{*}{$\psi^2 H^2 D$}}   & $Q_{Hq}^{(1,3)}$ & 4              &              & 2              & 2              & 2                &                &                 &                 &                  &                 &                    &                   &                &                &                &                &                  &                 &                  &                &                  &                  & 2                & 2                \\
\multicolumn{1}{|c|}{}                                  & $Q_{Hu}$                      & 2              &              &                &                &                  &                &                 &                 &                  &                 & 1                  & 1                 &                &                &                &                &                  &                 &                  &                &                  &                  &                  &                  \\
\multicolumn{1}{|c|}{}                                  & $Q_{Hd}$                      & 1              &              &                &                &                  &                &                 &                 &                  &                 &                    &                   &                &                &                &                &                  &                 & 1                &                &                  &                  &                  &                  \\
\multicolumn{1}{|c|}{}                                  & $Q_{Hud} $                 &                &              &                &                &                  &                &                 &                 &                  &                 &                    &                   & 1              & 1              &                &                &                  &                 &                  &                & 1                & 1                &                  &                  \\ \hline
\multicolumn{1}{|c|}{$(LL)(LL)$}                        & $Q_{qq}^{(1,3)}$ & 10             &              & 6              & 6              & 10               & 2              & 2               & 2               &                  &                 &                    &                   &                &                &                &                &                  &                 &                  &                &                  &                  & 6                & 6                \\ \hline
\multicolumn{1}{|c|}{\multirow{3}{*}{$(RR)(RR)$}}       & $Q_{uu}$                      & 5              &              &                &                &                  &                &                 &                 &                  &                 & 3                  & 3                 &                &                &                &                &                  &                 &                  &                &                  &                  &                  &                  \\
\multicolumn{1}{|c|}{}                                  & $Q_{dd}$                      & 2              &              &                &                &                  &                &                 &                 &                  &                 &                    &                   &                &                &                &                &                  &                 & 2                &                &                  &                  &                  &                  \\
\multicolumn{1}{|c|}{}                                  & $Q_{ud}^{(1,8)}$ & 4              &              &                &                &                  &                &                 &                 &                  &                 & 2                  & 2                 &                &                &                &                &                  &                 & 4                &                &                  &                  &                  &                  \\ \hline
\multicolumn{1}{|c|}{\multirow{2}{*}{$(LL)(RR)$}}       & $Q_{qu}^{(1,8)}$              & 8              &              & 4              & 4              & 4                &                &                 &                 & 2                & 2               & 6                  & 6                 &                &                &                &                &                  &                 &                  &                &                  &                  & 4                & 4                \\
\multicolumn{1}{|c|}{}                                  & $Q_{qd}^{(1,8)}$              & 4              &              & 2              & 2              & 2                &                &                 &                 &                  &                 &                    &                   &                &                &                &                &                  &                 & 4                &                &                  &                  & 4                & 4                \\ \hline
\multicolumn{1}{|c|}{$(LR)(LR)$}                     & $Q_{quqd}^{(1,8)}$            &                &              &                &                &                  &                &                 &                 &                  &                 &                    &                   & 2              & 2              & 4              & 4              & 4                & 4               &                  &                & 6                & 6                &                  &                  \\ \hline\hline
\multicolumn{2}{|c|}{Total}                                                             & 44             & 4            & 18             & 18             & 18               & 2              & 2               & 2               & 6                & 6               & 16                 & 16                & 7              & 7              & 8              & 8              & 8                & 8               & 11               &                & 11               & 11               & 20               & 20               \\ \hline
\end{tabular}}
\caption{Counting of the pure quark SMEFT operators (see Appendix \ref{app:Warsaw}) assuming $\U(2)_q \times \U(2)_u \times \U(3)_d$ symmetry in the quark sector. The counting is performed taking up to three insertions of $V_q$, one insertion of $\Delta_u$ or $\Sigma_d$, and two insertions of $\Lambda_d$ as well as one insertion of $\Delta_u V_q$, $V_q \Lambda_d$, $V_q \Sigma_d$, and $\Lambda_d \Sigma_d$ products each. The left (right) entry in each column gives the number of CP even (odd) coefficients at the given order in spurion counting.}
\label{tab:quark2U2xU3}
\end{table}

\subsubsection*{Decomposition of bilinear structures}
For the $\cO(1)$ bilinear structures, since $d^i\sim(\rep{1},\rep{1},\rep{3})$, there is only one $\cO(1)$ bilinear with two appearances of $d$ given as $(\bar{d}d)$ and four with two appearances of $q$ and $u$: $(\bar{q}q), (\bar{q}_3 q_3)$, $(\bar{u}u)$ and $(\bar{u}_3 u_3)$. One additional $\cO(1)$ structure can be formed with two singlets: $(\bar{q}_3 u_3)$.

We proceed with the bilinear structures containing one insertion of the $V_{q}$ or $\Lambda_d$ spurion. Since $V_q\sim(\rep{2},\rep{1},\rep{1})$, the only possible singlet is formed contracting $V_q$ and the quark doublet. There are two such structures of the form $(\bar{q}V_q q_3)$ and $(\bar{q}V_q u_3)$. Similar reasoning holds for $\Lambda_d$, yielding two more bilinear structures: $(\bar{q}_3 \Lambda_d^\dag d)$ and $(\bar{u}_3 \Lambda_d^\dag d)$.

For bilinear structures at $\cO(V^2)$ and $\cO(\Lambda_d^2)$, we have two structures formed by contracting the insertion of the spurion with the appropriate field: $(\bar{q}V_q V_q^\dag q)$ and $(\bar{d}\Lambda_d \Lambda_d^\dag d)$. At $\cO(\Delta)$ there is only one structure allowed: $(\bar{q}\Delta_u u)$, and similar conclusion holds for $\cO(\Sigma_d)$ with one structure: $(\bar{q}\Sigma_dd)$.

In order to construct structures with two insertions of different spurions, let us first note the transformation properties of the relevant spurion products. At $\cO(\Lambda_d \Sigma_d)$, we have $\Sigma_d \Lambda_d  \sim (\rep{2},\rep{1},\rep{1})$ transformation, which, being the same as for the $V_q$ spurion, gives two structures: $(\bar{q}\Sigma_d \Lambda_d  q_3)$ and $(\bar{q} \Sigma_d \Lambda_d u_3)$. The analogous product we have is $V_q^\dag \Sigma_d \sim(\rep{1},\rep{1},\repbar{3})$, which can be utilized to construct two more structures of the form $(\bar{q}_3 V_q^\dag \Sigma_d d)$ and $(\bar{u}_3 V_q^\dag \Sigma_d d)$. There is also one structure at $\cO(V_q\Lambda_d)$ given as $(\bar{q}V_q \Lambda_d^\dag d)$. Finally, we note the transformation $V_q^\dag \Delta_u\sim(\rep{1},\repbar{2},\rep{1})$, which can be used to construct the remaining $\cO(\Delta V_q)$ structures: $(\bar{u}\Delta_u^\dag V_q u_3)$ and $(\bar{q}_3 V_q^\dag \Delta_u u)$. The complete list of bilinears is presented below:  \\\\
$\boxed{(\bar{q}q)}$
\begin{equation}
\label{eq246}
\begin{alignedat}{2}
    &\cO(1):~ (\bar{q}q)~, \qquad (\bar{q}_3 q_3)~,\qquad\ord{V_q}:~ (\bar{q} V_q q_3 )~, \quad \hermc~,\qquad \ord{V_q^2}:~ (\bar{q}  V_q V_q^\dagger q )~, \\
	&\cO(\Lambda_d \Sigma_d):~(\bar{q} \Sigma_d \Lambda_d  q_3)~,\quad \hermc~.
\end{alignedat}
\end{equation}
$\boxed{(\bar{u}u)}$
	\begin{equation}
	\begin{alignedat}{4}
	& \ord{1}:~ (\bar{u} u)~, \quad (\bar{u}_3 u_3)~,\qquad\ord{\Delta V_q}:~ (\bar{u} \Delta_u^\dagger V_q u_3 )~,\quad    \hermc~.
	\end{alignedat}
	\end{equation}
$\boxed{(\bar{d}d)}$
\begin{equation}
\begin{alignedat}{2}
        &\cO(1):~ (\bar{d}d)~,\qquad \cO(\Lambda_d^2):~(\bar{d} \Lambda_d \Lambda_d^\dagger d)~.
\end{alignedat}
\end{equation}
$\boxed{(\bar{u}d)}$\\
\begin{equation}
    \begin{alignedat}{2}
            &\cO(\Lambda_d):~ (\bar{u}_3 \Lambda_d^\dag d)~,\qquad\cO(V_q \Sigma_d):~ (\bar{u}_3 V_q^\dag \Sigma_d d)~.
    \end{alignedat}
\end{equation}
$\boxed{(\bar{q}u)}$
	\begin{equation}
	\begin{alignedat}{4}
	 & \ord{1}:~ (\bar{q}_3 u_3)~,\qquad\ord{V_q}:~ (\bar{q} V_q u_3 )~,\qquad\ord{\Delta}: ~ (\bar{q} \Delta_u u )~,\qquad \ord{\Delta V_q}: ~ (\bar{q}_3 V_q^\dagger \Delta_u u )~,\\
	 &\cO(\Lambda_d \Sigma_d):~ (\bar{q} \Sigma_d \Lambda_d u_3)~.
	\end{alignedat}
	\end{equation}
$\boxed{(\bar{q}d)}$
\begin{equation}
\label{eq251}
\begin{alignedat}{2}
&\cO(\Lambda_d):~ {(\bar{q}_3 \Lambda_d^\dag d)}~,\qquad\cO(\Sigma_d):~ {(\bar{q}\Sigma_d d)}~,\qquad\cO(V_q \Lambda_d):~ {(\bar{q}V_q \Lambda_d^\dag d)}~,\\
&\cO(V_q \Sigma_d):~ {(\bar{q}_3 V_q^\dag \Sigma_d d)}~.
\end{alignedat}
\end{equation}

\subsubsection*{Decomposition of quartic structures}
At $\cO(1)$ there is only one unique structure containing four appearances of $d$, $(\bar{d}_i d^j)(\bar{d}_j d^i)$, two with four instances of $q$, $(\bar{q}_a q^b)(\bar{q}_b q^a)$ and $(\bar{q}_a q_3)(\bar{q}_3 q^a)$, and two similar ones with $u$. One insertion of $V_q$ yields only one unique $\cO(V_q)$ quartic structure $(\bar{q}_a q_3)(\bar{q}V_q q^a)$, and there are no structures with one insertion of $\Lambda_d$ only. There is one $\cO(\Delta)$ structure $(\bar{q}_a q_3)(\Delta_u)\ud{a}{b}(\bar{u}_3 u^b)$ and one at $\cO(\Sigma_d)$ given as $(\bar{q}_a u_3)(\Sigma_d)\ud{a}{j}(\bar{q}_3 d^j)$.

With two insertions of $V_q$ or $\Lambda_d$, there are only two unique structures: $(\bar{q}_a V_q^\dag q)(\bar{q}V_q q^a)$ and ${(\bar{d}\Lambda_d d^j)(\bar{d}_j \Lambda_d^\dag d)}$. We also get two structures with one insertion of the $\Lambda_d \Sigma_d$ product, ${(\bar{q} \Sigma_d \Lambda_d q^b)(\bar{q}_b q_3)}$ and ${(\bar{q}_a q_3)(\Sigma_d)^a_{\phantom{j}j}(\bar{d} \Lambda_d d^j)}$, and one at $\cO(V_q \Sigma_d)$ given as ${(\bar{q}_a u_3)(\Sigma_d)^a_{\phantom{j}j}(\bar{q}V_q d^j)}$. At $\cO(\Delta V_q)$ we have two structures, $(\bar{u}_a u_3) (\bar{u} \Delta_u^\dagger V_q u^a)$ and $(\bar{q}_a V_q^\dagger q) (\Delta_u)\ud{a}{b} (\bar{u}_3 u^b)$, and there is one at $\cO(\Delta \Lambda_d)$: ${(\bar{q}_3 u^a)(\Delta_u)\ud{b}{a} (\bar{q}_b \Lambda_d^\dag d)}$. We list all the quartic structures below:\\\\
$\boxed{(\bar{q}q)(\bar{q}q)}$
	\begin{equation}
	\label{eq252}
	\begin{alignedat}{4}
	& \ord{1}:~ (\bar{q}_a q^b) (\bar{q}_b q^a)~, \quad (\bar{q}_a q_3) (\bar{q}_3 q^a)~,\qquad\ord{V_q}:~ (\bar{q}_a q_3) (\bar{q} V_q q^a)~,\quad \hermc~,\\ 
	& \ord{V_q^2}: ~  (\bar{q}_a V_q^\dagger q) (\bar{q} V_q q^a)~,\qquad\cO(\Lambda_d \Sigma_d):~ {(\bar{q} \Sigma_d \Lambda_d q^b)(\bar{q}_b q_3)}~,\quad \hermc~.
	\end{alignedat}
	\end{equation}
$\boxed{(\bar{u}u)(\bar{u}u)}$
	\begin{equation}
	\begin{alignedat}{4}
	& \ord{1}:~ (\bar{u}_a u^b) (\bar{u}_b u^a)~, \quad (\bar{u}_a u_3) (\bar{u}_3 u^a)~,\qquad\ord{\Delta V_q}:~ (\bar{u}_a u_3) (\bar{u} \Delta_u^\dagger V_q u^a)~, \quad \hermc~.
	\end{alignedat}
	\end{equation}
$\boxed{(\bar{d}d)(\bar{d}d)}$
\begin{equation}
\begin{alignedat}{2}
        &\mathcal{O}(1):~ (\bar{d}_i d^j)(\bar{d}_j d^i)~,\qquad\cO(\Lambda_d^2):~ {(\bar{d}\Lambda_d d^j)(\bar{d}_j \Lambda_d^\dag d)}~.
\end{alignedat}
\end{equation}
$\boxed{(\bar{q}q)(\bar{u}u)}$
	\begin{equation}
	\begin{alignedat}{4}
	& \ord{\Delta}:~ (\bar{q}_a q_3) (\Delta_u)\ud{a}{b} (\bar{u}_3 u^b)~,\quad \hermc~,\qquad\ord{\Delta V_q}:~ (\bar{q}_a V_q^\dagger q) (\Delta_u)\ud{a}{b} (\bar{u}_3 u^b)~,\quad \hermc~.
	\end{alignedat}
	\end{equation}
\noindent$\boxed{(\bar{q}q)(\bar{d}d)}$
\begin{equation}
    \begin{alignedat}{2}
        \cO(\Lambda_d \Sigma_d):~ {(\bar{q}_a q_3)(\Sigma_d)^a_{\phantom{j}j}(\bar{d} \Lambda_d d^j)}~,\quad \hermc~.
    \end{alignedat}
\end{equation}
\noindent $\boxed{(\bar{q}u)(\bar{q}d)}$
\begin{equation}
\label{eq257}
    \begin{alignedat}{2}
        &\cO(\Sigma_d):~{(\bar{q}_a u_3)(\Sigma_d)^a_{\phantom{j}j}(\bar{q}_3 d^j) }~,\qquad\cO(V_q \Sigma_d):~{(\bar{q}_a u_3)(\Sigma_d)^a_{\phantom{j}j}(\bar{q}V_q d^j)}~,\quad\\
        &\cO(\Delta \Lambda_d):~ {(\bar{q}_3 u^a)(\Delta_u)\ud{b}{a} (\bar{q}_b \Lambda_d^\dag d)}~.
    \end{alignedat}
\end{equation}

\subsection{$\text{MFV}_Q$ symmetry}
Minimal flavor violation assumes that the only spurions of the $G_Q = \U(3)_q \times \U(3)_u \times \U(3)_d $ symmetry in the quark sector are the SM Yukawa couplings. The quarks transform as 
\begin{equation}
    q \sim (\boldsymbol3,\boldsymbol1,\boldsymbol1),\quad u\sim(\boldsymbol1,\boldsymbol3,\boldsymbol1),\quad d\sim(\boldsymbol1,\boldsymbol1,\boldsymbol3)
\end{equation}
under $ G_Q $. 
As the Yukawa couplings are the sources of the symmetry breaking, they are promoted into spurions with the transformations assigned as
\begin{equation}
    Y_u \sim (\boldsymbol3,\boldsymbol{\bar{3}},\boldsymbol1),\quad Y_d \sim(\boldsymbol3,\boldsymbol1,\boldsymbol{\bar{3}}).
\end{equation}

Fixing the parameters of the SM, i.e., the values of the $ Y_{u,d,e} $ spurions, breaks $ G_Q $. 
With no degenerate or vanishing eigenvalues nor any accidental alignment of $ Y_u $ and $ Y_d $, $ Y_u $ can be parametrized exclusively with the diagonal matrix of its singular values, $ \hat{Y}_u$: 
\begin{align}
	&Y_u \longrightarrow \hat{Y}_u: \qquad &&\U(3)_q \times \U(3)_u \longrightarrow \U(1)^3_{q+u}.\\
\intertext{The remaining quark sector symmetry can then be used to partially diagonalize $ Y_d $, writing}
	&Y_d \longrightarrow V \hat{Y}_d: \qquad &&\U(1)^3_{q+u} \times \U(3)_d \longrightarrow \U(1)_B. 
\end{align}
Here $ V $ is a special unitary matrix with 3 rotation angles but only 1 phase, as the others have been successfully factored out: $ V $ is nothing but the CKM matrix. Only the vectorial baryon number symmetry $ \U(1)_B $ remains unbroken after the inclusion of the quark Yukawa couplings.
Only 9 real parameters and 1 phase are physical; a total of $ 26 $ unphysical parameters have been removed.
The remnant flavor symmetry of the quark sector is $ \U(1)_B $, which is consistent with $ 26 $ broken generators. No additional phases can be removed from the baryon number--conserving SMEFT operators with the remnant symmetry.

The spurion counting of the pure quark operators is presented in Table \ref{quarkMFVtable}, while the decompositions of the bilinear and quartic structures are listed in Eqs.~(\ref{eq262}--\ref{eq267}) and Eqs.~(\ref{eq268}--\ref{eq273}).

\begin{table}[t]
\centering
\scalebox{0.80}{
\begin{tabular}{|cc|cc|cc|cc|cc|cc|cc|cc|cc|}
\hline
\multicolumn{2}{|c|}{$\text{MFV}_Q$}                                              & \multicolumn{2}{c|}{$\cO(1)$} & \multicolumn{2}{c|}{$\cO(Y_u)$} & \multicolumn{2}{c|}{$\cO(Y_u^2)$} & \multicolumn{2}{c|}{$\cO(Y_d)$} & \multicolumn{2}{c|}{$\cO(Y_d^2)$} & \multicolumn{2}{c|}{$\cO(Y_u Y_d)$} & \multicolumn{2}{l|}{$\cO(Y_u^2 Y_d,Y_d^2 Y_u)$} & \multicolumn{2}{l|}{$\cO(Y_u^3,Y_d^3)$} \\ \hline
\multicolumn{1}{|c|}{\multirow{2}{*}{$\psi^2 H^3 $}}  & $Q_{uH}$           &                &              & 1              & 1              &                  &                &                &                &                  &                &                  &                  & 1                      & 1                      & 1                  & 1                  \\
\multicolumn{1}{|c|}{}                                & $Q_{dH}$           &                &              &                &                &                  &                & 1              & 1              &                  &                &                  &                  & 1                      & 1                      & 1                  & 1                  \\ \hline
\multicolumn{1}{|c|}{\multirow{2}{*}{$\psi^2 XH $}}   & $Q_{u(G,W,B)}$     &                &              & 3              & 3              &                  &                &                &                &                  &                &                  &                  & 3                      & 3                      & 3                  & 3                  \\
\multicolumn{1}{|c|}{}                                & $Q_{d(G,W,B)}$     &                &              &                &                &                  &                & 3              & 3              &                  &                &                  &                  & 3                      & 3                      & 3                  & 3                  \\ \hline
\multicolumn{1}{|c|}{\multirow{4}{*}{$\psi^2 H^2 D$}} & $Q_{Hq}^{(1,3)}$   & 2              &              &                &                & 2                &                &                &                & 2                &                &                  &                  &                        &                        &                    &                    \\
\multicolumn{1}{|c|}{}                                & $Q_{Hu}$           & 1              &              &                &                & 1                &                &                &                &                  &                &                  &                  &                        &                        &                    &                    \\
\multicolumn{1}{|c|}{}                                & $Q_{Hd}$           & 1              &              &                &                &                  &                &                &                & 1                &                &                  &                  &                        &                        &                    &                    \\
\multicolumn{1}{|c|}{}                                & $Q_{Hud} $         &                &              &                &                &                  &                &                &                &                  &                & 1                & 1                &                        &                        &                    &                    \\ \hline
\multicolumn{1}{|c|}{$(LL)(LL)$}                      & $Q_{qq}^{(1,3)}$   & 4              &              &                &                & 4                &                &                &                & 4                &                &                  &                  &                        &                        &                    &                    \\ \hline
\multicolumn{1}{|c|}{\multirow{3}{*}{$(RR)(RR)$}}     & $Q_{uu} $          & 2              &              &                &                & 2                &                &                &                &                  &                &                  &                  &                        &                        &                    &                    \\
\multicolumn{1}{|c|}{}                                & $Q_{dd}$           & 2              &              &                &                &                  &                &                &                & 2                &                &                  &                  &                        &                        &                    &                    \\
\multicolumn{1}{|c|}{}                                & $Q_{ud}^{(1,8)}$   & 2              &              &                &                & 2                &                &                &                & 2                &                &                  &                  &                        &                        &                    &                    \\ \hline
\multicolumn{1}{|c|}{$(LL)(RR)$}                      & $Q_{qu}^{(1,8)}$   & 2              &              &                &                & 6                &                &                &                & 2                &                &                  &                  &                        &                        &                    &                    \\
\multicolumn{1}{|c|}{}                                & $Q_{qd}^{(1,8)}$   & 2              &              &                &                & 2                &                &                &                & 6                &                &                  &                  &                        &                        &                    &                    \\ \hline
\multicolumn{1}{|c|}{$(LR)(LR) $}                     & $Q_{quqd}^{(1,8)}$ &                &              &                &                &                  &                &                &                &                  &                & 4                & 4                &                        &                        &                    &                    \\ \hline\hline
\multicolumn{2}{|c|}{Total}                                                & 18             &              & 4              & 4              & 19               &                & 4              & 4              & 19               &                & 5                & 5                & 8                      & 8                      & 8                  & 8                  \\ \hline
\end{tabular}}
\caption{Counting of the pure quark SMEFT operators (see Appendix \ref{app:Warsaw}) assuming MFV symmetry in the quark sector and taking the combinations up to three insertions of $Y_u$ and $Y_d$ spurions in the decompositions. The left (right) entry in each column gives the number of CP even (odd) coefficients at the given order in spurion counting.}
\label{quarkMFVtable}
\end{table}

\subsubsection*{Decomposition of bilinear structures}
We present the decompositions of the bilinear structures with up to three insertions of the spurions. $\cO(1)$ structures can be formed only by contracting a field with its conjugate. This gives the three distinct structures $(\bar{q}q)$, $(\bar{u}u)$, and $(\bar{d}d)$.

At $\cO(Y_{u,d})$, we get only two bilinears: $(\bar{q}Y_u u)$ and $(\bar{q}Y_d d)$. Meanwhile, the $\cO(Y^2)$ structures can be obtained by contracting one index of the Yukawas with each other and the remaining two open indices with fields. There are five such structures: $(\bar{q}Y_u Y_u^\dag q)$, $(\bar{q}Y_d Y_d^\dag q)$, $(\bar{u} Y_u^\dag Y_u u)$, $( \bar{d} Y_d^\dag Y_d d)$, and $( \bar{u} Y_u^\dag Y_d d )$. Note that the singlets formed by tracing over the Yukawas, e.g., $(Y_u^\dag Y_u)(\bar{q}q)$ drop from the counting, since contractions like this give bilinears, which are structurally the same as the $\cO(1)$ ones.

Lastly, there are four $\cO(Y^3)$ structures, which can be formed by inserting the products $ Y_{u,d} Y_{u,d}^\dagger $ and $ Y_{u,d}^\dagger Y_{u,d} $ in the $\cO(Y)$ bilinears. These four structures are $(\bar{q}Y_d Y_d^\dag Y_u u)$, $( \bar{q}Y_u Y_u^\dag Y_u u)$, $( \bar{q}Y_u Y_u^\dag Y_d d)$ and $( \bar{q} Y_d Y_d^\dag Y_d d)$. The full list of bilinears follows below:\\\\
\noindent
$\boxed{(\bar{q}q)}$
\begin{equation}
\label{eq262}
\cO(1):~(\bar{q}q)~,\qquad \cO(Y_u^2): ~(\bar{q}Y_u Y_u^\dag q)~,\qquad \cO(Y_d^2):~(\bar{q}Y_d Y_d^\dag q)~.
\end{equation}
$\boxed{(\bar{u}u)}$
\begin{equation}
\begin{alignedat}{2}
        &\mathcal{O}(1):~ &&(\Bar{u} u)~,\qquad 
        &\mathcal{O}(Y_u^2):~ && (\bar{u} Y_u^\dag Y_u u)~.
\end{alignedat}
\end{equation}
$\boxed{(\bar{d}d)}$
\begin{equation}
\begin{alignedat}{2}
            &\cO(1):~ && ( \bar{d}d )~,\qquad
    &\cO(Y_d^2):~ &&( \bar{d} Y_d^\dag Y_d d)~.
\end{alignedat}
\end{equation}
$\boxed{(\bar{u}d)}$
\begin{equation}
\begin{alignedat}{2}
    &\cO(Y_u Y_d):~ && ( \bar{u} Y_u^\dag Y_d d )~.
\end{alignedat}
\end{equation}
$\boxed{(\bar{q}u)}$
\begin{equation}
\begin{alignedat}{2}
    &\cO(Y_u):~  ( \bar{q}Y_u u )~,\qquad \cO(Y_d^2 Y_u):~ (\bar{q}Y_d Y_d^\dag Y_u u)~,\qquad \cO(Y_u^3):~ ( \bar{q}Y_u Y_u^\dag Y_u u)~.
\end{alignedat}
\end{equation}
$\boxed{(\bar{q}d)}$
\begin{equation}
\label{eq267}
\begin{alignedat}{2}
    &\cO(Y_d):~  ( \bar{q}Y_d d )~,\qquad \cO(Y_u^2 Y_d):~ ( \bar{q}Y_u Y_u^\dag Y_d d)~,\qquad \cO(Y_d^3):~ ( \bar{q} Y_d Y_d^\dag Y_d d)~.
\end{alignedat}
\end{equation}

\subsubsection*{Decomposition of quartic structures}
At $\cO(1)$ in the spurion counting, there are only three unique structures that can be formed: $(\bar{q}_iq^j)(\bar{q}_j q^i)$, $(\bar{u}_i u^j)(\bar{u}_j u^i)$ and $(\bar{d}_i d^j)(\bar{d}_j d^i)$. All the remaining unique quartic structures contain two insertions of $Y_{u,d}$ spurions and can be formed accounting for the transformation properties of the bilinears, e.g., $ \bar{q}_i \lzm Y_u \dzm^i_j q^k  \sim(\rep{3},\repbar{3},\rep{1})$ and $\bar{q}_k (Y_u^\dag) \ud{j}{\ell} q^\ell \sim(\repbar{3},\rep{3},\rep{1})$, and contracting the two. Examining and counting these combinations, we conclude that there are in total seven unique $\cO(Y^2)$ quartic structures, which we list below:\\

\noindent
$\boxed{(\bar{q}q)(\bar{q}q)}$
\begin{equation}
\label{eq268}
\begin{alignedat}{2}
    &\cO(1):~ (\bar{q}_iq^j)(\bar{q}_j q^i)~,\\
    &\cO(Y_u^2):~  (\bar{q}_i q^j)(Y_u Y_u^\dag)\ud{i}{k}(\bar{q}_j q^k)~,\qquad \cO(Y_d^2):~  (\bar{q}_i q^j)(Y_d Y_d^\dag)\ud{i}{k}(\bar{q}_j q^k)~.
\end{alignedat}
\end{equation}
$\boxed{(\bar{u}u)(\bar{u}u)}$
\begin{equation}
\begin{alignedat}{2}
        &\mathcal{O}(1):~ &&(\bar{u}_i u^j)(\bar{u}_j u^i)~,\qquad
        &\mathcal{O}(Y_u^2):~ &&  (\bar{u}_i u^j)(Y_u^\dag Y_u)\ud{i}{k}(\bar{u}_j u^k)~.
\end{alignedat}
\end{equation}
$\boxed{(\bar{d}d)(\bar{d}d)}$
\begin{equation}
\begin{alignedat}{2}
        &\mathcal{O}(1):~ (\bar{d}_i d^j)(\bar{d}_j d^i)~,\qquad \cO(Y_d^2):~  (\bar{d}_id^j)(Y_d^\dag Y_d)\ud{i}{k}(\bar{d}_jd^k)~.
\end{alignedat}
\end{equation}
$\boxed{(\bar{q}q)(\bar{u}u)}$
\begin{equation}
\begin{alignedat}{2}
    &\cO(Y_u^2):~ ( Y_u )\ud{i}{\ell} ( Y_u^\dag )\ud{k}{j}  ( \bar{q}_i  q^j )( \bar{u}_k  u^\ell )~.\\
\end{alignedat}
\end{equation}
$\boxed{(\bar{q}q)(\bar{d}d)}$
\begin{equation}
\begin{alignedat}{2}
    &\cO(Y_d^2):~ && ( Y_d)\ud{i}{k} ( Y_d^\dag )\ud{\ell}{j} ( \bar{q}_i  q^j ) ( \bar{d}_\ell  d^k)~.\\
\end{alignedat}
\end{equation}
$\boxed{(\bar{q}u)(\bar{q}d)}$
\begin{equation}
\label{eq273}
\begin{split}
    &\cO(Y_u Y_d):~ ( Y_d )\ud{i}{j} ( Y_u )\ud{\ell}{k} ( \bar{q}_i  u^k ) ( \bar{q}_\ell  d^j )~.
\end{split}
\end{equation}

\section{Lepton Sector}
\label{sec:leptonic}

Similarly to quarks, leptons come in three flavors, allowing for flavor transformations, which leaves physics unchanged. The lepton kinetic terms are symmetric under flavor transformations from the group $ G_L = \U(3)_\ell \times \U(3)_e $. In the SM, this symmetry is broken explicitly by the Yukawa term
	\begin{equation}
	\mathcal{L} \supset - \bar{\ell}_\LL Y_e e_\RR H \hc,   
	\end{equation}
leaving an accidental $ \U(1)^3 $ symmetry conserving the individual lepton numbers. Clearly, the observation of neutrino oscillations indicates that BSM physics must necessarily violate this accidental symmetry at some level. On the other hand, the non-observation of any charged lepton flavor--violating decays indicates that TeV-scale NP must suppress contribution to such processes. 

In contrast to the quark Yukawa matrices, from which derives the CKM matrix, the charged lepton Yukawa matrix is not fixed by the PMNS mixing matrix, which could come from the neutrino sector. Neutrino masses in the SMEFT originate from the lepton number-violating operators, such as the dimension-5 Weinberg operator. All flavor symmetries we consider here contain the usual lepton number symmetry as a subgroup, and this symmetry is preserved by the spurions populating the Yukawa matrix. The inclusion of  neutrino masses, therefore, necessitates a new lepton number--violating spurions. Due to the smallness of the neutrino masses, these additional spurions must be vanishingly small (for TeV-scale NP) and can be neglected.\footnote{Furthermore, for the lepton number--conserving operators we consider here, the neutrino mass spurions must combine in lepton number--conserving combinations requiring at least two such. This further suppresses the relevance of such insertions.} In extreme examples where the neutrino mass operators are highly suppressed by loop factors and/or operator dimension rather than the size of the spurions, the relevant spurions can simply be included in the symmetries. 

In this work, we identify several viable scenarios for the lepton flavor structure of the SMEFT that can accommodate hierarchical Yukawa couplings while suppressing charged lepton flavor--violating contributions from the dimension-6 operators. Accordingly, we consider a variety of different options for a flavor symmetry $ G \subset G_L $:
\begin{enumerate}[i)]
\item $ G= \U(1)^3 $ vectorial provides lepton flavor conservation but does not allow for any spurions providing a perturbative suppression of the electron mass; 
\item $ G= \U(1)^6 $ is also compatible with exact lepton number conservation but allows for a controlled expansion in lepton masses; 
\item $ G= \U(2) $ vectorial symmetry gives the additional correlation between the light leptons but no LFV;
\item $ G= \U(2)^2 $ decouples the 3rd generation from the first two generations of leptons while completely forbidding LFV depending on the minimal set of spurions; 
\item $ G= \U(2)^2 \times \U(1)^2 $ decouples the 3rd generation from the first two generations of leptons and allows for a perturbative expansion in all lepton masses;
\item $ G= \U(3) $ vectorial symmetry is compatible with exact lepton flavor conservation; 
\item $ G= \U(3)^2 $, linearly realized MFV symmetry gives lepton flavor conservation but only SM-like violation of lepton flavor universality. 
\end{enumerate}

\subsection{$\U(1)^3$ vectorial symmetry}
The first symmetry of the lepton sector we consider is the vectorial $ G= \U(1)^3 = \U(1)_{e} \times \U(1)_\mu \times \U(1)_\tau \subset G_L $ symmetry, under which the $\ell$ and $e$ fields decompose as
    \begin{equation}\label{eq:3.2.}
    \ell= \begin{bmatrix}
	\ell_1\sim (1,0,0)\\
	\ell_2\sim (0,1,0)\\
	\ell_3\sim (0,0,1)
	\end{bmatrix},\quad e= \begin{bmatrix}
	e_1\sim (1,0,0)\\
	e_2\sim (0,1,0)\\
	e_3\sim (0,0,1)
	\end{bmatrix},
    \end{equation}
with the $\U(1)$ charges indicated in the brackets. Since the $\U(1)^3$ vectorial symmetry is exact, there are no spurions present. The Yukawa matrix can be written as 
\begin{equation}
    Y_e = \begin{bmatrix} y_e &&\\&y_\mu&\\&&y_\tau \end{bmatrix}.
\end{equation}
These coefficients break the axial $\U(1)^3_A \subset G_L $ symmetry, which can be used to rotate away unphysical phases from $y_{e,\mu,\tau}$. Accordingly, they can be chosen to be real and positive. The numerical values of these parameters are
\begin{equation}
    y_e=\num{2.793e-6},\quad y_\mu=\num{5.884e-4},\quad y_\tau=\num{9.994e-3}.
\end{equation}
The smallness ($y_\ell \ll 1$) and the hierarchy among different generations is an open question. In Section~\ref{sec:3.2}, these parameters are actually spurions, which explains their smallness but not the hierarchy. The motivation to consider this flavor structure comes from stringent experimental constraints on charged lepton flavor--violating processes. 

The neutrino masses can be minimally accounted for by the dimension-5 Weinberg operator and three additional spurions with the opposite charge to the leptons in Eq.~\eqref{eq:3.2.}. The PMNS mixing matrix is accommodated by assuming no hierarchy in the three spurions. Note that, for the TeV-scale cutoff, these spurions take extremely small values in order to reproduce the observed neutrino masses. Therefore, their effect on charged lepton flavor--violating processes is negligible. For these reasons, we omit them from the counting of dimension-6 operators. 

The counting of the pure lepton SMEFT operators is presented in Table~\ref{lep3U1table} and the decompositions of the bilinear and unique quartic structures are listed in Eqs.~(\ref{eq32}--\ref{eq36}).

\begin{table}[t]
\centering
\begin{tabular}{|cc|cc|}
\hline
\multicolumn{2}{|c|}{$\U(1)_e \times \U(1)_\mu \times \U(1)_\tau $}                                           & \multicolumn{2}{c|}{$\cO(1)$} \\ \hline
\multicolumn{1}{|c|}{$\psi^2 H^3 $}                   & $Q_{eH}$            & 3             & 3             \\ \hline
\multicolumn{1}{|c|}{$\psi^2 XH $}                    & $Q_{e(W,B)}$  & 6             & 6             \\ \hline
\multicolumn{1}{|c|}{\multirow{2}{*}{$\psi^2 H^2 D$}} & $Q_{H\ell}^{(1,3)}$ & 6             &               \\
\multicolumn{1}{|c|}{}                                & $Q_{He}$            & 3             &               \\ \hline
\multicolumn{1}{|c|}{$(LL)(LL)$}                      & $Q_{\ell\ell} $     & 9             &               \\ \hline
\multicolumn{1}{|c|}{$(RR)(RR)$}                      & $Q_{ee} $           & 6             &               \\ \hline
\multicolumn{1}{|c|}{$(LL)(RR)$}                      & $Q_{\ell e} $       & 12            & 3             \\ \hline\hline
\multicolumn{2}{|c|}{Total}                                                 & 45            & 12            \\ \hline
\end{tabular}
\caption{Counting of the pure lepton SMEFT operators (see Appendix \ref{app:Warsaw}) assuming $\U(1)_V^3$ symmetry in the lepton sector. Since the $\U(1)_V^3$ symmetry is exact (no spurions), the counting is presented for the $\cO(1)$ operators only. The left (right) entry in each column gives the number of CP even (odd) coefficients at the given order in spurion counting.}
\label{lep3U1table}
\end{table}

\subsubsection*{Decomposition of bilinear and quartic structures}
Constructing bilinear and quartic structures is straightforward in the case of $\U(1)^3$ vectorial symmetry since only $\cO(1)$ structures are present. These are given by $(\bar{\ell}_i \ell_i)$, $(\bar{e}_i e_i)$ and $(\bar{\ell}_i e_i)$. Similarly, the only possible non-factorizable quartic structures are given by $(\bar{\ell}_i \ell_j)(\bar{\ell}_j \ell_i)$ and $(\bar{\ell}_i \ell_j)(\bar{e}_j e_i)$.

The structures of the form $(\bar{e}_ie_j)(\bar{e}_je_i)$ are identical to $(\bar{e}_ie_i)(\bar{e}_je_j)$ due to the Fierz identity for vector currents in the underlying operators. This is valid not only here but for all the lepton symmetries we consider. That is, there are no unique $(\bar{e}e)(\bar{e}e)$ structures. Thus, the quartic structures with four insertions of $e$ are formed solely by multiplying $(\bar{e}e)$ bilinears. 
We list the decompositions of the bilinear and unique quartic structures below:\\  

\noindent $\boxed{(\bar{\ell}e)}$
\begin{equation}
\label{eq32}
    \mathcal{O}(1):~ (\Bar{\ell}_1 e_1)~,\quad (\Bar{\ell}_2 e_2)~,\quad (\Bar{\ell}_3 e_3)~.
\end{equation}
$\boxed{(\bar{\ell}\ell)}$
\begin{equation}
    \mathcal{O}(1):~ (\Bar{\ell}_1 \ell_1)~,\quad (\Bar{\ell}_2 \ell_2)~,\quad (\Bar{\ell}_3 \ell_3)~.
\end{equation}
$\boxed{(\bar{e}e)}$
\begin{equation}
    \mathcal{O}(1):~ (\Bar{e}_1 e_1)~,\quad (\Bar{e}_2 e_2)~,\quad (\Bar{e}_3 e_3)~.
\end{equation}
$\boxed{(\bar{\ell}\ell)(\bar{\ell}\ell)}$
\begin{equation}
\begin{alignedat}{1}
        \mathcal{O}(1):~ (\bar{\ell}_1 \ell_2)(\bar{\ell}_2 \ell_1)~,\quad (\bar{\ell}_1 \ell_3)(\bar{\ell}_3 \ell_1)~,\quad (\bar{\ell}_2 \ell_3)(\bar{\ell}_3 \ell_2)~.
\end{alignedat}
\end{equation}
$\boxed{(\bar{\ell}\ell)(\bar{e}e)}$
\begin{equation}
\label{eq36}
    \begin{split}
        \mathcal{O}(1):~(\bar{\ell}_1 \ell_2)(\bar{e}_2 e_1)~,\quad (\bar{\ell}_2 \ell_3)(\bar{e}_3 e_2)~,\quad (\bar{\ell}_3 \ell_1)(\bar{e}_1 e_3)~,\quad \hermc~.
    \end{split}
\end{equation}
$\boxed{(\bar{e}e)(\bar{e}e)}$
\begin{equation}
\text{No unique structures present due to Fierz identities}.
\nonumber
\end{equation}


\subsection{$\U(1)^6$ symmetry}
\label{sec:3.2}
For a $ G= \U(1)^6 = \U(1)_{\ell_1 }\times \U(1)_{e_1} \times \U(1)_{\ell_2} \times \U(1)_{e_2} \times \U(1)_{\ell_3}\times \U(1)_{e_3} \subset G_L $ symmetry of the lepton sector, we have the field decompositions and $ \U(1)^6 $ charge assignments
    \begin{equation}
        \ell = \begin{bmatrix}
\ell_1\sim (1,0,0,0,0,0)\\
\ell_2\sim (0,0,1,0,0,0)\\
\ell_3\sim (0,0,0,0,1,0)
\end{bmatrix},\quad e= \begin{bmatrix}
e_1 \sim (0,1,0,0,0,0)\\
e_2 \sim (0,0,0,1,0,0)\\
e_3 \sim (0,0,0,0,0,1)
\end{bmatrix}.
    \end{equation}
The minimal set of spurions required to write the Yukawa matrix is 
    \begin{equation}
        y_{e} \sim (1,-1,0,0,0,0),\quad y_{\mu} \sim (0,0,1,-1,0,0),\quad y_{\tau} \sim (0,0,0,0,1,-1),
    \end{equation}
which we use to write the Yukawa matrix
\begin{equation}
Y_e =\begin{bmatrix} y_e&&\\&y_\mu&\\&&y_\tau \end{bmatrix},
\end{equation}
The flavor symmetry breaking pattern is given by:
\begin{equation}
    \begin{alignedat}{2}
        &y_{e}:\quad \U(1)_{e_\LL}\times \U(1)_{e_\RR}&&\longrightarrow \U(1)_{L_e},\\
        &y_{\mu}:\quad \U(1)_{\mu_\LL}\times \U(1)_{\mu_\RR}&&\longrightarrow \U(1)_{L_\mu},\\
        &y_{\tau}:\quad \U(1)_{\tau_\LL}\times \U(1)_{\tau_\RR}&&\longrightarrow \U(1)_{L_\tau},\\
    \end{alignedat}
\end{equation}
yielding a total of 3 broken generators. 
This allows for the elimination of redundant phases from $y_{e,\mu,\tau}$, which can all be taken to be positive, real numbers. 

We present the spurion counting of the leptonic operators assuming $\U(1)^6$ flavor symmetry in Table \ref{tab:lep6U1}. The decompositions are listed in Eqs.~(\ref{eq312}--\ref{eq315}).
\begin{table}[t]
\centering
\begin{tabular}{|cc|cc|cc|}
\hline
\multicolumn{2}{|c|}{$ \U(1)^6 $}                                           & \multicolumn{2}{c|}{$\cO(1)$} & \multicolumn{2}{c|}{$\cO(y)$} \\ \hline
\multicolumn{1}{|c|}{$\psi^2 H^3 $}                   & $Q_{eH}$            &                &              & 3             & 3             \\ \hline
\multicolumn{1}{|c|}{$\psi^2 XH $}                    & $Q_{e(W,B)}$  &                &              & 6             & 6             \\ \hline
\multicolumn{1}{|c|}{\multirow{2}{*}{$\psi^2 H^2 D$}} & $Q_{H\ell}^{(1,3)}$ & 6              &             &               &               \\
\multicolumn{1}{|c|}{}                                & $Q_{He}$            & 3              &             &               &               \\ \hline
\multicolumn{1}{|c|}{$(LL)(LL)$}                      & $Q_{\ell\ell} $     & 9              &              &               &               \\ \hline
\multicolumn{1}{|c|}{$(RR)(RR)$}                      & $Q_{ee} $           & 6              &              &               &               \\ \hline
\multicolumn{1}{|c|}{$(LL)(RR)$}                      & $Q_{\ell e} $       & 9              &              &               &               \\ \hline\hline
\multicolumn{2}{|c|}{Total}                                                 & 33             &             & 9             & 9             \\ \hline
\end{tabular}
\caption{Counting of the pure lepton SMEFT operators (see Appendix \ref{app:Warsaw}) assuming $\U(1)^6$ symmetry in the lepton sector. The counting is performed taking up to one insertion of $y_{e,\mu,\tau}$ spurion. The left (right) entry in each column gives the number of CP even (odd) coefficients at the given order in spurion counting.}
\label{tab:lep6U1}
\end{table}

\subsubsection*{Decomposition of bilinear and quartic structures}
In this case, the $\cO(1)$ bilinear structures can only be constructed with two appearances of the same field. These structures are $(\bar{\ell}_i \ell_i)$ and $(\bar{e}_i e_i)$. 
Matching the charges, the bilinear structure with one insertion of a spurion $(Y_e)_{ii}$ can only be of the form $[\bar{\ell}_i (Y_e)_{ii} e_i]$. The set of unique quartic structures is comprised of three $\cO(1)$ structures of the form $(\Bar{\ell}_1 \ell_2)(\Bar{\ell}_2 \ell_1)$, $(\Bar{\ell}_2 \ell_3)(\Bar{\ell}_3 \ell_2)$ and $(\Bar{\ell}_3 \ell_1)(\Bar{\ell}_1 \ell_3)$, whereas there are no unique $(\bar{\ell}\ell)(\bar{e}e)$ structures present. We list the decompositions below:\\\\
$\boxed{(\bar{\ell}e)}$
\begin{equation}
\label{eq312}
    \mathcal{O}(y):~ (\Bar{\ell}_1 y_{e} e_1)~,\quad  (\Bar{\ell}_2 y_{\mu} e_2)~,\quad  (\Bar{\ell}_3 y_{\tau} e_3)~.
\end{equation}
$\boxed{(\bar{\ell}\ell)}$
\begin{equation}
    \mathcal{O}(1):~ (\Bar{\ell}_1 \ell_1)~,\quad (\Bar{\ell}_2 \ell_2)~,\quad (\Bar{\ell}_3 \ell_3)~.
\end{equation}
$\boxed{(\bar{e}e)}$
\begin{equation}
    \mathcal{O}(1):~ (\Bar{e}_1 e_1)~,\quad (\Bar{e}_2 e_2)~,\quad (\Bar{e}_3 e_3)~.
\end{equation}
$\boxed{(\bar{\ell}\ell)(\bar{\ell}\ell)}$
\begin{equation}
\label{eq315}
\begin{split}
        \mathcal{O}(1):~ (\Bar{\ell}_1 \ell_2)(\Bar{\ell}_2 \ell_1)~,\quad (\Bar{\ell}_2 \ell_3)(\Bar{\ell}_3 \ell_2)~,\quad (\Bar{\ell}_3 \ell_1)(\Bar{\ell}_1 \ell_3)~.
\end{split}
\end{equation}
$\boxed{(\bar{\ell}\ell)(\bar{e}e)}$
\begin{equation}
\text{No unique structures present}.
\nonumber
\end{equation}
$\boxed{(\bar{e}e)(\bar{e}e)}$
\begin{equation}
\text{No unique structures present due to Fierz identites}.
\nonumber
\end{equation}

\subsection{$\U(2)$ vectorial symmetry}
Next, we consider a $\U(2) \subset G_L$ vectorial flavor symmetry, under which the fields decompose as
\begin{equation}
    \begin{alignedat}{2}
            \ell&=\begin{bmatrix} \ell^a\sim\rep{2} \\ \ell_3\sim\rep{1} \end{bmatrix},\quad
            e &= \begin{bmatrix} e^a\sim\rep{2}\\e_3 \sim\rep{1} \end{bmatrix}.
    \end{alignedat}
\end{equation}
The minimal choice of spurion necessary to produce a realistic Yukawa coupling is $\Delta_\ell\sim\rep{3}$, which we take to be real.\footnote{A slightly less minimal choice would be to introduce a $ V_\ell \sim \rep{2} $ in place of $\Delta_\ell $.} We use the simplifying notation
\begin{equation}
    \Delta^I_\ell (T^I)\ud{a}{b} = (\Delta_\ell)\ud{a}{b}.
\end{equation}
With this spurion, the Yukawa matrix generically takes the form 
\begin{equation} \label{eq:U2V_Yuk}
    Y_e = \begin{bmatrix} \Delta_\ell+ s_\ell \mathds{1} & 0\\0 & y_\tau \end{bmatrix}.
\end{equation}
$\Delta_\ell$ breaks the $\U(2)_V$ symmetry as
\begin{equation}
    \begin{alignedat}{2}
        &\Delta_\ell\longrightarrow\begin{bmatrix}
      -\delta_\ell & 0 \\
      0 & \delta_\ell
    \end{bmatrix}:\qquad\qquad &&\U(2)_V\longrightarrow \U(1)^2_{\ell+e},\\
    \end{alignedat}
\end{equation}
where we use the general properties of the special unitary matrices along with the flavor symmetry breaking pattern to parametrize the $\Delta_\ell$ spurion with 1 real parameter. The Yukawa matrix~\eqref{eq:U2V_Yuk} also preserves an accidental $ \U(1)_{\ell_3+e_3} $ symmetry, while it breaks axial $ \U(1)'$s, which can be used to remove phases from the coefficients $ s_\ell$ and $ y_\tau $. %
The numerical values of the relevant parameters are
\begin{equation}
s_\ell =\num{2.956e-4},\quad 
\delta_\ell = \num{2.928e-4},\quad 
y_\tau = \num{9.994e-3}.
\end{equation}
This flavor structure fails to explain the hierarchy among generations. Furthermore, a tuning is needed between $s_\ell$ and (the symmetry-breaking) $\delta_\ell$ in order to accommodate for the observed $e$ and $\mu$ masses.

The spurion counting of the pure lepton operators assuming $\U(2)_V$ symmetry is given in Table \ref{tab:lepU2diag} and we list the decompositions in Eqs.~(\ref{eq323}--\ref{eq327}).

\begin{table}[t]
\centering
\begin{tabular}{|cc|cc|cc|}
\hline
\multicolumn{2}{|c|}{$\U(2)_V$}                                   & \multicolumn{2}{c|}{$\cO(1)$} & \multicolumn{2}{c|}{$\cO(\Delta_\ell)$} \\ \hline
\multicolumn{1}{|c|}{$\psi^2 H^3 $}                   & $Q_{eH}$            & 2              & 2            & 1                   & 1                 \\ \hline
\multicolumn{1}{|c|}{$\psi^2 XH $}                    & $Q_{e(W,B)}$        & 4              & 4            & 2                   & 2                 \\ \hline
\multicolumn{1}{|c|}{\multirow{2}{*}{$\psi^2 H^2 D$}} & $Q_{H\ell}^{(1,3)}$ & 4              &              & 2                   &                   \\
\multicolumn{1}{|c|}{}                                & $Q_{He}$            & 2              &              & 1                   &                   \\ \hline
\multicolumn{1}{|c|}{$(LL)(LL)$}                      & $Q_{\ell\ell} $     & 5              &              & 3                   &                   \\ \hline
\multicolumn{1}{|c|}{$(RR)(RR)$}                      & $Q_{ee} $           & 3              &              & 2                   &                   \\ \hline
\multicolumn{1}{|c|}{$(LL)(RR)$}                      & $Q_{\ell e} $       & 6              & 1            & 5                   & 2                 \\ \hline\hline
\multicolumn{2}{|c|}{Total}                                                 & 26             & 7            & 16                  & 5                 \\ \hline
\end{tabular}
\caption{Counting of the pure lepton SMEFT operators (see Appendix \ref{app:Warsaw}) assuming $\U(2)$ vectorial symmetry in the lepton sector. The counting is performed taking up to one insertion of $\Delta_\ell$ spurion. The left (right) entry in each column gives the number of CP even (odd) coefficients at the given order in spurion counting.}
\label{tab:lepU2diag}
\end{table}

\subsubsection*{Decomposition of bilinear and quartic structures}
$\cO(1)$ bilinear structures can only be constructed by directly contracting the doublets or the singlets of the $\ell$ and $e$ fields. There are six such structures: $(\bar{\ell}e)$, $(\bar{\ell}_3 e_3)$, $(\bar{\ell}\ell)$, $(\bar{\ell}_3 \ell_3)$, $(\bar{e}e)$, and $(\bar{e}_3 e_3)$. To form the $\cO(\Delta_\ell)$ bilinears, the $\Delta_\ell$ has to be contracted to $\ell^a$ and $e^a$, giving three bilinears: $(\bar{\ell}\Delta_\ell e)$, $(\bar{\ell}\Delta_\ell \ell)$, and $(\bar{e}\Delta_\ell e)$.

Analyzing the possible contractions involving four field appearances in an analogous way, we obtain four $\cO(1)$ quartic structures: $(\bar{\ell}_a \ell^b)(\bar{\ell}_b \ell^a)$, $(\bar{\ell}_a \ell_3)(\bar{\ell}_3 \ell^a)$, $(\bar{\ell}_a \ell)(\bar{e} e^a)$, and $ (\bar{\ell}_a \ell_3)(\bar{e}_3 e^a) $. Similarly, we find that there are three $\cO(\Delta_\ell)$ structures given by $(\bar{\ell}_a \ell_3)\lzm \Delta_\ell \dzm^a_{\phantom{b}b}(\bar{\ell}_3 \ell^b)$, $(\bar{\ell}_a \ell^b)\lzm \Delta_\ell \dzm^{a}_{\phantom{b}c} (\bar{e}_b e^c)$, and $(\bar{\ell}_a \ell_3)\lzm \Delta_\ell \dzm_{\phantom{b}b}^{a} (\bar{e}_3 e^b)$.

We remark that there is an overcounting if we include all the  $(\bar{\ell}\ell)(\bar{e}e)$ factorizing quartic structures contain only the $\U(2)_V$ doublets $\ell$ and $e$. Having the $\cO(1)$ and $\cO(\Delta_\ell)$ decompositions of the bilinear structures, the factorizing $\cO(\Delta_\ell)$ quartic structures are formed simply by multiplying the corresponding bilinears. Applying this recipe trivially gives two $\cO(\Delta_\ell)$ structures of the form $(\bar{\ell}\ell)(\bar{e} \Delta_\ell e)$, $(\bar{\ell} \Delta_\ell \ell) (\bar{e}e)$. It turns out that due to the group identity \ref{dsixident}, one of these structures can be expressed in terms of the other factorizing quartic structure and the unique contraction $(\bar{\ell}_a \ell^b)\lzm \Delta_\ell \dzm^{a}_{\phantom{b}c} (\bar{e}_b e^c)$ and its Hermitian conjugate: 
\begin{equation}
    \begin{alignedat}{2}
    (\bar{\ell}\ell)(\bar{e}\Delta_\ell e)&= \Delta_\ell^I (t^I)\ud{a}{b}\delta\ud{c}{d} (\bar{\ell}_c \ell^d)(\bar{e}_a e^b) = 
    \Delta_\ell^I \lzs (t^I)\ud{c}{b}\delta\ud{a}{d}-(t^I)\ud{c}{d}\delta\ud{a}{b}+(t^I)\ud{a}{d}\delta\ud{c}{b} \dzs(\bar{\ell}_c \ell^d)(\bar{e}_a e^b)\\
    &=\lzs (\bar{\ell}_a \ell^b)\lzm \Delta_\ell \dzm^{a}_{\phantom{b}c} (\bar{e}_b e^c) +\hermc \dzs - (\bar{\ell}\Delta_\ell \ell)(\bar{e}e).
    \end{alignedat}
\end{equation}
The $(\bar{\ell}\ell)(\bar{e}\Delta_\ell e)$ structure, therefore, drops from the counting. We list the complete decompositions below:\\\\
$\boxed{(\bar{\ell}e)}$
\begin{equation}
\label{eq323}
\begin{alignedat}{2}
        &\mathcal{O}(1):~(\bar{\ell}e)~,\quad(\bar{\ell}_3 e_3)~,\qquad\mathcal{O}(\Delta_\ell):~(\bar{\ell} \Delta_\ell e)~.
\end{alignedat}
\end{equation}
$\boxed{(\bar{\ell}\ell)}$
\begin{equation}
\begin{alignedat}{2}
        &\mathcal{O}(1):~(\bar{\ell}\ell)~,\quad(\bar{\ell}_3 \ell_3)~,\qquad\mathcal{O}(\Delta_\ell):~(\bar{\ell} \Delta_\ell \ell)~.
\end{alignedat}
\end{equation}
$\boxed{(\bar{e}e)}$
\begin{equation}
\begin{alignedat}{2}
        &\mathcal{O}(1):~(\bar{e}e)~,\quad(\bar{e}_3 e_3)~,\qquad\mathcal{O}(\Delta_\ell):~(\bar{e} \Delta_\ell e)~.
\end{alignedat}
\end{equation}
$\boxed{(\bar{\ell}\ell)(\bar{\ell}\ell)}$
\begin{equation}
\begin{alignedat}{2}
&\mathcal{O}(1):~(\bar{\ell}_a \ell^b)(\bar{\ell}_b \ell^a)~,\quad (\bar{\ell}_a \ell_3)(\bar{\ell}_3 \ell^a)~,\qquad\mathcal{O}(\Delta_\ell):~(\bar{\ell}_a \ell_3)\lzm \Delta_\ell \dzm^a_{\phantom{b}b}(\bar{\ell}_3 \ell^b)~.
\end{alignedat}
\end{equation}
$\boxed{(\bar{\ell}\ell)(\bar{e}e)}$
\begin{equation}
\label{eq327}
{\small
\begin{alignedat}{2}
        &\mathcal{O}(1):~&&(\bar{\ell}_a \ell)(\bar{e} e^a)~,\quad \lzs (\bar{\ell}_a \ell_3)(\bar{e}_3 e^a)~,\quad\hermc \dzs~,\qquad\\
        &\mathcal{O}(\Delta_\ell):~&&(\bar{\ell}_a \ell^b)\lzm \Delta_\ell \dzm^{a}_{\phantom{b}c} (\bar{e}_b e^c)~,\quad(\bar{\ell}_a \ell_3)\lzm \Delta_\ell \dzm_{\phantom{b}b}^{a} (\bar{e}_3 e^b)~,\quad\hermc~.
            \color{black}
\end{alignedat}}
\end{equation}
$\boxed{(\bar{e}e)(\bar{e}e)}$
\begin{equation}
\text{No unique structures present due to Fierz identities}.
\nonumber
\end{equation}


\subsection{$\U(2)^2$ symmetry}
We consider the case where NP is invariant under $G=  \U(2)_\ell \times \U(2)_e \subset G_L$. In this case, the fields decompose as 
\begin{equation}
    \begin{alignedat}{2}
            &\ell= 
            \begin{bmatrix}
                \ell^a\sim (\boldsymbol2,\boldsymbol1) \\
                \ell_3\sim(\boldsymbol1,\boldsymbol1)
            \end{bmatrix},\quad e=
            \begin{bmatrix}
                e^a\sim(\boldsymbol1,\boldsymbol2) \\
                e_3\sim (\boldsymbol1,\boldsymbol1)
            \end{bmatrix}.
    \end{alignedat}
\end{equation}
In order to write a realistic lepton Yukawa matrix, a spurion
\begin{equation}
    \Delta_e \sim (\boldsymbol2,\boldsymbol{\bar{2}})
\end{equation}
is required. It is then possible to write the Yukawa matrix
\begin{equation}
    Y_e = \begin{bmatrix}
        \Delta_e & 0 \\
         0& y_\tau
    \end{bmatrix}.
\end{equation}
We also allow for the non-minimal inclusion of the spurion $V_\ell \sim (\boldsymbol2,\boldsymbol1)$ to allow for mixing third-generation leptons with the light generations. Note that $V_\ell$ will be included in the decompositions (SMEFT operators), but it is absent from the Yukawa matrix since the inclusion of $V_\ell$ leads to a non-minimal parametrization of Yukawa. 

We determine the breaking pattern of the $\U(2)^2$ flavor symmetry by the spurions. Fixing the $\Delta_e$ bi-doublet, we have
\begin{align}
    &\Delta_e \longrightarrow \begin{bmatrix}
    \delta_\ell & 0\\
    0 & \delta'_\ell
    \end{bmatrix}: &&\U(2)_{\ell}\times \U(2)_{e}\longrightarrow \U(1)^2_{\ell+e}.
\intertext{In the second step, $V_\ell$ breaks}
    &V_\ell \longrightarrow     \begin{bmatrix}
        \epsilon_\ell\\ \epsilon'_\ell
    \end{bmatrix}: &&\U(1)^2_{\ell+e} \longrightarrow \emptyset,
\end{align}
for $ \epsilon^{(\prime)}_\ell > 0 $.
Thus, we conclude that these spurions completely break the flavor $\U(2)^2$ symmetry, making it possible to remove 8 unphysical parameters from the parametrization of the spurions. $ y_\tau $ breaks a third generation axial $ \U(1) $ symmetry, and its unphysical phase can be removed. The numerical values of the relevant parameters for the Yukawa matrix are
\begin{equation}
    \delta_\ell= \num{2.793e-6},\quad \delta_\ell'= \num{5.884e-4},\quad y_\tau= \num{9.994e-3}.
\end{equation}
This flavor structure provides a rationale for why the tau is much heavier than the other two leptons. The light lepton masses can be accommodated without tuning in contrast to the previous section; however, the hierarchy among them is left unexplained.

The spurion counting of the pure lepton operators assuming $\U(2)^2$ symmetry of the lepton sector is presented in Table \ref{tab:lep2U2}. The decompositions of bilinear and quartic structures are listed in Eqs.~(\ref{eq334}--\ref{eq338}).
\begin{table}[t]
\centering
\begin{tabular}{|cc|cc|cc|cc|cc|cc|cc|}
\hline
\multicolumn{2}{|c|}{$\U(2)_\ell \times \U(2)_e $}                              & \multicolumn{2}{c|}{$\cO(1)$} & \multicolumn{2}{c|}{$\cO(V)$} & \multicolumn{2}{c|}{$\cO(V^2)$} & \multicolumn{2}{c|}{$\cO(V^3)$} & \multicolumn{2}{c|}{$\cO(\Delta)$} & \multicolumn{2}{c|}{$\cO(\Delta V)$} \\ \hline
\multicolumn{1}{|c|}{$\psi^2 H^3 $}                   & $Q_{eH}$            & 1              & 1            & 1             & 1             &                &                &                &                & 1                & 1               & 1                 & 1                \\ \hline
\multicolumn{1}{|c|}{$\psi^2 XH $}                    & $Q_{e(W,B)}$  & 2              & 2            & 2             & 2             &                &                &                &                & 2                & 2               & 2                 & 2                \\ \hline
\multicolumn{1}{|c|}{\multirow{2}{*}{$\psi^2 H^2 D$}} & $Q_{H\ell}^{(1,3)}$ & 4              &              & 2             & 2             & 2              &                &                &                &                  &                 &                   &                  \\
\multicolumn{1}{|c|}{}                                & $Q_{He}$            & 2              &              &               &               &                &                &                &                &                  &                 & 1                 & 1                \\ \hline
\multicolumn{1}{|c|}{$(LL)(LL)$}                      & $Q_{\ell\ell} $     & 5              &              & 3             & 3             & 5              & 1              & 1              & 1              &                  &                 &                   &                  \\ \hline
\multicolumn{1}{|c|}{$(RR)(RR)$}                      & $Q_{ee} $           & 3              &              &               &               &                &                &                &                &                  &                 & 2                 & 2                \\ \hline
\multicolumn{1}{|c|}{$(LL)(RR)$}                      & $Q_{\ell e} $       & 4              &              & 2             & 2             & 2              &                &                &                & 1                & 1               & 3                 & 3                \\ \hline\hline
\multicolumn{2}{|c|}{Total}                                                 & 21             & 3             & 10            & 10            & 9              & 1              & 1              & 1              & 4                & 4               & 9                 & 9                \\ \hline
\end{tabular}
\caption{Counting of the pure lepton SMEFT operators (see Appendix \ref{app:Warsaw}) assuming $\U(2)_\ell \times \U(2)_e $ symmetry in the lepton sector. Analogously to the counting performed in the quark sector assuming the same symmetry (see Table \ref{tab:quark3U2}), we once again take up to three insertions of $V_\ell$ spurion, one insertion of $\Delta_{e}$ and one insertion of the $\Delta_{e}V_\ell$ spurion product. The left (right) entry in each column gives the number of CP even (odd) coefficients at the given order in spurion counting.}
\label{tab:lep2U2}
\end{table}

\subsubsection*{Decomposition of bilinear and quartic structures}
Forming the structures invariant under $\U(2)^2$ symmetry follows the same approach as in the case of $\U(2)^3$ symmetry in the quark sector (see Eqs.~(\ref{eq212}--\ref{eq223})). Therefore, in order to obtain the invariant structures, it is sufficient to take the corresponding structures in the quark sector (either $q$ and $u$ or $q$ and $d$ structures) and do a relabeling $q \to \ell$ and $u/d \to e$. The decompositions of the bilinear and quartic structures are listed below:\\\\
$\boxed{(\bar{\ell}e)}$
\begin{equation}
\label{eq334}
\begin{alignedat}{2}
        &\mathcal{O}(1):~ (\bar{\ell}_3 e_3)~,\qquad\mathcal{O}(V):~ (\bar{\ell} V_\ell e_3)~,\qquad\mathcal{O}(\Delta):~ (\bar{\ell} \Delta_e e)~,\qquad\mathcal{O}(\Delta V):~ [ \bar{\ell}_3 (V_\ell^\dag \Delta_e) e ]~.
        \end{alignedat}
\end{equation}
$\boxed{(\bar{\ell}\ell)}$
\begin{equation}
    \begin{alignedat}{2}
        &\mathcal{O}(1):~ (\bar{\ell}\ell)~,\quad (\bar{\ell}_3 \ell_3)~,\qquad\mathcal{O}(V):~ (\bar{\ell}V_\ell \ell_3)~,\quad \hermc~,\qquad\mathcal{O}(V^2):~ ( \bar{\ell} V_\ell V_\ell^\dag \ell )~.\\
    \end{alignedat}
\end{equation}
$\boxed{(\bar{e}e)}$
\begin{equation}
        \begin{alignedat}{2}
        &\mathcal{O}(1):~ (\bar{e}e)~,\quad (\bar{e}_3 e_3)~,\qquad\mathcal{O}(\Delta V):~  (\bar{e} \Delta_e^\dag V_\ell e_3)~,\quad \hermc~.
    \end{alignedat} 
\end{equation}
$\boxed{(\bar{\ell}\ell)(\bar{\ell}\ell)}$
\begin{equation}
\begin{alignedat}{2}
        &\mathcal{O}(1):~ (\bar{\ell}_a \ell^b)(\bar{\ell}_b \ell^a)~,\quad (\bar{\ell}\ell_3)(\bar{\ell}_3 \ell)~,\qquad\mathcal{O}(V):~ (\bar{\ell} V_\ell \ell^b)(\bar{\ell}_b \ell_3)~,\quad \hermc~,\\
        &\mathcal{O}(V^2):~ (\bar{\ell} V_\ell \ell^b)(\bar{\ell}_b V_\ell^\dag \ell)~. \\
\end{alignedat}
\end{equation}
$\boxed{(\bar{\ell}\ell)(\bar{e}e)}$
\begin{equation}
\label{eq338}
\begin{alignedat}{2}
        &\mathcal{O}(\Delta):~ (\bar{\ell}_a \ell_3)(\Delta_e)^a_{\phantom{b}b}(\bar{e}_3 e^b)~,\quad \hermc~,\qquad\mathcal{O}(\Delta V):~ (\bar{\ell} V_\ell \ell^b)(\Delta_e)\ud{c}{b}(\bar{e}_c e_3)~,\quad \hermc~.
\end{alignedat}
\end{equation}
$\boxed{(\bar{e}e)(\bar{e}e)}$
\begin{equation}
\text{No unique structures present due to Fierz identities}.
\nonumber
\end{equation}


\subsection{$\U(2)^2 \times \U(1)^2$ symmetry}
If we wish to include the $ \tau $ Yukawa as a spurion in our expansion, we can consider extending the lepton symmetry to $ G= \U(2)^2 \times \U(1)^2 \subset G_L$ (a similar construction is possible with one $ \U(1) $ factor).
Under this symmetry, the fields decompose as\footnote{The representations under the flavor group are indicated in the $(\U(2)_\ell,\U(2)_e \lvert \U(1)_{\ell_3},\U(1)_{e_3})$ format.}
\begin{equation}
    \ell= \begin{bmatrix}\ell^a\sim (\boldsymbol2,\boldsymbol1\lvert 0,0)\\\ell_3\sim (\boldsymbol1,\boldsymbol1\lvert 1,0)\end{bmatrix},\quad e= \begin{bmatrix}e^a\sim(\boldsymbol1,\boldsymbol2\lvert 0,0)\\e_3\sim (\boldsymbol1,\boldsymbol1\lvert 0,1)\end{bmatrix},
\end{equation}
and the minimal set of spurions required to produce a realistic Yukawa matrix is given by
\begin{equation}
        \Delta_e \sim (\boldsymbol2,\boldsymbol{\Bar{2}} \lvert 0,0),\quad X_\ell\sim(\boldsymbol1,\boldsymbol1\lvert 1,-1).
\end{equation}
With these spurions, the Yukawa matrix can be written as 
\begin{equation}
    Y_e = \begin{bmatrix}
    \Delta_e & 0\\
    0 & X_\ell
    \end{bmatrix}.
\end{equation}
We will also include $V_\ell \sim (\boldsymbol2,\boldsymbol1 \lvert 0,-1)$ and $V_e\sim(\boldsymbol1,\boldsymbol2\lvert -1,0)$ spurions in the decompositions of the bilinear and quartic structures, but these spurions can be set to 0 in the minimal parametrization of the Yukawa matrix (see Section~\ref{sec:quark_U2^3} for details).

Let us take a look at how this set of spurions breaks the flavor symmetry. First, $\Delta_e$ breaks
\begin{align}
    &\Delta_e\longrightarrow\begin{bmatrix}
        \delta_\ell &0\\0&\delta'_\ell
    \end{bmatrix}:&&\U(2)_\ell\times \U(2)_e\longrightarrow\U(1)^2_{\ell+e}.\\
\intertext{In the next step, $X_\ell$ breaks}
        &X_\ell \longrightarrow \chi_\ell :&&\U(1)_{\ell_3}\times\U(1)_{e_3}\longrightarrow\U(1)_{\ell_3+e_3}.
\intertext{Furthermore, inclusion of $V_\ell$ breaks the flavor symmetry down to global lepton number}
    &V_\ell\longrightarrow\begin{bmatrix}\epsilon_\ell \\ \epsilon'_\ell\end{bmatrix},
    &&\U(1)^2_{\ell+e}\times \U(1)_{\ell_3+e_3}\longrightarrow\U(1)_L,
\end{align}
with $ \epsilon_\ell^{(\prime)} >0$. 
$V_e$ does not break the symmetry further and has 2 complex parameters. 
We are left with 5 real, positive parameters and 2 complex parameters in the spurion parametrization. The numerical values of the spurion parameters appearing in the Yukawa matrix are given by
\begin{equation}
    \delta_\ell= \num{2.793e-6},\quad \delta_\ell'= \num{5.884e-4},\quad \chi_\ell= \num{9.994e-3}.
\end{equation}
The added advantage with respect to the previous section is the explanation of the small $y_\tau$.

The flavor spurion counting of the leptonic operators assuming $\U(2)^2\times \U(1)^2$ symmetry is presented in Table \ref{tab:lep2U2x2U1}, and the flavor decompositions of the bilinear and unique quartic structures are listed in Eqs.~(\ref{eq346}--\ref{eq350}).
\begin{table}[t]
\centering
\scalebox{0.92}{\begin{tabular}{|cc|cc|cc|cc|cc|cc|cc|cc|cc|}
\hline
\multicolumn{2}{|c|}{$\U(2)^2 \times \U(1)^2$}                              & \multicolumn{2}{c|}{$\cO(1)$} & \multicolumn{2}{c|}{$\cO(V)$} & \multicolumn{2}{c|}{$\cO(V^2)$} & \multicolumn{2}{c|}{$\cO(\Delta)$} & \multicolumn{2}{c|}{$\cO(X)$} & \multicolumn{2}{c|}{$\cO(XV)$} & \multicolumn{2}{c|}{$\cO(\Delta V)$} & \multicolumn{2}{c|}{$\cO(XV\Delta)$} \\ \hline
\multicolumn{1}{|c|}{$\psi^2 H^3 $}                   & $Q_{eH}$            &                &              & 2             & 2             &                 &               & 1                & 1               & 1             & 1             &                &               &                   &                  & 2                 & 2                \\ \hline
\multicolumn{1}{|c|}{$\psi^2 XH $}                    & $Q_{e(W,B)}$  &                &              & 4             & 4             &                 &               & 2                & 2               & 2             & 2             &                &               &                   &                  & 4                 & 4                \\ \hline
\multicolumn{1}{|c|}{\multirow{2}{*}{$\psi^2 H^2 D$}} & $Q_{H\ell}^{(1,3)}$ & 4              &              &               &               & 2               &               &                  &                 &               &               & 2              & 2             & 2                 & 2                &                   &                  \\
\multicolumn{1}{|c|}{}                                & $Q_{He}$            & 2              &              &               &               & 1               &               &                  &                 &               &               & 1              & 1             & 1                 & 1                &                   &                  \\ \hline
\multicolumn{1}{|c|}{$(LL)(LL)$}                      & $Q_{\ell\ell} $     & 5              &              &               &               &  4&               &                  &                 &               &               & 3            & 3             &     3              &  3                &                   &                  \\ \hline
\multicolumn{1}{|c|}{$(RR)(RR)$}                      & $Q_{ee} $           & 3              &              &               &               &  2              &               &                  &                 &               &               & 2              & 2             & 2                 & 2                &                   &                  \\ \hline
\multicolumn{1}{|c|}{$(LL)(RR)$}                      & $Q_{\ell e} $       & 4              &              &               &               & 5               & 1              &  {\color{teal} }                &    {\color{teal} }            &               &               & 4              & 4             & 6                 & 6                &                   &                  \\ \hline\hline
\multicolumn{2}{|c|}{Total}                                                 & 18             &              & 6             & 6             & 14              &  1             & 3                & 3               & 3             & 3             & 12             & 12            & 14                & 14               & 6                 & 6                \\ \hline
\end{tabular}}
\caption{Counting of the pure lepton SMEFT operators (see Appendix \ref{app:Warsaw}) assuming $\U(2)^2 \times \U(1)^2$ symmetry in the lepton sector. The counting is performed up to two insertions of $V$ and one insertion of $\Delta_e$ and $X_\ell$ spurion. Moreover, the counting is presented taking two ($X_\ell V_{e,\ell}$, $\Delta_e V_{e,\ell}$) and three ($X_\ell V_{e,\ell}\Delta_e$) insertions in the spurion product. The left (right) entry in each column gives the number of CP even (odd) coefficients at the given order in spurion counting.}
\label{tab:lep2U2x2U1}
\end{table}

\subsubsection*{Decomposition of bilinear and quartic structures}
Decomposing $\cO(1)$ bilinear structures proceeds in a similar way as in the case of quark $\U(2)^3$ symmetry. We obtain following $\cO(1)$ structures: $(\bar{\ell}\ell)$, $(\bar{e}e)$, $(\bar{\ell}_3 \ell_3)$ and $(\bar{e}_3 e_3)$. At $\cO(\Delta)$ there is only one bilinear structure allowed: $(\bar{\ell}\Delta_\ell e)$. There are two $\cO(V)$ bilinears of the form $(\bar{\ell}V_\ell e_3)$ and $(\bar{\ell}_3 V_e^\dag e)$. Two doublets of $\ell$ or $e$ can also be contracted to two insertions of $V_\ell$ or $V_e$ respectively, giving $(\bar{\ell}V_\ell V_\ell^\dag \ell)$ and $(\bar{e}V_e V_e^\dag e)$ structures. We also get one $\cO(X)$ structure of the form $(\bar{\ell}_3 X_\ell e_3)$.

More interesting bilinears emerge for two or three spurion insertions. Two $\cO(XV)$ structures can be written based on $X_\ell^* V_\ell \sim(\rep{2},\rep{1}\lvert -1,0)$ and $X_\ell V_e \sim (\rep{2},\rep{1}\lvert 0,-1)$: $[ \bar{\ell} (X_\ell^* V_\ell) \ell_3 ]$ and $[ \bar{e} (X_\ell V_e) e_3 ]$. Similarly, at order $\cO(\Delta V)$, we have $\Delta_e V_e \sim (\rep{2},\rep{1}\lvert -1,0)$ and $\Delta_e^\dag V_\ell \sim (\rep{1},\rep{2}\lvert 0,-1)$, giving two bilinears of the form $[ \bar{\ell}\lzm \Delta_e V_e \dzm \ell_3 ]$ and $[ \bar{\ell}\lzm \Delta_e V_e \dzm \ell_3 ]$. At order $\Delta XV$, we have the transformations $X_\ell V_\ell^\dag \Delta_e \sim (\rep{1},\repbar{2}\lvert 1,0)$ and $X_\ell \Delta_e V_e \sim (\rep{2},\rep{1}\lvert 0,-1)$, yielding two additional bilinears: $\big[ \Bar{\ell}_3 (X_\ell V_\ell^{\dag}\Delta_e)   e \big]$ and $[ \bar{\ell} (X_\ell \Delta_e V_e) e_3 ]$.

The unique $\cO(1)$ quartic structures with four instances of $\ell$ include $(\bar{\ell}_a \ell^b)(\bar{\ell}_b \ell^a)$ and $(\bar{\ell}\ell_3)(\bar{\ell}_3 \ell)$. Using the transformation properties of the $X_\ell^* V_\ell$ product, we can construct one more structure: $(\bar{\ell}_a \ell_3) [ \bar{\ell}_b (X_\ell^* V_\ell)^b \ell^a]$. Analogously, we deduce that there are three $\cO(\Delta V)$ structures, $[\bar{\ell}_a (V_e \Delta_e)^a \ell^b](\bar{\ell}_b  \ell_3 )$, $(\Bar{\ell} V_\ell \ell^a)\lzm \Delta_e^{*} \dzm\du{a}{b}(\Bar{e}_b e_3)$, and $(\bar{\ell}_a \ell^3)\lzm \Delta_e \dzm\ud{a}{b}(\bar{e}V_e e^b)$, and three $\cO(V^2)$ ones, $(\Bar{\ell}_a V_\ell^{\dag} \ell)(\Bar{\ell} V_\ell \ell^a)$, $(\bar{\ell}_3 V_\ell^\dag \ell)(\bar{\ell}V_\ell \ell_3)$, and $\lzm \bar{\ell}V_\ell \ell_3 \dzm\lzm \bar{e} V_e e_3 \dzm$. We present the complete list below:\\\\ 
$\boxed{(\bar{\ell}e)}$
\begin{equation}
\label{eq346}
\begin{alignedat}{2}
    &\mathcal{O}(\Delta):~ (\Bar{\ell} \Delta_e e)~,\qquad\mathcal{O}(V):~(\Bar{\ell} V_\ell e_3)\color{black}~,\quad (\bar{\ell}_3 V_e^\dag e)\color{black}~,\qquad  \mathcal{O}(X):~(\Bar{\ell}_3 X_\ell e_3)~, \\
    &\mathcal{O}(\Delta X V):~\big[ \Bar{\ell}_3 (X_\ell V_\ell^{\dag}\Delta_e)   e \big]\color{black}~,\quad \lzs \bar{\ell} (X_\ell \Delta_e V_e) e_3 \dzs \color{black}~.
\end{alignedat}
\end{equation}
$\boxed{(\bar{\ell}\ell)}$
\begin{equation}
{\small
\begin{alignedat}{2}
        &\mathcal{O}(1):~ (\Bar{\ell} \ell)~,\quad (\Bar{\ell}_3 \ell_3)~,\qquad\mathcal{O}(XV):~\lzs \Bar{\ell} (X_\ell^* V_\ell) \ell_3 \dzs~,\quad \hermc~,\qquad \color{black}\cO(\Delta V):~ \lzs \bar{\ell}\lzm \Delta_e V_e \dzm \ell_3 \dzs~,\quad \hermc~, \\
        &\mathcal{O}(V^2):~ ( \Bar{\ell} V_\ell V_\ell^\dag \ell )~.
\end{alignedat}}
\end{equation}
$\boxed{(\bar{e}e)}$
\begin{equation}
{\small
\begin{alignedat}{2}
        &\mathcal{O}(1):~(\Bar{e} e)~,\quad (\Bar{e}_3 e_3)~,\qquad\color{black}\cO(X V):~\lzs \bar{e}(X_\ell V_e)e_3 \dzs~,\quad \hermc~,\qquad\mathcal{O}(\Delta V):~\lzs \Bar{e} (\Delta_e^\dag V_\ell) e_3 \dzs ~,\quad \hermc~,\\
        &\color{black}\cO(V^2):~\lzm \bar{e} V_e V_e^\dag e \dzm ~.
\end{alignedat}}
\end{equation}
$\boxed{(\bar{\ell}\ell)(\bar{\ell}\ell)}$
\begin{equation}
\begin{alignedat}{2}
        &\mathcal{O}(1):~ (\Bar{\ell}_a \ell^b)(\Bar{\ell}_b \ell^a)~,\quad(\Bar{\ell} \ell_3)(\Bar{\ell}_3 \ell)~,\qquad\mathcal{O}(V^2):~ (\Bar{\ell}_a V_\ell^{\dag} \ell)(\Bar{\ell} V_\ell \ell^a)~,\quad (\bar{\ell}_3 V_\ell^\dag \ell)(\bar{\ell}V_\ell \ell_3)~, \\
        &\mathcal{O}(\Delta V):~[\bar{\ell}_a (V_e \Delta_e)^a \ell^b](\bar{\ell}_b  \ell_3 )~,\quad  \hermc~,\qquad \cO(XV):~ (\bar{\ell}_a \ell_3) [ \bar{\ell}_b (X_\ell^* V_\ell)^b \ell^a]~,\quad \hermc~.\\
\end{alignedat}
\end{equation}
$\boxed{(\bar{\ell}\ell)(\bar{e}e)}$
\begin{equation}
\label{eq350}
{\small
\begin{alignedat}{2}
    &\mathcal{O}(V^2):~\lzm \bar{\ell}V_\ell \ell_3 \dzm\lzm \bar{e} V_e e_3 \dzm~,\quad \hermc~,\qquad\mathcal{O}(\Delta V):~(\Bar{\ell} V_\ell \ell^b)\lzm \Delta_e^{*} \dzm\du{b}{a}(\Bar{e}_a e_3) ~,\quad (\bar{\ell}_a \ell^3)\lzm \Delta_e \dzm\ud{a}{b}(\bar{e}V_e e^b)~,\quad \hermc~.\\
\end{alignedat}}
\end{equation}
$\boxed{(\bar{e}e)(\bar{e}e)}$
\begin{equation}
\begin{alignedat}{2}
\text{No unique structures present due to Fierz identities}.
\end{alignedat}
\nonumber
\end{equation}


\subsection{$\U(3)$ vectorial symmetry}
In this section, we take a look at the $G= \U(3) \subset G_L$ vectorial symmetry, under which the fields transform as
    \begin{equation}
    \ell \sim \rep{3},\quad e\sim \rep{3}.
    \end{equation}
The minimal spurion required in this case, similar to the $\U(2)_V$ case, is $\Delta_\ell\sim\rep{8}$, which we take to be real. For simplicity, we once again use the implicit contraction with the generator:
\begin{equation}
    \Delta^A_\ell (T^A)\ud{i}{j} = (\Delta_{\ell})\ud{i}{j}.
\end{equation}

With the $\Delta_\ell$ spurion, the Yukawa matrix is parametrized by 
\begin{equation}
    Y_e = \Delta_\ell + x_\ell \mathds{1}.
\end{equation}
$\Delta_\ell$ breaks the $\U(3)_V$ flavor symmetry as
\begin{equation}
    \Delta_\ell \longrightarrow \begin{bmatrix} -\delta_{\ell}&&\\&-\delta'_{\ell}&\\&&\delta_{\ell}+\delta'_{\ell} \end{bmatrix}:\qquad\qquad \U(3)_V \to \U(1)^3_{\ell+e}.
\end{equation}
The flavor symmetry, thus, removes 6 unphysical parameters from $ \Delta_\ell$, leaving $ \delta^{(\prime)}_\ell>0 $. Furthermore, $ x_\ell $ breaks a $ \U(1) $ axial symmetry, which allows us to remove an unphysical phase. We observe that the realistic lepton masses require a high degree of tuning between $ x_\ell $ and $ \delta^{(\prime)}_\ell $, naturally of the order of the tau Yukawa coupling, to produce the small electron and muon Yukawa couplings. The numerical values of the relevant parameters are given by
\begin{equation}
\delta_\ell =\num{3.526e-3},\quad
\delta'_\ell= \num{2.940e-3},\quad 
x_\ell= \num{3.529e-3}.
\end{equation}
Note that one needs a tuning among parameters to accommodate for the smallness of $y_e$ and $y_\mu$ with respect to $y_\tau$.

We present the spurion counting in Table \ref{tab:lepU3diag} and list the flavor decompositions of the bilinear and unique quartic structures in Eqs.~(\ref{eq356}--\ref{eq360}). 
\begin{table}[t]
\centering
\begin{tabular}{|cc|cc|cc|}
\hline
\multicolumn{2}{|c|}{$ \U(3)_V$}                               & \multicolumn{2}{c|}{$\cO(1)$} & \multicolumn{2}{c|}{$\cO(\Delta_\ell)$} \\ \hline
\multicolumn{1}{|c|}{$\psi^2 H^3 $}                   & $Q_{eH}$            & 1              & 1            & 1                 & 1                \\ \hline
\multicolumn{1}{|c|}{$\psi^2 XH $}                    & $Q_{e(W,B)}$  & 2              & 2            & 2                 & 2                \\ \hline
\multicolumn{1}{|c|}{\multirow{2}{*}{$\psi^2 H^2 D$}} & $Q_{H\ell}^{(1,3)}$ & 2              &              & 2                 &                 \\
\multicolumn{1}{|c|}{}                                & $Q_{He}$            & 1              &              & 1                 &                 \\ \hline
\multicolumn{1}{|c|}{$(LL)(LL)$}                      & $Q_{\ell\ell} $     & 2              &              & 2                 &                 \\ \hline
\multicolumn{1}{|c|}{$(RR)(RR)$}                      & $Q_{ee} $           & 1              &              & 1                 &                \\ \hline
\multicolumn{1}{|c|}{$(LL)(RR)$}                      & $Q_{\ell e} $       & 2              &              & 3                 & 1                \\ \hline\hline
\multicolumn{2}{|c|}{Total}                                                 &  11            & 3            & 12                & 4               \\ \hline
\end{tabular}
\caption{Counting of the pure lepton SMEFT operators (see Appendix \ref{app:Warsaw}) assuming $\U(3)$ vectorial symmetry in the lepton sector. The counting is performed up to one insertion of $\Delta_\ell$ spurion. The left (right) entry in each column gives the number of CP even (odd) coefficients at the given order in spurion counting.}
\label{tab:lepU3diag}
\end{table}

\subsubsection*{Decomposition of bilinear and quartic structures}
The $\cO(1)$ bilinears are given by $(\bar{\ell}\ell)$, $(\bar{e}e)$ or $(\bar{\ell}e)$. Bilinears containing one insertion of $\Delta_\ell$ are formed similarly to the $\U(2)_V$ case and are given by $(\bar{\ell}\Delta_\ell \ell)$, $(\bar{e}\Delta_\ell e)$ and $(\bar{\ell}\Delta_\ell e)$. The $\cO(1)$ unique quartic structures we obtain are $(\bar{\ell}_i \ell^j)(\bar{\ell}_j \ell^i)$ and $(\bar{\ell}_i \ell^j)(\bar{e}_j e^i)$. Considering quartic structures with one insertion of $\Delta_\ell$, there are two allowed structures: $(\bar{\ell}_i \ell^j)\lzm \Delta_\ell \dzm\ud{i}{k}(\bar{\ell}_j \ell^k)$ and $(\bar{\ell}_i \ell^j)\lzm \Delta_\ell \dzm\ud{i}{k} (\bar{e}_j e^k)$. Allowed $\cO(1)$ and $\cO(\Delta_\ell)$ structures are presented below:\\\\
$\boxed{(\bar{\ell}e)}$
\begin{equation}
\label{eq356}
\begin{alignedat}{2}
        &\mathcal{O}(1):~ (\bar{\ell}e)~,\qquad \mathcal{O}(\Delta_\ell):~(\bar{\ell} \Delta_\ell e)~.
\end{alignedat}
\end{equation}
$\boxed{(\bar{\ell}\ell)}$
\begin{equation}
\begin{alignedat}{2}
        &\mathcal{O}(1):~ (\bar{\ell}\ell)~,\qquad\mathcal{O}(\Delta_\ell):~(\bar{\ell} \Delta_\ell \ell)~.
\end{alignedat}
\end{equation}
$\boxed{(\bar{e}e)}$
\begin{equation}
\begin{alignedat}{2}
                &\mathcal{O}(1):~(\bar{e}e)~,\qquad\mathcal{O}(\Delta_\ell):~(\bar{e} \Delta_\ell e)~.     
\end{alignedat}
\end{equation}
$\boxed{(\bar{\ell}\ell)(\bar{\ell}\ell)}$
\begin{equation}
\begin{alignedat}{2}
        &\mathcal{O}(1):~(\bar{\ell}_i \ell^j)(\bar{\ell}_j \ell^i)~,\qquad\mathcal{O}(\Delta_\ell):~(\bar{\ell}_i \ell^j)\lzm \Delta_\ell \dzm\ud{i}{k}(\bar{\ell}_j \ell^k)~.
\end{alignedat}
\end{equation}
$\boxed{(\bar{\ell}\ell)(\bar{e}e)}$
\begin{equation}
\label{eq360}
\begin{alignedat}{2}
        &\mathcal{O}(1):~(\bar{\ell}_i \ell^j)(\bar{e}_j e^i)~,\qquad\mathcal{O}(\Delta_\ell):~\color{black} (\bar{\ell}_i \ell^j)\lzm \Delta_\ell \dzm\ud{i}{k} (\bar{e}_j e^k)~,\quad \hermc~.
\end{alignedat}
\end{equation}
$\boxed{(\bar{e}e)(\bar{e}e)}$
\begin{equation}
\text{No unique structures present due to Fierz identities}.
\nonumber
\end{equation}


\subsection{$\text{MFV}_L$ symmetry}
Lastly, let us take a look at minimal flavor violation in the lepton sector, with the full symmetry $ G =\U(3)_\ell \times \U(3)_e = G_L$. The leptons are in the representations 
\begin{equation}
    \ell\sim (\rep{3},\rep{1}),\quad e\sim(\rep{1},\rep{3}),
\end{equation}
and the leptonic Yukawa matrix, which serves as the sole spurion transforms as
\begin{equation}
    Y_e \sim (\boldsymbol3,\boldsymbol{\bar{3}}).
\end{equation}
As in the SM, the Yukawa matrix breaks the symmetry according to 
\begin{equation}
    Y_e\longrightarrow\hat{Y}_e:\qquad\qquad\U(3)_\ell \times \U(3)_e \longrightarrow \U(1)_{\ell+e}^3,
\end{equation}
where $ \hat{Y}_e $ is a real, positive, diagonal matrix. 15 unphysical parameters are removed from the spurion in this manner, and the remnant symmetry ensures conservation of individual lepton numbers.

The flavor spurion counting of the pure lepton operators is presented in Table \ref{tab:lepMFV} and the decompositions are listed in Eqs.~(\ref{eq365}--\ref{eq368}).
\begin{table}[t]
\centering
\begin{tabular}{|cc|cc|cc|}
\hline
\multicolumn{2}{|c|}{$\text{MFV}_L$}                         & \multicolumn{2}{c|}{$\cO(1)$} & \multicolumn{2}{c|}{$\cO(Y_e)$} \\ \hline
\multicolumn{1}{|c|}{$\psi^2 H^3 $}                   & $Q_{eH}$            &                &              & 1              & 1              \\ \hline
\multicolumn{1}{|c|}{$\psi^2 XH $}                    & $Q_{e(W,B)}$  &                &              & 2              & 2              \\ \hline
\multicolumn{1}{|c|}{\multirow{2}{*}{$\psi^2 H^2 D$}} & $Q_{H\ell}^{(1,3)}$ & 2              &              &                &                \\
\multicolumn{1}{|c|}{}                                & $Q_{He}$            & 1              &              &                &                \\ \hline
\multicolumn{1}{|c|}{$(LL)(LL)$}                      & $Q_{\ell\ell} $     & 2              &              &                &                \\ \hline
\multicolumn{1}{|c|}{$(RR)(RR)$}                      & $Q_{ee} $           & 1              &              &                &                \\ \hline
\multicolumn{1}{|c|}{$(LL)(RR)$}                      & $Q_{\ell e} $       & 1              &              &                &                \\ \hline \hline
\multicolumn{2}{|c|}{Total}                                                 & 7              &              & 3              & 3              \\ \hline
\end{tabular}
\caption{Counting of the pure lepton SMEFT operators (see Appendix \ref{app:Warsaw}) assuming $\text{MFV}_L$ symmetry in the lepton sector. The counting is performed up to one insertion of $Y_e$ spurion. The left (right) entry in each column gives the number of CP even (odd) coefficients at the given order in spurion counting.}
\label{tab:lepMFV}
\end{table}
\subsubsection*{Decomposition of bilinear and quartic structures}
At $\cO(1)$, we get two structures of the form $(\bar{\ell}_i \ell^i)$ and $(\bar{e}_i e^i)$. Using one insertion of $Y_e$ we obtain one structure given by $(\bar{\ell}Y_e e)$. Regarding the non-factorizing quartic structures, there is only one that can be formed with four appearances of $\ell$ of the form $(\bar{\ell}_i \ell^j)(\bar{\ell}_j \ell^i)$. There are no unique structures of the form $(\bar{\ell}\ell)(\bar{e}e)$. Taking these remarks into account, the list of the $\cO(1)$ and $\cO(Y_e)$ structures is presented below:
\\\\
$\boxed{(\bar{\ell}e)}$
\begin{equation}
\label{eq365}
    \mathcal{O}(Y_e):~ (\Bar{\ell} Y_e e)~.
\end{equation}
$\boxed{(\bar{\ell}\ell)}$
\begin{equation}
    \mathcal{O}(1):~ (\Bar{\ell}\ell) ~.
\end{equation}
$\boxed{(\bar{e}e)}$
\begin{equation}
    \mathcal{O}(1):~ (\Bar{e}e) ~.
\end{equation}
$\boxed{(\bar{\ell}\ell)(\bar{\ell}\ell)}$
\begin{equation}
\label{eq368}
    \mathcal{O}(1):~ (\Bar{\ell}_i \ell^j)(\Bar{\ell}_j \ell^i)~.
\end{equation}
$\boxed{(\bar{\ell}\ell)(\bar{e}e)}$
\begin{equation}
\text{No unique structures present}.
\nonumber
\end{equation}
$\boxed{(\bar{e}e)(\bar{e}e)}$
\begin{equation}
\text{No unique structures present due to Fierz identities}.
\nonumber
\end{equation}

\section{Conclusions}
\label{sec:conc}

The hierarchical pattern of charged fermion masses and mixings observed in nature craves an explanation: the dimension-4 Yukawa interactions in the SM provide only a parametrization but not an understanding of flavor. To make progress in addressing this long-standing puzzle, we must uncover new, flavored interactions beyond the SM. The SMEFT is a powerful framework that can capture the low-energy physics of a high-energy model. There are 2499 leading dimension-6 baryon and lepton number--conserving operators, the great majority of which are flavorful. The hope is that experiments will observe some of these interactions and start clarifying their flavor patterns. This might provide a crucial clue to solving the puzzle. Perhaps the ongoing flavor anomalies are the first step in this direction. 

Patterns are closely related to symmetries. In this paper, we systematically explored the flavor structure of $\Delta B = 0$ dimension-6 SMEFT operators using flavor symmetries as an organizing principle in an extension of Ref.~\cite{Faroughy:2020ina}. Our underlying assumption is that short-distance physics will leave a global flavor symmetry and a breaking pattern in the effective operators at low energies. From the IR perspective, postulating different flavor structures in the effective field theory means mapping the space of physics beyond the SM into the universality classes. These assumptions impose correlations among operators, which can be tested in experiments, providing a systematic way to learn about BSM.

Concretely, working in the Warsaw basis (Appendix~\ref{app:Warsaw}), we imposed a variety of different global flavor symmetry assumptions in both the quark and the lepton sectors. The symmetries are carefully chosen to allow for new physics at (not far beyond) the TeV scale while respecting the experimental bounds from FCNC, LFV, and EDMs. In particular, they allow for potential new physics effects in the high-$p_T$ experiments and motivate global SMEFT fits in the top, Higgs, and EW sectors. The symmetry-breaking spurions are non-dynamical objects formally transforming in a non-trivial representation of the imposed flavor group. 
For each flavor structure, we construct the basis of dimension-6 operators compatible with the flavor symmetry and breaking spurions.

As a supplement to this work, we also provide a {\tt Mathematica} package {\tt SMEFTflavor} for automatic generation of the operators should the user have a different symmetry group or breaking spurions in mind (Appendix~\ref{sec:tool}).

As shown in Table~\ref{tab:intro}, the number of leading flavor-symmetric operators without spurion insertions, which are important for the high-$p_T$ fits, is significantly reduced from the initial 2499 when no symmetries are imposed. In Section~\ref{sec:quark}, we explicitly construct independent operators for $\U(2)^3$, $\U(2)^3\times \U(1)$, $\U(2)^2\times \U(3)$, and $\U(3)^3$ quark flavor symmetries, focusing on the quark-only operators. We pay special attention to the $\SU(2)$ subgroup invariants. We also derive minimal spurion parametrizations, which can be directly employed in phenomenological studies. The  novelty here is the use of the full $\U(3)^3$ field redefinitions consistent with the imposed symmetry to avoid overparametrizations often found in the literature. In Section~\ref{sec:leptonic}, we repeat the analysis for the lepton-only operators and lepton symmetries $\U(1)^6$, $\U(1)_V^3$, $\U(2)_{V}$,  $\U(2)^2$, $\U(2)^2 \times \U(1)^2$, $\U(3)_{V}$, and $\U(3)^2$. The counting of the mixed quark-lepton operators is worked out in Appendix~\ref{sec:mixed} for each combination of the four quark and the seven lepton symmetries. 

Our methodology can be extended to the dimension-8 operators in the SMEFT, which will be presented in a separate publication. We hope the flavor structures proposed in this work will find their place in the future phenomenological studies of low and high-$p_T$ data.

\section*{Acknowledgments}

We acknowledge with thanks the discussions held within the \href{https://lpcc.web.cern.ch/lhc-eft-wg}{{\color{blue} LHC EFT WG}}. We also thank Gino Isidori, Felix Wilsch, and Javier Fuentes-Martín for useful discussions. This work received funding from the Swiss National Science Foundation (SNF) through the Eccellenza Professorial Fellowship ``Flavor Physics at the High Energy Frontier'' project number 186866. AG is also partially supported by the European Research Council (ERC) under the European Union’s Horizon 2020 research and innovation program, grant agreement 833280 (FLAY).

\newpage
\appendix

\section{Warsaw basis}
\label{app:Warsaw}
Here we list the $\Delta B = 0$ dimension-6 fermionic SMEFT operators in the Warsaw basis \cite{Grzadkowski:2010es} with division into classes as presented in \cite{Alonso:2013hga}. 

\newcommand{\OpScale}{.85}
\begin{table}[h]
	\centering
	\renewcommand{\arraystretch}{1.5}
	\scalebox{\OpScale}{
	\centering
	\begin{tabular}{| lc || lc | lc | lc |}
		\multicolumn{8}{c}{5--7: Fermion Bilinears} \\[.1cm] \hline
		\multicolumn{8}{|c|}{non-hermitian $(\bar L R)$} \\ \hline
		\multicolumn{2}{|c||}{5: $\psi^2 H^3$} & \multicolumn{6}{c|}{6: $\psi^2 X H $}  \\ \hline
		$Q_{eH}$ & $(H^\dagger H)(\bar\ell_p e_r H)$ &	
		$Q_{eW}$ & $(\bar\ell_p \sigma^{\mu\nu}e_r)\tau^I H W_{\mu\nu}^I$ &
		$Q_{uG}$ & $(\bar q_p \sigma^{\mu\nu}T^A u_r)\tilde{H}G_{\mu\nu}^A$ &
		$Q_{dG}$ & $(\bar q_p \sigma^{\mu\nu}T^A d_r)H G_{\mu\nu}^A$ \\ 
		$Q_{uH}$ & $(H^\dagger H)(\bar q_p u_r \tilde{H})$ &
		$Q_{eB}$ & $(\bar\ell_p \sigma^{\mu\nu}e_r) H B_{\mu\nu}$ &
		$Q_{uW}$ & $(\bar q_p \sigma^{\mu\nu}u_r)\tau^I \tilde{H}W_{\mu\nu}^I$ &
		$Q_{dW}$ & $(\bar q_p \sigma^{\mu\nu}d_r)\tau^I H W_{\mu\nu}^I$ \\ 
		$Q_{dH}$ & $(H^\dagger H)(\bar q_p d_r H)$ &		
		& & 
		$Q_{uB}$ & $(\bar q_p \sigma^{\mu\nu}u_r)\tilde{H}B_{\mu\nu}$ &
		$Q_{dB}$ & $(\bar q_p \sigma^{\mu\nu}d_r) H B_{\mu\nu}$ \\ \hline
		\end{tabular}
	}
	\newline\centering
	\scalebox{\OpScale}{
	\begin{tabular}{| lc | lc | lc |}
		\hline \multicolumn{6}{|c|}{hermitian (+ $Q_{Hud}$) $\quad \sim \quad$ 7: $\psi^2 H^2 D$} \\ \hline
		\multicolumn{2}{|c|}{$(\bar L L)$} & 
		\multicolumn{2}{c|}{$(\bar R R)$} & 
		\multicolumn{2}{c|}{$(\bar R R^\prime)$} \\ \hline
		$Q_{H\ell}^{(1)}$ & $(H^\dagger i \overleftrightarrow{D}_\mu H)(\bar\ell_p \gamma^\mu \ell_r)$ & 
		$Q_{H e}$ & $(H^\dagger i \overleftrightarrow{D}_\mu H)(\bar e_p \gamma^\mu e_r)$ & 
		$Q_{Hud}$ & $i(\tilde{H}^\dagger D_\mu H)(\bar u_p \gamma^\mu d_r)$ \\
		$Q_{H\ell}^{(3)}$ & $(H^\dagger i \overleftrightarrow{D}_\mu^I H)(\bar\ell_p \tau^I\gamma^\mu \ell_r)$ & 
		$Q_{H u}$ & $(H^\dagger i \overleftrightarrow{D}_\mu H)(\bar u_p \gamma^\mu u_r)$ & 
		& \\
		$Q_{Hq}^{(1)}$ & $(H^\dagger i \overleftrightarrow{D}_\mu H)(\bar q_p \gamma^\mu q_r)$ & 
		$Q_{H d}$ & $(H^\dagger i \overleftrightarrow{D}_\mu H)(\bar d_p \gamma^\mu d_r)$ & 
		& \\
		$Q_{Hq}^{(3)}$ & $(H^\dagger i \overleftrightarrow{D}_\mu^I H)(\bar q_p \tau^I\gamma^\mu q_r)$ & 
		& & 
		& \\ \hline
	\end{tabular}
	}
	\vspace{0.5cm}
	\\ \centering
	\scalebox{\OpScale}{
	\begin{tabular}{| llc | llc | llc |}
		\multicolumn{9}{c}{8: Fermion Quadrilinears} \\[.1cm] \hline
		\multicolumn{9}{|c|}{hermitian} \\ \hline
		\multicolumn{3}{|c|}{$(\bar L L)(\bar L L)$} & 
		\multicolumn{3}{c|}{$(\bar R R)(\bar R R)$} & 
		\multicolumn{3}{c|}{$(\bar L L)(\bar R R)$} \\ \hline
		$Q_{\ell\ell}$ &  & $(\bar\ell_p \gamma_\mu \ell_r)(\bar\ell_s \gamma^\mu \ell_t)$ & 
		$Q_{ee}$ &  & $(\bar e_p \gamma_\mu e_r)(\bar e_s \gamma^\mu e_t)$ & 
		$Q_{\ell e}$ &  & $(\bar\ell_p \gamma_\mu \ell_r)(\bar e_s \gamma^\mu e_t)$ \\
		$Q_{qq}^{(1)}$ &  & $(\bar q_p \gamma_\mu q_r)(\bar q_s \gamma^\mu q_t)$ & 
		$Q_{uu}$ &  & $(\bar u_p \gamma_\mu u_r)(\bar u_s \gamma^\mu u_t)$ & 
		$Q_{\ell u}$ &  & $(\bar\ell_p \gamma_\mu \ell_r)(\bar u_s \gamma^\mu u_t)$ \\
		$Q_{qq}^{(3)}$ &  & $(\bar q_p \gamma_\mu \tau^I q_r)(\bar q_s \gamma^\mu \tau^I q_t)$ & 
		$Q_{dd}$ &  & $(\bar d_p \gamma_\mu d_r)(\bar d_s \gamma^\mu d_t)$ & 
		$Q_{\ell d}$ &  & $(\bar\ell_p \gamma_\mu \ell_r)(\bar d_s \gamma^\mu d_t)$ \\
		$Q_{\ell q}^{(1)}$ &  & $(\bar\ell_p \gamma_\mu \ell_r)(\bar q_s \gamma^\mu q_t)$ & 
		$Q_{eu}$ &  & $(\bar e_p \gamma_\mu e_r)(\bar u_s \gamma^\mu u_t)$ & 
		$Q_{q e}$ &  & $(\bar q_p \gamma_\mu q_r)(\bar e_s \gamma^\mu e_t)$ \\
		$Q_{\ell q}^{(3)}$ &  & $(\bar\ell_p \gamma_\mu \tau^I \ell_r)(\bar q_s \gamma^\mu \tau^I q_t)$ & 
		$Q_{ed}$ &  & $(\bar e_p \gamma_\mu e_r)(\bar d_s \gamma^\mu d_t)$ & 
		$Q_{qu}^{(1)}$ &  & $(\bar q_p \gamma_\mu q_r)(\bar u_s \gamma^\mu u_t)$ \\
		& & & 
		$Q_{ud}^{(1)}$ &  & $(\bar u_p \gamma_\mu u_r)(\bar d_s \gamma^\mu d_t)$ & 
		$Q_{qu}^{(8)}$ &  & $(\bar q_p \gamma_\mu T^A q_r)(\bar u_s \gamma^\mu T^A u_t)$ \\
		& & & 
		$Q_{ud}^{(8)}$ &  & $(\bar u_p \gamma_\mu T^A u_r)(\bar d_s \gamma^\mu T^A d_t)$ & 
		$Q_{qd}^{(1)}$ &  & $(\bar q_p \gamma_\mu q_r)(\bar d_s \gamma^\mu d_t)$ \\
		& & & 
		&  & & 
		$Q_{qd}^{(8)}$ &  & $(\bar q_p \gamma_\mu T^A q_r)(\bar d_s \gamma^\mu T^A d_t)$ \\ \hline
	\end{tabular}
	}
	\newline\centering
	\scalebox{\OpScale}{
	\begin{tabular}{| llc | llc |}
		\hline \multicolumn{6}{|c|}{non-hermitian} \\ \hline
		\multicolumn{3}{|c|}{$(\bar L R)(\bar R L)$} & 
		\multicolumn{3}{c|}{$(\bar L R)(\bar L R)$} \\ \hline
		$Q_{\ell e d q }$ &  & $(\bar \ell _p^j e_r)(\bar d_s q_{tj})$ & 
		$Q_{quqd}^{(1)}$ &  & $(\bar q_p^j u_r)\varepsilon_{jk}(\bar q_s^k d_t)$ \\ 
		& & & 
		$Q_{quqd}^{(8)}$ &  & $(\bar q_p^j T^A u_r)\varepsilon_{jk}(\bar q_s^k T^A d_t)$ \\ 
		& & & 
		$Q_{\ell equ}^{(1)}$ &  & $(\bar \ell_p^j e_r)\varepsilon_{jk}(\bar q_s^k u_t)$ \\ 
		& & & 
		$Q_{\ell equ}^{(3)}$ &  & $(\bar \ell_p^j \sigma_{\mu\nu} e_r)\varepsilon_{jk}(\bar q_s^k \sigma^{\mu\nu} u_t)$ \\ \hline 
	\end{tabular}
	}
	\vglue -2 true cm
\end{table}

\newpage

\section{{\tt SMEFTflavor}}
\label{sec:tool}

In this section we briefly outline the details of the \texttt{Mathematica} package developed during this project, which has played an important, twofold role: First, it has been used to perform the cross-checks of all the decompositions and counting tables. 
Second, the code has been developed to allow for implementation of other flavor symmetries than those analyzed here to determine the SMEFT operators and corresponding spurion counting tables. 
The package can be downloaded from the \texttt{github} page \url{https://github.com/aethomsen/SMEFTflavor}. 
It can then be run from a notebook located in the base directory of the package (the one with the `Tutorial.nb' notebook) by setting the directory to that of the notebook with \coloredbox{white!92!black}{blueRef}{SetDirectory@ NotebookDirectory[]} and then running \coloredbox{white!92!black}{blueRef}{<< SMEFTflavor\,$\grave{}$}.

Let us now present the main functions of the program and describe their output. 
\texttt{SMEFTflavor} comes with implementations of the 4 quark and 7 lepton symmetries considered in this paper ready to use, and more symmetries can be added by the user.
The details of the implemented symmetries can be found in the association \coloredbox{white!92!black}{blueRef}{\$flavorSymmetries}.
The keys of this association, \coloredbox{white!92!black}{blueRef}{Keys@ \$flavorSymmetries}, are all the names of symmetries (strings) with the indicative labels, e.g., `3U2' for $\U(2)^3$, `6U1' for $\U(1)^6$, `2U2xU3' for $\U(2)^2\times\U(3)$ etc. 
To access the details (groups, representations, spurions) of a particular combination of quark and lepton symmetry assignment, one simply picks out the element of the association, e.g.,  \coloredbox{white!92!black}{blueRef}{\$flavorSymmetries@ \{"quark:MFV", "lep:2U2"\}} for the MFV in quark and $\U(2)^2$ in the lepton sector. 
Similarly, details of the SMEFT operators (Warsaw basis) are found in the association \coloredbox{white!92!black}{blueRef}{\$smeftOperators} and the keys  \coloredbox{white!92!black}{blueRef}{Keys@ \$smeftOperators} constitute a list of all the operators.
As the pure quark, pure lepton and mixed quark-lepton operators are analyzed separately in this work, these subsets of the SMEFT operators are listed in the program using \coloredbox{white!92!black}{blueRef}{leptonicOperators}, \coloredbox{white!92!black}{blueRef}{quarkOperators} and \coloredbox{white!92!black}{blueRef}{semiLeptonicOperators}

The main function of the package is \coloredbox{white!92!black}{blueRef}{DetermineOperatorBasis[symmetry, options]} and it is used to construct the operator basis provided the symmetry (input as a string or a list of strings if mixed quark-lepton operators are considered) along with other options. The options this function can take are \coloredbox{white!92!black}{blueRef}{SpurionCount} (taking an integer value---default 3---and used to indicate the order in the counting of the spurions based on the corresponding input in the \texttt{FlavorSymmetries}) and \coloredbox{white!92!black}{blueRef}{SMEFToperators} (defining the set SMEFT operators to consider---default \coloredbox{white!92!black}{blueRef}{All}---e.g., one of the three subsets of the SMEFT operators mentioned above). An illustrative example of this command is\\[3pt] \scalebox{0.99}{\coloredbox{white!92!black}{blueRef}{\parbox{\linewidth}{
    DetermineOperatorBasis["lep:2U2", \\
    \hspace*{2em}SpurionCount$\to$1, \\
    \hspace*{2em}SMEFToperators$\to$ leptonicOperators]}}}\\[3pt]
The output of this function consists of an association of the pure lepton operator basis (organized by SMEFT operator and spurion insertions) up to order 1 in the specified spurion counting. In order to present operators in a more legible form with the contractions explicitly indicated, \coloredbox{white!92!black}{blueRef}{//OpForm} can be added to the end of the previous line. \coloredbox{white!92!black}{blueRef}{OpForm} generally formats all the operators to make them legible to humans.  

The second important function we point out is the \scalebox{0.95}{\coloredbox{white!92!black}{blueRef}{CountingTable[symmetry, options]}} function, which also takes \coloredbox{white!92!black}{blueRef}{SpurionCount} and \coloredbox{white!92!black}{blueRef}{SMEFToperators} for options. 
This function returns the spurion counting table for particular symmetry based on the operator basis. 
Let us present three illustrative examples of this function. In order to return the spurion counting table for the pure quark MFV case (Table~\ref{quarkMFVtable}) up to order 3 in the spurion counting one can call \coloredbox{white!92!black}{blueRef}{CountingTable["quark:MFV", SpurionCount$\to$ 3]}. 
Similarly, to get the spurion counting table for the pure lepton $\U(1)^3$ case (Table~\ref{lep3U1table}) one can run the analogous command  \coloredbox{white!92!black}{blueRef}{CountingTable["lep:3U1", SpurionCount$\to$ 1]} 
Lastly, to obtain the full spurion counting table for all operators assuming, MFV in the quark and $\U(1)^3$ in the lepton sector one can run {\scalebox{1.00}{\coloredbox{white!92!black}{blueRef}{CountingTable[\{"quark:MFV", "lep:3U1"\}, SpurionCount$\to$3]}}}

The last function we would like to point out is \coloredbox{white!92!black}{blueRef}{AddSMEFTSymmetry}, which enables user to implement a new quark, lepton, or mixed quark-lepton symmetry. The syntax of this function is \scalebox{0.94}{\coloredbox{white!92!black}{blueRef}{AddSMEFTSymmetry["Type", "Name"$\to$ GroupInfoAssociation]}} The first argument can either be \coloredbox{white!92!black}{blueRef}{"Lepton"}, \coloredbox{white!92!black}{blueRef}{"Quark"}, or \coloredbox{white!92!black}{blueRef}{"Mixed"}, designating which sector the newly added symmetry is associated to. \texttt{Name} refers to the string introduced as the name of the added symmetry and the \texttt{GroupInfoAssociation} associates all the symmetry properties (similar to those contained in  \coloredbox{white!92!black}{blueRef}{\$flavorSymmetries}) to the new symmetry. 
To illustrate how this function is used, let us imagine that we would like to introduce a $\U(2)_\text{diag}$ symmetry group in the lepton sector with a real spurion transforming in the adjoint:\\[3pt]
\scalebox{0.99}{\coloredbox{white!92!black}{blueRef}{\parbox{\linewidth}{\small{AddSMEFTSymmetry["Lepton", "U2diag"-> <|\\
\hspace*{2em} Groups-> <|"U2l"-> SU@ 2|>,\\
\hspace*{2em} Spurions-> {"$\Delta$l"},\\
\hspace*{2em} Charges-> <|"l12"-> {1}, "l3"-> 0, "e12"-> {1}, "e3"-> 0, "$\Delta$l"-> 0, "Vl"-> {1}|>,\\
\hspace*{2em} Representations-> <|"l12"-> {"U2l"@ fund}, "e12"-> {"U2l"@ fund},\\
\hspace*{4em} "$\Delta$l"-> {"U2l"@ adj}|>,\\
\hspace*{2em} FieldSubstitutions-> <|"l"-> \{"l12", "l3"\}, "e"-> \{"e12", "e3"\}|>,\\
\hspace*{2em} SpurionCounting-> <|"$\Delta$l"-> 2|>,\\
\hspace*{2em} SelfConjugate-> {"$\Delta$l"}\\
|>]}}}}\\[3pt]
Note that only $ \SU(2) $ and $ \SU(3) $ factors are supported. This should cover the vast majority of cases. 

All the aforementioned functions and their outputs along with additional practical examples have been presented in the tutorial notebook provided in the package.

\newpage
\section{Mixed quark-lepton operators}
\label{sec:mixed}

The mixed quark-lepton four-fermion operators change for every combination of the 4 quark and 7 lepton flavor structures we have considered (for a total of 28 unique cases). In each case, the flavor structure factorizes straight-forwardly into a quark and a lepton bilinear, all of which we have presented in the main text.  
Here we report in tables below the counting for all 28 cases while the exhaustive results, including the explicit forms for the operators, can be generated using the {\tt SMEFTflavor} package.

\subsection*{$\text{MFV}_Q \times \text{MFV}_L$  }
\begin{table}[h]
\centering
\begin{tabular}{|cc|cc|cc|cc|cc|cc|}
\hline
\multicolumn{2}{|c|}{$\text{MFV}_Q \times \text{MFV}_L$}                                 & \multicolumn{2}{c|}{$\cO(1)$} & \multicolumn{2}{c|}{$\cO(Y_u^2)$} & \multicolumn{2}{c|}{$\cO(Y_d^2)$} & \multicolumn{2}{c|}{$\cO(Y_e Y_u)$} & \multicolumn{2}{c|}{$\cO(Y_e Y_d)$} \\ \hline
\multicolumn{1}{|c|}{$(LL)(LL)$}                  & $Q_{\ell q}^{(1,3)}$   & 2              &              & 2                &                & 2                &                &                  &                  &                  &                  \\ \hline
\multicolumn{1}{|c|}{\multirow{2}{*}{$(RR)(RR)$}} & $Q_{eu}$               & 1              &              & 1                &                &                  &                &                  &                  &                  &                  \\
\multicolumn{1}{|c|}{}                            & $Q_{ed}$               & 1              &              &                  &                & 1                &                &                  &                  &                  &                  \\ \hline
\multicolumn{1}{|c|}{\multirow{3}{*}{$(LL)(RR)$}} & $Q_{\ell u}$           & 1              &              & 1                &                &                  &                &                  &                  &                  &                  \\
\multicolumn{1}{|c|}{}                            & $Q_{\ell d}$           & 1              &              &                  &                & 1                &                &                  &                  &                  &                  \\
\multicolumn{1}{|c|}{}                            & $Q_{qe}$               & 1              &              & 1                &                & 1                &                &                  &                  &                  &                  \\
\hline
\end{tabular}
\end{table}

\subsection*{$\text{MFV}_Q \times \U(3)_{V,L}$}
\begin{table}[h]
\centering
\scalebox{0.75}{
\begin{tabular}{|cc|cc|cc|cc|cc|cc|cc|cc|cc|cc|cc|}
\hline
\multicolumn{2}{|c|}{$\text{MFV}_Q \times \U(3)_{V,L}$}          & \multicolumn{2}{c|}{$\cO(1)$} & \multicolumn{2}{c|}{$\cO(Y_u)$} & \multicolumn{2}{c|}{$\cO(Y_u^2)$} & \multicolumn{2}{c|}{$\cO(Y_d)$} & \multicolumn{2}{c|}{$\cO(Y_d^2)$} & \multicolumn{2}{c|}{$\cO(Y_u^2 Y_d)$} & \multicolumn{2}{c|}{$\cO(Y_u Y_d^2)$} & \multicolumn{2}{c|}{$\cO(\Delta_\ell)$} & \multicolumn{2}{c|}{$\cO(\Delta_\ell Y_u)$} & \multicolumn{2}{c|}{$\cO(\Delta_\ell Y_d)$} \\ \hline
\multicolumn{1}{|c|}{$(LL)(LL)$}                  & $Q_{\ell q}^{(1,3)}$   & 2              &              &                &                & 2                &                &                &                & 2                &                &                   &                   &                   &                   & 2                   &                   &                      &                      &                      &                      \\ \hline
\multicolumn{1}{|c|}{\multirow{2}{*}{$(RR)(RR)$}} & $Q_{eu}$               & 1              &              &                &                & 1                &                &                &                &                  &                &                   &                   &                   &                   & 1                   &                   &                      &                      &                      &                      \\
\multicolumn{1}{|c|}{}                            & $Q_{ed}$               & 1              &              &                &                &                  &                &                &                & 1                &                &                   &                   &                   &                   & 1                   &                   &                      &                      &                      &                      \\
\hline
\end{tabular}}
\end{table}

\newpage
\subsection*{$\text{MFV}_Q \times (\U(2)^2 \times \U(1)^2)_L$  }
\begin{table}[h]
\centering
\scalebox{0.69}{
\begin{tabular}{|cc|cc|cc|cc|cc|cc|cc|cc|cc|cc|}
\hline
\multicolumn{2}{|c|}{$\text{MFV}_Q \times (\U(2)^2\times \U(1)^2)_L$}          & \multicolumn{2}{c|}{$\cO(1)$} & \multicolumn{2}{c|}{$\cO(Y_u^2)$} & \multicolumn{2}{c|}{$\cO(Y_d^2)$} & \multicolumn{2}{c|}{$\cO(XV_{\ell,e})$} & \multicolumn{2}{c|}{$\cO(\Delta V_{\ell,e})$} & \multicolumn{2}{c|}{$\cO(V_{\ell,e}^2)$} & \multicolumn{2}{c|}{{\scalebox{0.93}{$\cO(\Delta Y_d,\Delta Y_u)$}}} & \multicolumn{2}{c|}{{\scalebox{0.93}{$\cO(V_{\ell,e}Y_d,V_{\ell,e}Y_u)$}}} & \multicolumn{2}{c|}{{\scalebox{0.93}{$\cO(XY_d,XY_u)$}}} \\ \hline
\multicolumn{1}{|c|}{$(LL)(LL)$}                  & $Q_{\ell q}^{(1,3)}$   & 4              &              & 4                &                & 4                &                & 2              & 2             & 2                 & 2                & 2               &               &                         &                         &                   &                   &                   &                   \\ \hline
\multicolumn{1}{|c|}{\multirow{2}{*}{$(RR)(RR)$}} & $Q_{eu}$               & 2              &              & 2                &                &                  &                & 1              & 1             & 1                 & 1                & 1               &               &                         &                         &                   &                   &                   &                   \\
\multicolumn{1}{|c|}{}                            & $Q_{ed}$               & 2              &              &                  &                & 2                &                & 1              & 1             & 1                 & 1                & 1               &               &                         &                         &                   &                   &                   &                   \\
\hline
\end{tabular}}
\end{table}

\subsection*{$\text{MFV}_Q \times \U(2)^2_L$  }
\begin{table}[h]
\centering
\scalebox{0.76}{
\begin{tabular}{|cc|cc|cc|cc|cc|cc|cc|cc|cc|cc|cc|}
\hline
\multicolumn{2}{|c|}{$\text{MFV}_Q \times \U(2)^2_L$}                 & \multicolumn{2}{c|}{$\cO(1)$} & \multicolumn{2}{c|}{$\cO(Y_u)$} & \multicolumn{2}{c|}{$\cO(Y_d)$} & \multicolumn{2}{c|}{$\cO(Y_u^2)$} & \multicolumn{2}{c|}{$\cO(Y_d^2)$} & \multicolumn{2}{c|}{$\cO(V)$} & \multicolumn{2}{c|}{$\cO(V^2)$} & \multicolumn{2}{c|}{$\cO(\Delta V)$} & \multicolumn{2}{c|}{$\cO(VY_d,VY_u)$} & \multicolumn{2}{c|}{$\cO(\Delta Y_u,\Delta Y_d)$} \\ \hline
\multicolumn{1}{|c|}{$(LL)(LL)$}                  & $Q_{\ell q}^{(1,3)}$   & 4              &              &                &                &                &                & 4                &                & 4                &                & 2             & 2             & 2               &               &                   &                  &                   &                   &                         &                         \\ \hline
\multicolumn{1}{|c|}{\multirow{2}{*}{$(RR)(RR)$}} & $Q_{eu}$               & 2              &              &                &                &                &                & 2                &                &                  &                &               &               &                 &               & 1                 & 1                &                   &                   &                         &                         \\
\multicolumn{1}{|c|}{}                            & $Q_{ed}$               & 2              &              &                &                &                &                &                  &                & 2                &                &               &               &                 &               & 1                 & 1                &                   &                   &                         &                         \\
\hline
\end{tabular}}
\end{table}

\subsection*{$\text{MFV}_Q \times \U(2)_{V,L}$}
\begin{table}[h]
\centering
\scalebox{0.973}{
\begin{tabular}{|cc|cc|cc|cc|cc|cc|cc|cc|cc|}
\hline
\multicolumn{2}{|c|}{$\text{MFV}_Q \times \U(2)_{V,L}$}          & \multicolumn{2}{c|}{$\cO(1)$} & \multicolumn{2}{c|}{$\cO(Y_u)$} & \multicolumn{2}{c|}{$\cO(Y_u^2)$} & \multicolumn{2}{c|}{$\cO(Y_d)$} & \multicolumn{2}{c|}{$\cO(Y_d^2)$} & \multicolumn{2}{c|}{$\cO(\Delta_\ell)$} & \multicolumn{2}{c|}{$\cO(\Delta_\ell Y_u)$} & \multicolumn{2}{c|}{$\cO(\Delta_\ell Y_d)$} \\ \hline
\multicolumn{1}{|c|}{$(LL)(LL)$}                  & $Q_{\ell q}^{(1,3)}$   & 4              &              &                &                & 4                &                &                &                & 4                &                & 2                   &                   &                      &                      &                      &                      \\ \hline
\multicolumn{1}{|c|}{\multirow{2}{*}{$(RR)(RR)$}} & $Q_{eu}$               & 2              &              &                &                & 2                &                &                &                &                  &                & 1                   &                   &                      &                      &                      &                      \\
\multicolumn{1}{|c|}{}                            & $Q_{ed}$               & 2              &              &                &                &                  &                &                &                & 2                &                & 1                   &                   &                      &                      &                      &                      \\ \hline
\multicolumn{1}{|c|}{\multirow{3}{*}{$(LL)(RR)$}} & $Q_{\ell u}$           & 2              &              &                &                & 2                &                &                &                &                  &                & 1                   &                   &                      &                      &                      &                      \\
\hline
\end{tabular}}
\end{table}

\newpage
\subsection*{$\text{MFV}_Q \times \U(1)^6_L$  }
\begin{table}[h]
\centering
\begin{tabular}{|cc|cc|cc|cc|cc|cc|}
\hline
\multicolumn{2}{|c|}{$\text{MFV}_Q \times \U(1)^6_L$}                 & \multicolumn{2}{c|}{$\cO(1)$} & \multicolumn{2}{c|}{$\cO(Y_u^2)$} & \multicolumn{2}{c|}{$\cO(Y_d^2)$} & \multicolumn{2}{c|}{$\cO(y Y_u)$} & \multicolumn{2}{c|}{$\cO(y Y_d)$} \\ \hline
\multicolumn{1}{|c|}{$(LL)(LL)$}                  & $Q_{\ell q}^{(1,3)}$   & 6              &              & 6                &                & 6                &                &                 &                 &                 &                 \\ \hline
\multicolumn{1}{|c|}{\multirow{2}{*}{$(RR)(RR)$}} & $Q_{eu}$               & 3              &              & 3                &                &                  &                &                 &                 &                 &                 \\
\multicolumn{1}{|c|}{}                            & $Q_{ed}$               & 3              &              &                  &                & 3                &                &                 &                 &                 &                 \\ \hline
\multicolumn{1}{|c|}{\multirow{3}{*}{$(LL)(RR)$}} & $Q_{\ell u}$           & 3              &              & 3                &                &                  &                &                 &                 &                 &                 \\
\multicolumn{1}{|c|}{}                            & $Q_{\ell d}$           & 3              &              &                  &                & 3                &                &                 &                 &                 &                 \\
\multicolumn{1}{|c|}{}                            & $Q_{qe}$               & 3              &              & 3                &                & 3                &                &                 &                 &                 &                 \\
\hline
\end{tabular}
\end{table}

\subsection*{$\text{MFV}_Q \times \U(1)^3_L$  }
\begin{table}[h]
\centering
\begin{tabular}{|cc|cc|cc|cc|cc|cc|cc|cc|}
\hline
\multicolumn{2}{|c|}{$\text{MFV}_Q \times \U(1)^3_L$}                 & \multicolumn{2}{c|}{$\cO(1)$} & \multicolumn{2}{c|}{$\cO(Y_u)$} & \multicolumn{2}{c|}{$\cO(Y_d)$} & \multicolumn{2}{c|}{$\cO(Y_u^2)$} & \multicolumn{2}{c|}{$\cO(Y_d^2)$} & \multicolumn{2}{c|}{$\cO(Y_u^2 Y_d)$} & \multicolumn{2}{c|}{$\cO(Y_d^2 Y_u)$} \\ \hline
\multicolumn{1}{|c|}{$(LL)(LL)$}                  & $Q_{\ell q}^{(1,3)}$   & 6              &              &                &                &                &                & 6                &                & 6                &                &                   &                   &                   &                   \\ \hline
\multicolumn{1}{|c|}{\multirow{2}{*}{$(RR)(RR)$}} & $Q_{eu}$               & 3              &              &                &                &                &                & 3                &                &                  &                &                   &                   &                   &                   \\
\multicolumn{1}{|c|}{}                            & $Q_{ed}$               & 3              &              &                &                &                &                &                  &                & 3                &                &                   &                   &                   &                   \\ \hline
\multicolumn{1}{|c|}{\multirow{3}{*}{$(LL)(RR)$}} & $Q_{\ell u}$           & 3              &              &                &                &                &                & 3                &                &                  &                &                   &                   &                   &                   \\
\hline
\end{tabular}
\label{quarkMFVlep3U1}
\end{table}

\subsection*{$(\U(2)^2\times \U(3))_Q\times\text{MFV}_L $}
\begin{table}[h]
\centering
\scalebox{0.59}{
\begin{tabular}{|cc|cc|cc|cc|cc|cc|cc|cc|cc|cc|cc|cc|cc|cc|}
\hline
\multicolumn{2}{|c|}{$(\U(2)^2\times \U(3))_Q\times\text{MFV}_L$}          & \multicolumn{2}{c|}{$\cO(1)$} & \multicolumn{2}{c|}{$\cO(V_q)$} & \multicolumn{2}{c|}{$\cO(Y_e)$} & \multicolumn{2}{c|}{$\cO(\Lambda_d^2)$} & \multicolumn{2}{c|}{$\cO(V_q^2)$} & \multicolumn{2}{c|}{$\cO(\Lambda_d \Sigma_d)$} & \multicolumn{2}{c|}{$\cO(\Lambda_d Y_e)$} & \multicolumn{2}{c|}{$\cO(V_q Y_e)$} & \multicolumn{2}{c|}{$\cO(V_q \Delta)$} & \multicolumn{2}{c|}{$\cO(\Sigma_d^2)$} & \multicolumn{2}{c|}{$\cO(\Sigma_d Y_e)$} & \multicolumn{2}{c|}{$\cO(Y_e^2)$} & \multicolumn{2}{c|}{$\cO(Y_e \Delta)$} \\ \hline
\multicolumn{1}{|c|}{$(LL)(LL)$}                  & $Q_{\ell q}^{(1,3)}$   & 4              &              & 2              & 2              &                &                &                  &                & 2                &                & 2                & 2                &                  &                  &                  &                  &                    &                   & 2                &                &                  &                  & 4                &                &                    &                   \\ \hline
\multicolumn{1}{|c|}{\multirow{2}{*}{$(RR)(RR)$}} & $Q_{eu}$,              & 2              &              &                &                &                &                &                  &                &                  &                &                  &                  &                  &                  &                  &                  & 1                  & 1                 &                  &                &                  &                  & 2                &                &                    &                   \\
\hline
\end{tabular}}
\end{table}

\newpage
\subsection*{$(\U(2)^2\times \U(3))_Q\times\U(3)_{V,L}$}
\begin{table}[h]
\centering
\scalebox{0.655}{
\begin{tabular}{|cc|cc|cc|cc|cc|cc|cc|cc|cc|cc|cc|}
\hline
\multicolumn{2}{|c|}{$(\U(2)^2\times \U(3))_Q\times\U(3)_{V,L}$} & \multicolumn{2}{c|}{$\cO(1)$} & \multicolumn{2}{c|}{$\cO(\Lambda_d)$} & \multicolumn{2}{c|}{$\cO(V_q)$} & \multicolumn{2}{c|}{$\cO(\Sigma_d)$} & \multicolumn{2}{c|}{$\cO(\Delta_\ell)$} & \multicolumn{2}{c|}{$\cO(\Delta)$} & \multicolumn{2}{c|}{$\cO(\Lambda_d^2,V_q^2,\Lambda_dV_q)$} & \multicolumn{2}{c|}{$\cO(\Lambda_d\Sigma_d,V_q\Sigma_d)$} & \multicolumn{2}{c|}{$\cO(\Lambda_d\Delta_\ell,V_q\Delta_\ell)$} & \multicolumn{2}{c|}{$\cO(V_q \Delta)$} \\ \hline
\multicolumn{1}{|c|}{$(LL)(LL)$}                  & $Q_{\ell q}^{(1,3)}$   & 4              &              &                   &                   & 2              & 2              &                   &                  & 4                   &                   &                  &                 & 2                            &                             & 2                           & 2                           & 2                              & 2                              &                    &                   \\ \hline
\multicolumn{1}{|c|}{\multirow{2}{*}{$(RR)(RR)$}} & $Q_{eu}$,              & 2              &              &                   &                   &                &                &                   &                  & 2                   &                   &                  &                 &                              &                             &                             &                             &                                &                                & 1                  & 1                 \\
\hline
\end{tabular}}
\end{table}

\subsection*{$(\U(2)^2\times \U(3))_Q\times(\U(2)^2\times \U(1)^2)_L$}
\begin{table}[h]
\centering
\scalebox{0.843}{
\begin{tabular}{|cc|cc|cc|cc|cc|cc|cc|}
\hline
\multicolumn{2}{|c|}{$(\U(2)^2\times \U(3))_Q\times(\U(2)^2\times \U(1)^2)_L$} & \multicolumn{2}{c|}{$\cO(1)$} & \multicolumn{2}{c|}{$\cO(V_{e,\ell,q})$} & \multicolumn{2}{c|}{$\cO(X)$} & \multicolumn{2}{c|}{$\cO(\Delta_e)$} & \multicolumn{2}{c|}{\scalebox{0.92}{$\cO(V^2_{e,\ell,q},\Lambda_d^2,\Lambda_d V_{e,\ell}, V_{e,\ell}V_q)$}} & \multicolumn{2}{c|}{$\cO(XV_{e,\ell,q})$} \\ \hline
\multicolumn{1}{|c|}{$(LL)(LL)$}                    & $Q_{\ell q}^{(1,3)}$     & 8              &              & 4                   & 4                  &               &               &                   &                  & 8                          &                          & 4                   & 4                   \\ \hline
\multicolumn{1}{|c|}{\multirow{2}{*}{$(RR)(RR)$}}   & $Q_{eu}$,                & 4              &              &                     &                    &               &               &                   &                  & 2                          &                          & 2                   & 2                   \\
\multicolumn{1}{|c|}{}                              & $Q_{ed}$                 & 2              &              &                     &                    &               &               &                   &                  & 3                          &                          & 1                   & 1                   \\ \hline
\multicolumn{1}{|c|}{\multirow{3}{*}{$(LL)(RR)$}}                   & $Q_{\ell u}$             & 4              &              &                     &                    &               &               &                   &                  & 2                          &                          & 2                   & 2                   \\
\hline
\end{tabular}}
\end{table}

\subsection*{$(\U(2)^2\times \U(3))_Q\times\U(2)^2_L$}
\begin{table}[h]
\centering
\scalebox{0.918}{
\begin{tabular}{|cc|cc|cc|cc|cc|cc|cc|}
\hline
\multicolumn{2}{|c|}{$(\U(2)^2\times \U(3))_Q\times\U(2)^2_L$}             & \multicolumn{2}{c|}{$\cO(1)$} & \multicolumn{2}{c|}{$\cO(\Lambda_d,V_{\ell,q})$} & \multicolumn{2}{c|}{$\cO(\Sigma_d)$} & \multicolumn{2}{c|}{$\cO(\Delta_{e,u})$} & \multicolumn{2}{c|}{$\cO(\Lambda_d^2,V^2_{\ell,q})$} & \multicolumn{2}{c|}{$\cO(\Lambda_dV_\ell,\Lambda_dV_q,V_\ell V_q)$} \\ \hline
\multicolumn{1}{|c|}{$(LL)(LL)$}                  & $Q_{\ell q}^{(1,3)}$   & 8              &              & 8                   & 8                  &                &                &                    &                   & 8                     &                    & 4                          & 4                          \\ \hline
\multicolumn{1}{|c|}{\multirow{2}{*}{$(RR)(RR)$}} & $Q_{eu}$,              & 4              &              &                     &                    &                &                &                    &                   &                       &                    &                            &                            \\
\multicolumn{1}{|c|}{}                            & $Q_{ed}$               & 2              &              &                     &                    &                &                &                    &                   & 2                     &                    &                            &                            \\ \hline
\multicolumn{1}{|c|}{\multirow{3}{*}{$(LL)(RR)$}}                 & $Q_{\ell u}$           & 4              &              & 2                   & 2                  &                &                &                    &                   & 2                     &                    &                            &                            \\
\hline
\end{tabular}}
\end{table}

\newpage
\subsection*{$(\U(2)^2\times \U(3))_Q\times\U(2)_{V,L}$}
\begin{table}[h]
\centering
\scalebox{0.931}{
\begin{tabular}{|cc|cc|cc|cc|cc|cc|cc|cc|}
\hline
\multicolumn{2}{|c|}{$(\U(2)^2\times \U(3))_Q\times\U(2)_{V,L}$} & \multicolumn{2}{c|}{$\cO(1)$} & \multicolumn{2}{c|}{$\cO(\Lambda_d)$} & \multicolumn{2}{c|}{$\cO(V_q)$} & \multicolumn{2}{c|}{$\cO(\Sigma_d)$} & \multicolumn{2}{c|}{$\cO(\Delta_\ell)$} & \multicolumn{2}{c|}{$\cO(\Delta)$} & \multicolumn{2}{c|}{{\scalebox{0.92}{$\cO(\Lambda_d^2,V_{q}^2,\Lambda_dV_{q})$}}} \\ \hline
\multicolumn{1}{|c|}{$(LL)(LL)$}                  & $Q_{\ell q}^{(1,3)}$   & 8              &              &                   &                   & 4              & 4              &                   &                  & 4                   &                   &                  &                 & 4                                                  &                                                   \\ \hline
\multicolumn{1}{|c|}{\multirow{2}{*}{$(RR)(RR)$}} & $Q_{eu}$,              & 4              &              &                   &                   &                &                &                   &                  & 2                   &                   &                  &                 &                                                    &                                                   \\
\multicolumn{1}{|c|}{}                            & $Q_{ed}$               & 2              &              &                   &                   &                &                &                   &                  & 1                   &                   &                  &                 & 2                                                  &                                                   \\
\hline
\end{tabular}}
\end{table}

\subsection*{$(\U(2)^2\times \U(3))_Q\times\U(1)^6_L$}
\begin{table}[h]
\centering
\scalebox{0.90}{
\begin{tabular}{|cc|cc|cc|cc|cc|cc|cc|cc|cc|}
\hline
\multicolumn{2}{|c|}{$(\U(2)^2\times \U(3))_Q\times\U(1)^6_L$}             & \multicolumn{2}{c|}{$\cO(1)$} & \multicolumn{2}{c|}{$\cO(V_q)$} & \multicolumn{2}{c|}{$\cO(y)$} & \multicolumn{2}{c|}{$\cO(\Lambda_d^2,V_{q}^2)$} & \multicolumn{2}{c|}{$\cO(y\Lambda_d)$} & \multicolumn{2}{c|}{$\cO(yV_q)$} & \multicolumn{2}{c|}{$\cO(\Lambda_d \Sigma_d)$} & \multicolumn{2}{c|}{$\cO(V_q \Delta)$} \\ \hline
\multicolumn{1}{|c|}{$(LL)(LL)$}                  & $Q_{\ell q}^{(1,3)}$   & 12             &              & 6              & 6              &               &               & 6                  &                  &                 &                &                 &                & 6                & 6                &                    &                   \\ \hline
\multicolumn{1}{|c|}{\multirow{2}{*}{$(RR)(RR)$}} & $Q_{eu}$,              & 6              &              &                &                &               &               &                    &                  &                 &                &                 &                &                  &                  & 3                  & 3                 \\
\multicolumn{1}{|c|}{}                            & $Q_{ed}$               & 3              &              &                &                &               &               & 3                  &                  &                 &                &                 &                &                  &                  &                    &                   \\ \hline
\multicolumn{1}{|c|}{\multirow{3}{*}{$(LL)(RR)$}}                 & $Q_{\ell u}$           & 6              &              &                &                &               &               &                    &                  &                 &                &                 &                &                  &                  & 3                  & 3                 \\
\hline
\end{tabular}}
\end{table}

\subsection*{$(\U(2)^2\times \U(3))_Q\times\U(1)^3_L$}
\begin{table}[h]
\centering
\scalebox{0.80}{
\begin{tabular}{|cc|cc|cc|cc|cc|cc|cc|cc|cc|}
\hline
\multicolumn{2}{|c|}{$(\U(2)^2\times \U(3))_Q\times\U(1)^3_L$}             & \multicolumn{2}{c|}{$\cO(1)$} & \multicolumn{2}{c|}{$\cO(\Lambda_d)$} & \multicolumn{2}{c|}{$\cO(V_q)$} & \multicolumn{2}{c|}{$\cO(\Sigma_d)$} & \multicolumn{2}{c|}{$\cO(\Delta)$} & \multicolumn{2}{c|}{$\cO(\Lambda_d^2,V_{q}^2,\Lambda_dV_q)$} & \multicolumn{2}{c|}{$\cO(\Lambda_d\Sigma_d,V_{q}\Sigma_d)$} & \multicolumn{2}{c|}{$\cO(V_q\Delta)$} \\ \hline
\multicolumn{1}{|c|}{$(LL)(LL)$}                  & $Q_{\ell q}^{(1,3)}$   & 12             &              &                &                & 6              & 6              &                &                &                  &                 & 6                     &                      & 6                  & 6                 &                   &                   \\ \hline
\multicolumn{1}{|c|}{\multirow{2}{*}{$(RR)(RR)$}} & $Q_{eu}$,              & 6              &              &                &                &                &                &                &                &                  &                 &                       &                      &                    &                   & 3                 & 3                 \\
\multicolumn{1}{|c|}{}                            & $Q_{ed}$               & 3              &              &                &                &                &                &                &                &                  &                 & 3                     &                      &                    &                   &                   &                   \\
\hline
\end{tabular}}
\end{table}
\newpage
\subsection*{$(\U(2)^3\times \U(1))_Q\times\text{MFV}_L$}
\begin{table}[h]
\centering
\scalebox{0.95}{
\begin{tabular}{|cc|cc|cc|cc|cc|cc|cc|cc|}
\hline
\multicolumn{2}{|c|}{$(\U(2)^3\times \U(1))_Q\times\text{MFV}_L$}          & \multicolumn{2}{c|}{$\cO(1)$} & \multicolumn{2}{c|}{$\cO(V_q)$} & \multicolumn{2}{c|}{$\cO(Y_e)$} & \multicolumn{2}{c|}{$\cO(V_q^2)$} & \multicolumn{2}{c|}{$\cO(V_q Y_e)$} & \multicolumn{2}{c|}{$\cO(V_q \Delta)$} & \multicolumn{2}{c|}{$\cO(X_b Y_e)$} \\ \hline
\multicolumn{1}{|c|}{$(LL)(LL)$}                  & $Q_{\ell q}^{(1,3)}$   & 4              &              & 2              & 2              &                &                & 2                &                &                  &                  &                    &                   &                  &                  \\ \hline
\multicolumn{1}{|c|}{\multirow{2}{*}{$(RR)(RR)$}} & $Q_{eu}$,              & 2              &              &                &                &                &                &                  &                &                  &                  & 1                  & 1                 &                  &                  \\
\multicolumn{1}{|c|}{}                            & $Q_{ed}$               & 2              &              &                &                &                &                &                  &                &                  &                  &                    &                   &                  &                  \\ \hline
\multicolumn{1}{|c|}{\multirow{3}{*}{$(LL)(RR)$}}                 & $Q_{\ell u}$           & 2              &              &                &                &                &                &                  &                &                  &                  & 1                  & 1                 &                  &                  \\
\hline
\end{tabular}}
\end{table}

\subsection*{$(\U(2)^3\times \U(1))_Q\times\U(3)_{V,L}$}
\begin{table}[h]
\centering
\scalebox{0.70}{
\begin{tabular}{|cc|cc|cc|cc|cc|cc|cc|cc|cc|cc|cc|}
\hline
\multicolumn{2}{|c|}{$(\U(2)^3\times \U(1))_Q\times\U(3)_{V,L}$} & \multicolumn{2}{c|}{$\cO(1)$} & \multicolumn{2}{c|}{$\cO(V_q)$} & \multicolumn{2}{c|}{$\cO(X_b)$} & \multicolumn{2}{c|}{$\cO(\Delta)$} & \multicolumn{2}{c|}{$\cO(\Delta_\ell)$} & \multicolumn{2}{c|}{$\cO(V_q^2)$} & \multicolumn{2}{c|}{$\cO(V_q X_b)$} & \multicolumn{2}{c|}{$\cO(V_q \Delta)$} & \multicolumn{2}{c|}{$\cO(V_q \Delta_\ell)$} & \multicolumn{2}{c|}{$\cO(X_b \Delta_\ell)$} \\ \hline
\multicolumn{1}{|c|}{$(LL)(LL)$}                  & $Q_{\ell q}^{(1,3)}$   & 4              &              & 2              & 2              &                &                &                  &                 & 4                   &                   & 2                &                &                  &                  &                    &                   & 2                    & 2                    &                      &                      \\ \hline
\multicolumn{1}{|c|}{\multirow{2}{*}{$(RR)(RR)$}} & $Q_{eu}$,              & 2              &              &                &                &                &                &                  &                 & 2                   &                   &                  &                &                  &                  & 1                  & 1                 &                      &                      &                      &                      \\
\multicolumn{1}{|c|}{}                            & $Q_{ed}$               & 2              &              &                &                &                &                &                  &                 & 2                   &                   &                  &                &                  &                  &                    &                   &                      &                      &                      &                      \\
\hline
\end{tabular}}
\end{table}

\subsection*{$(\U(2)^3\times \U(1))_Q\times(\U(2)^2\times \U(1)^2)_L$}
\begin{table}[h]
\centering
\scalebox{0.62}{
\begin{tabular}{|cc|cc|cc|cc|cc|cc|cc|cc|cc|cc|cc|}
\hline
\multicolumn{2}{|c|}{$(\U(2)^3\times \U(1))_Q\times(\U(2)^2\times \U(1)^2)_L$} & \multicolumn{2}{c|}{$\cO(1)$} & \multicolumn{2}{c|}{$\cO(V_{e,\ell,q})$} & \multicolumn{2}{c|}{$\cO(X)$} & \multicolumn{2}{c|}{$\cO(\Delta_e)$} & \multicolumn{2}{c|}{$\cO(V_e^2,V_eV_q)$} & \multicolumn{2}{c|}{$\cO(V_eX,V_eX_b)$} & \multicolumn{2}{c|}{$\cO(V_\ell^2,V_\ell V_q)$} & \multicolumn{2}{c|}{$\cO(V_\ell X, V_\ell X_b)$} & \multicolumn{2}{c|}{$\cO(V_q^2)$} & \multicolumn{2}{c|}{$\cO(XX_b)$} \\ \hline
\multicolumn{1}{|c|}{$(LL)(LL)$}                     & $Q_{\ell q}^{(1,3)}$     & 8              &              & 4                   & 4                  &               &               &                   &                  &                     &                    &                    &                    & 4                       &                       & 4                       & 4                      & 4                &                &                 &                \\ \hline
\multicolumn{1}{|c|}{\multirow{2}{*}{$(RR)(RR)$}}    & $Q_{eu}$,                & 4              &              &                     &                    &               &               &                   &                  & 2                   &                    & 2                  & 2                  &                         &                       &                         &                        &                  &                &                 &                \\
\multicolumn{1}{|c|}{}                               & $Q_{ed}$                 & 4              &              &                     &                    &               &               &                   &                  & 2                   &                    & 2                  & 2                  &                         &                       &                         &                        &                  &                &                 &                \\
\hline
\end{tabular}}
\end{table}

\newpage
\subsection*{$(\U(2)^3\times \U(1))_Q\times \U(2)^2_L$}

\begin{table}[h]
\centering
\scalebox{0.71}{
\begin{tabular}{|cc|cc|cc|cc|cc|cc|cc|cc|cc|cc|cc|}
\hline
\multicolumn{2}{|c|}{$(\U(2)^3\times \U(1))_Q\times(\U(2)^2)_L$}          & \multicolumn{2}{c|}{$\cO(1)$} & \multicolumn{2}{c|}{$\cO(V_\ell)$} & \multicolumn{2}{c|}{$\cO(V_q)$} & \multicolumn{2}{c|}{$\cO(X_b)$} & \multicolumn{2}{c|}{$\cO(\Delta)$} & \multicolumn{2}{c|}{$\cO(\Delta_e)$} & \multicolumn{2}{c|}{$\cO(V_\ell^2)$} & \multicolumn{2}{c|}{$\cO(V_q^2)$} & \multicolumn{2}{c|}{$\cO(V_\ell V_q)$} & \multicolumn{2}{c|}{$\cO(X_b V_{\ell,q})$} \\ \hline
\multicolumn{1}{|c|}{$(LL)(LL)$}                  & $Q_{\ell q}^{(1,3)}$   & 8              &              & 4                & 4               & 4              & 4              &                &                &                  &                 &                   &                  & 4                 &                  & 4                &                & 4                  & 4                 &                      &                     \\ \hline
\multicolumn{1}{|c|}{\multirow{2}{*}{$(RR)(RR)$}} & $Q_{eu}$,              & 4              &              &                  &                 &                &                &                &                &                  &                 &                   &                  &                   &                  &                  &                &                    &                   &                      &                     \\
\multicolumn{1}{|c|}{}                            & $Q_{ed}$               & 4              &              &                  &                 &                &                &                &                &                  &                 &                   &                  &                   &                  &                  &                &                    &                   &                      &                     \\
\hline
\end{tabular}}
\end{table}

\subsection*{$(\U(2)^3\times \U(1))_Q\times \U(2)_{V,L}$}
\begin{table}[h]
\centering
\scalebox{0.93}{\begin{tabular}{|cc|cc|cc|cc|cc|cc|cc|cc|}
\hline
\multicolumn{2}{|c|}{$(\U(2)^3\times \U(1))_Q\times \U(2)_{V,L}$} & \multicolumn{2}{c|}{$\cO(1)$} & \multicolumn{2}{c|}{$\cO(V_q)$} & \multicolumn{2}{c|}{$\cO(X_b)$} & \multicolumn{2}{c|}{$\cO(\Delta)$} & \multicolumn{2}{c|}{$\cO(\Delta_\ell)$} & \multicolumn{2}{c|}{$\cO(V_q^2)$} & \multicolumn{2}{c|}{$\cO(X_b V_q)$} \\ \hline
\multicolumn{1}{|c|}{$(LL)(LL)$}                   & $Q_{\ell q}^{(1,3)}$   & 8              &              & 4              & 4              &                &                &                  &                 & 4                   &                   & 4                &                &                  &                  \\ \hline
\multicolumn{1}{|c|}{\multirow{2}{*}{$(RR)(RR)$}}  & $Q_{eu}$               & 4              &              &                &                &                &                &                  &                 & 2                   &                   &                  &                &                  &                  \\
\multicolumn{1}{|c|}{}                             & $Q_{ed}$               & 4              &              &                &                &                &                &                  &                 & 2                   &                   &                  &                &                  &                  \\ \hline
\multicolumn{1}{|c|}{\multirow{3}{*}{$(LL)(RR)$}}  & $Q_{\ell u}$           & 4              &              &                &                &                &                &                  &                 & 2                   &                   &                  &                &                  &                  \\
\hline
\end{tabular}}
\end{table}

\subsection*{$(\U(2)^3\times \U(1))_Q\times \U(1)_L^6$}
\begin{table}[h]
\centering
\scalebox{0.89}{
\begin{tabular}{|cc|cc|cc|cc|cc|cc|cc|cc|}
\hline
\multicolumn{2}{|c|}{$(\U(2)^3\times \U(1))_Q\times  \U(1)^6_L$}           & \multicolumn{2}{c|}{$\cO(1)$} & \multicolumn{2}{c|}{$\cO(V_q)$} & \multicolumn{2}{c|}{$\cO(V_q^2)$} & \multicolumn{2}{c|}{$\cO(y)$} & \multicolumn{2}{c|}{$\cO(V_q y)$} & \multicolumn{2}{c|}{$\cO(yX_b)$} & \multicolumn{2}{c|}{\scalebox{0.94}{$\cO(V_q \Delta_u,V_q X_b\Delta_d)$}} \\ \hline
\multicolumn{1}{|c|}{$(LL)(LL)$}                  & $Q_{\ell q}^{(1,3)}$   & 12             &              & 6              & 6              & 6                &                &               &               &                 &                 &                 &                &                     &                    \\ \hline
\multicolumn{1}{|c|}{\multirow{2}{*}{$(RR)(RR)$}} & $Q_{eu}$,              & 6              &              &                &                &                  &                &               &               &                 &                 &                 &                & 3                   & 3                  \\
\multicolumn{1}{|c|}{}                            & $Q_{ed}$               & 6              &              &                &                &                  &                &               &               &                 &                 &                 &                & 3                    & 3                   \\ \hline
\multicolumn{1}{|c|}{\multirow{3}{*}{$(LL)(RR)$}}                 & $Q_{\ell u}$           & 6              &              &                &                &                  &                &               &               &                 &                 &                 &                & 3                   & 3                  \\
\hline
\end{tabular}}
\end{table}

\newpage
\subsection*{$(\U(2)^3\times \U(1))_Q\times \U(1)_L^3$}
\begin{table}[h]
\centering
\scalebox{0.93}{
\begin{tabular}{|cc|cc|cc|cc|cc|cc|cc|cc|}
\hline
\multicolumn{2}{|c|}{$(\U(2)^3\times \U(1))_Q\times  \U(1)^3_L$}           & \multicolumn{2}{c|}{$\cO(1)$} & \multicolumn{2}{c|}{$\cO(V_q)$} & \multicolumn{2}{c|}{$\cO(X_b)$} & \multicolumn{2}{c|}{$\cO(\Delta)$} & \multicolumn{2}{c|}{$\cO(V_q^2)$} & \multicolumn{2}{c|}{$\cO(X_b V_q)$} & \multicolumn{2}{c|}{$\cO(V_q \Delta)$} \\ \hline
\multicolumn{1}{|c|}{$(LL)(LL)$}                  & $Q_{\ell q}^{(1,3)}$   & 12             &              & 6              & 6              &                &                &                  &                 & 6                &                &                  &                  &                    &                   \\ \hline
\multicolumn{1}{|c|}{\multirow{2}{*}{$(RR)(RR)$}} & $Q_{eu}$,              & 6              &              &                &                &                &                &                  &                 &                  &                &                  &                  & 3                  & 3                 \\
\multicolumn{1}{|c|}{}                            & $Q_{ed}$               & 6              &              &                &                &                &                &                  &                 &                  &                &                  &                  &                    &                   \\ \hline
\multicolumn{1}{|c|}{\multirow{3}{*}{$(LL)(RR)$}}                 & $Q_{\ell u}$           & 6              &              &                &                &                &                &                  &                 &                  &                &                  &                  & 3                  & 3                 \\
\hline
\end{tabular}}
\end{table}


\subsection*{$\U(2)^3_Q \times \text{MFV}_L$  }
\begin{table}[h]
\centering
\scalebox{0.863}{
\begin{tabular}{|cc|cc|cc|cc|cc|cc|cc|cc|cc|}
\hline
\multicolumn{2}{|c|}{$\U(2)^3_Q \times \text{MFV}_L$}          & \multicolumn{2}{c|}{$\cO(1)$} & \multicolumn{2}{c|}{$\cO(V)$} & \multicolumn{2}{c|}{$\cO(V^2)$} & \multicolumn{2}{c|}{$\cO(\Delta V)$} & \multicolumn{2}{c|}{$\cO(Y_e)$} & \multicolumn{2}{c|}{$\cO(Y_e V)$} & \multicolumn{2}{c|}{$\cO(Y_e \Delta)$} & \multicolumn{2}{c|}{$\cO(Y_e \Delta V)$} \\ \hline
\multicolumn{1}{|c|}{$(LL)(LL)$}                  & $Q_{\ell q}^{(1,3)}$       & 4              &              & 2             & 2             & 2               &               &                   &                  &                &                &                 &                 &                    &                   &                     &                    \\ \hline
\multicolumn{1}{|c|}{$(RR)(RR)$}                  & $Q_{eu}$, $Q_{ed}$         & 4              &              &               &               &                 &               & 2                 & 2                &                &                &                 &                 &                    &                   &                     &                    \\ \hline
\multicolumn{1}{|c|}{\multirow{2}{*}{$(LL)(RR)$}} & $Q_{\ell u}$, $Q_{\ell d}$ & 4              &              &               &               &                 &               & 2                 & 2                &                &                &                 &                 &                    &                   &                     &                    \\
\multicolumn{1}{|c|}{}                            & $Q_{qe}$                   & 2              &              & 1             & 1             & 1               &               &                   &                  &                &                &                 &                 &                    &                   &                     &                    \\
\hline
\end{tabular}}
\end{table}

\subsection*{$\U(2)^3_Q \times \U(3)_{V,L}$}
\begin{table}[h]
\centering
\scalebox{0.88}{\begin{tabular}{|cc|cc|cc|cc|cc|cc|cc|cc|cc|}
\hline
\multicolumn{2}{|c|}{$\U(2)^3_Q \times \U(3)_{V,L}$}                 & \multicolumn{2}{c|}{$\cO(1)$} & \multicolumn{2}{c|}{$\cO(V)$} & \multicolumn{2}{c|}{$\cO(V^2)$} & \multicolumn{2}{c|}{$\cO(\Delta)$} & \multicolumn{2}{c|}{$\cO(\Delta V)$} & \multicolumn{2}{c|}{$\cO(\Delta_\ell)$} & \multicolumn{2}{c|}{$\cO(\Delta_\ell V)$} & \multicolumn{2}{c|}{$\cO(\Delta_\ell \Delta)$} \\ \hline
\multicolumn{1}{|c|}{$(LL)(LL)$}                  & $Q_{\ell q}^{(1,3)}$       & 4              &              & 2             & 2             & 2               &               &                  &                 &                   &                  & 4                   &                   & 2                   & 2                   &                        &                       \\ \hline
\multicolumn{1}{|c|}{$(RR)(RR)$}                  & $Q_{eu}$, $Q_{ed}$         & 4              &              &               &               &                 &               &                  &                 & 2                 & 2                & 4                   &                   &                     &                     &                        &                       \\ \hline
\multicolumn{1}{|c|}{\multirow{2}{*}{$(LL)(RR)$}} & $Q_{\ell u}$, $Q_{\ell d}$ & 4              &              &               &               &                 &               &                  &                 & 2                 & 2                & 4                   &                   &                     &                     &                        &                       \\
\hline
\end{tabular}}
\end{table}

\newpage
\subsection*{$\U(2)^3_Q \times (\U(2)^2 \times \U(1)^2)_L$  }
\begin{table}[h]
\centering
\scalebox{0.77}{
\begin{tabular}{|cc|cc|cc|cc|cc|cc|cc|cc|cc|cc|cc|}
\hline
\multicolumn{2}{|c|}{$\U(2)^3_Q \times (\U(2)^2 \times \U(1)^2)_L$}                          & \multicolumn{2}{c|}{$\cO(1)$} & \multicolumn{2}{c|}{$\cO(V)$} & \multicolumn{2}{c|}{$\cO(V^2)$} & \multicolumn{2}{c|}{$\cO(V^3)$} & \multicolumn{2}{c|}{$\cO(\Delta)$} & \multicolumn{2}{c|}{$\cO(X)$} & \multicolumn{2}{c|}{$\cO(\Delta V)$} & \multicolumn{2}{c|}{$\cO(XV)$} & \multicolumn{2}{c|}{$\cO(X\Delta)$} & \multicolumn{2}{c|}{$\cO(XV\Delta)$} \\ \hline
\multicolumn{1}{|c|}{$(LL)(LL)$}                  & $Q_{\ell q}^{(1,3)}$       & 8              &              & 4             & 4             & 8               &               & 2              & 2              &                  &                 &               &               & 4                 & 4                & 4              & 4             &                  &                  &                   &                  \\ \hline
\multicolumn{1}{|c|}{$(RR)(RR)$}                  & $Q_{eu}$, $Q_{ed}$         & 8              &              &               &               & 4               &               &                &                &                  &                 &               &               & 8                 & 8                & 4              & 4             &                  &                  &                   &                  \\ \hline
\multicolumn{1}{|c|}{\multirow{2}{*}{$(LL)(RR)$}} & $Q_{\ell u}$, $Q_{\ell d}$ & 8              &              &               &               & 4               &               &                &                &                  &                 &               &               & 8                 & 8                & 4              & 4             &                  &                  &                   &                  \\
\hline
\end{tabular}}
\end{table}

\subsection*{$\U(2)^3_Q \times \U(2)^2_L$  }
\begin{table}[h]
\centering
\begin{tabular}{|cc|cc|cc|cc|cc|cc|cc|}
\hline
\multicolumn{2}{|c|}{$\U(2)^3_Q \times \U(2)^2_L$}                             & \multicolumn{2}{c|}{$\cO(1)$} & \multicolumn{2}{c|}{$\cO(V)$} & \multicolumn{2}{c|}{$\cO(V^2)$} & \multicolumn{2}{c|}{$\cO(V^3)$} & \multicolumn{2}{c|}{$\cO(\Delta)$} & \multicolumn{2}{c|}{$\cO(\Delta V)$} \\ \hline
\multicolumn{1}{|c|}{$(LL)(LL)$}                  & $Q_{\ell q}^{(1,3)}$       & 8              &              & 8             & 8             & 12              & 4             & 4              & 4              &                  &                 &                   &                  \\ \hline
\multicolumn{1}{|c|}{$(RR)(RR)$}                  & $Q_{eu}$, $Q_{ed}$         & 8              &              &               &               &                 &               &                &                &                  &                 & 8                 & 8                \\ \hline
\multicolumn{1}{|c|}{\multirow{2}{*}{$(LL)(RR)$}} & $Q_{\ell u}$, $Q_{\ell d}$ & 8              &              & 4             & 4             & 4               &               &                &                &                  &                 & 4                 & 4                \\
\multicolumn{1}{|c|}{}                            & $Q_{qe}$                   & 4              &              & 2             & 2             & 2               &               &                &                &                  &                 & 2                 & 2                \\ \hline
\multicolumn{1}{|c|}{$(LR)(RL)$}                  & $Q_{\ell edq}$             & 1              & 1            & 2             & 2             & 1               & 1             &                &                & 2                & 2               & 4                 & 4                \\
\hline
\end{tabular}
\end{table}

\subsection*{$\U(2)^3_Q \times \U(2)_{V,L}$}
\begin{table}[h]
\centering
\scalebox{0.87}{\begin{tabular}{|cc|cc|cc|cc|cc|cc|cc|cc|}
\hline
\multicolumn{2}{|c|}{$\U(2)^3_Q \times \U(2)_{V,L}$}                 & \multicolumn{2}{c|}{$\cO(1)$} & \multicolumn{2}{c|}{$\cO(V_q)$} & \multicolumn{2}{c|}{$\cO(\Delta)$} & \multicolumn{2}{c|}{$\cO(\Delta_\ell)$} & \multicolumn{2}{c|}{$\cO(V_q^2)$} & \multicolumn{2}{c|}{$\cO(V_q\Delta)$} & \multicolumn{2}{c|}{$\cO(V_q\Delta_\ell)$} \\ \hline
\multicolumn{1}{|c|}{$(LL)(LL)$}                  & $Q_{\ell q}^{(1,3)}$       & 8              &              & 4              & 4              &                  &                 & 4                   &                   & 4                &                &                   &                   & 2                     & 2                    \\ \hline
\multicolumn{1}{|c|}{$(RR)(RR)$}                  & $Q_{eu}$, $Q_{ed}$         & 8              &              &                &                &                  &                 & 4                   &                   &                  &                & 4                 & 4                 &                       &                      \\ \hline
\multicolumn{1}{|c|}{\multirow{2}{*}{$(LL)(RR)$}} & $Q_{\ell u}$, $Q_{\ell d}$ & 8              &              &                &                &                  &                 & 4                   &                   &                  &                & 4                 & 4                 &                       &                      \\
\multicolumn{1}{|c|}{}                            & $Q_{qe}$                   & 4              &              & 2              & 2              &                  &                 & 2                   &                   & 2                &                &                   &                   & 1                     & 1                    \\
\hline
\end{tabular}}
\end{table}

\newpage
\subsection*{$\U(2)^3_Q \times \U(1)^6_L$  }
\begin{table}[h]
\centering
\scalebox{0.88}{
\begin{tabular}{|cc|cc|cc|cc|cc|cc|cc|cc|cc|cc|}
\hline
\multicolumn{2}{|c|}{$\U(2)^3_Q \times \U(1)^6_L$}                          & \multicolumn{2}{c|}{$\cO(1)$} & \multicolumn{2}{c|}{$\cO(V)$} & \multicolumn{2}{c|}{$\cO(V^2)$} & \multicolumn{2}{c|}{$\cO(\Delta)$} & \multicolumn{2}{c|}{$\cO(\Delta V)$} & \multicolumn{2}{c|}{$\cO(y)$} & \multicolumn{2}{c|}{$\cO(yV)$} & \multicolumn{2}{c|}{$\cO(y\Delta)$} & \multicolumn{2}{c|}{$\cO(y\Delta V)$} \\ \hline
\multicolumn{1}{|c|}{$(LL)(LL)$}                  & $Q_{\ell q}^{(1,3)}$       & 12             &              & 6             & 6             & 6               &               &                  &                 &                   &                  &               &               &                &               &                  &                  &                   &                   \\ \hline
\multicolumn{1}{|c|}{$(RR)(RR)$}                  & $Q_{eu}$, $Q_{ed}$         & 12             &              &               &               &                 &               &                  &                 & 6                 & 6                &               &               &                &               &                  &                  &                   &                   \\ \hline
\multicolumn{1}{|c|}{\multirow{2}{*}{$(LL)(RR)$}} & $Q_{\ell u}$, $Q_{\ell d}$ & 12             &              &               &               &                 &               &                  &                 & 6                 & 6                &               &               &                &               &                  &                  &                   &                   \\
\hline
\end{tabular}}
\end{table}

\subsection*{$\U(2)^3_Q \times \U(1)^3_L$  }
\begin{table}[h]
\centering
%
\end{table}


\section{Group identities}
\label{app:d}

In $ \SU(2) $ the following identities hold:
	\begin{equation}
	\begin{split}
	\varepsilon^{ij} \varepsilon_{k\ell} &= \delta\ud{i}{\ell} \delta\ud{j}{k} - \delta\ud{i}{k} \delta\ud{j}{\ell}
	\end{split}
	\end{equation}
using the convention $ \varepsilon_{12} = -\varepsilon^{12} $.

In $ \SU(N) $ the following identities hold:
	\begin{align}
	t\ud{ai}{j} t\ud{ak}{\ell} &= \dfrac{1}{2} \delta\ud{i}{\ell}  \delta\ud{k}{j} - \dfrac{1}{2N} \delta\ud{i}{j}  \delta\ud{k}{\ell}, \\
	f^{abc} t\ud{bi}{j} t\ud{ck}{\ell} &= \dfrac{i}{2} \big(t\ud{ai}{\ell} \delta\ud{k}{j} - t\ud{ak}{j} \delta\ud{i}{\ell} \big), \\
	d^{abc} t\ud{bi}{j} t\ud{ck}{\ell} 	&= \dfrac{1}{2} \big( t\ud{ai}{\ell} \delta\ud{k}{j} + t\ud{ak}{j} \delta\ud{i}{\ell} \big) - \dfrac{1}{N} \big(t\ud{ai}{j} \delta\ud{k}{\ell} + t\ud{ak}{\ell} \delta\ud{i}{j} \big), \label{eq:d_2tens}
	\end{align}
where the defining identity for the symmetric tensor is 
	\begin{equation}\label{eq:def_sym_tens}
	t^{a} t^{b} = \dfrac{1}{2} \left[\dfrac{1}{N} \delta^{ab} \mathds{1} + (d^{abc} + i  f^{abc}) t^{c} \right].
	\end{equation}
In the case of $ \SU(2) $ there is no 3-index symmetric tensor and Eq.~\eqref{eq:d_2tens} implies the identity
    \begin{equation}
    \label{dsixident}
    t\ud{ai}{\ell} \delta\ud{k}{j} + t\ud{ak}{j} \delta\ud{i}{\ell} = t\ud{ai}{j} \delta\ud{k}{\ell} + t\ud{ak}{\ell} \delta\ud{i}{j}. 
    \end{equation}

\bibliographystyle{JHEP}
\bibliography{References}

\providecommand{\href}[2]{#2}\begingroup\raggedright\begin{thebibliography}{100}

\bibitem{Faroughy:2020ina}
D.~A. Faroughy, G.~Isidori, F.~Wilsch and K.~Yamamoto, \emph{{Flavour
  symmetries in the SMEFT}},
  \href{https://doi.org/10.1007/JHEP08(2020)166}{\emph{JHEP} {\bfseries 08}
  (2020) 166}, [\href{https://arxiv.org/abs/2005.05366}{{\ttfamily
  2005.05366}}].

\bibitem{Weinberg:1968de}
S.~Weinberg, \emph{{Nonlinear realizations of chiral symmetry}},
  \href{https://doi.org/10.1103/PhysRev.166.1568}{\emph{Phys. Rev.} {\bfseries
  166} (1968) 1568--1577}.

\bibitem{Wilson:1969zs}
K.~G. Wilson, \emph{{Nonlagrangian models of current algebra}},
  \href{https://doi.org/10.1103/PhysRev.179.1499}{\emph{Phys. Rev.} {\bfseries
  179} (1969) 1499--1512}.

\bibitem{Wilson:1970ag}
K.~G. Wilson, \emph{{The Renormalization Group and Strong Interactions}},
  \href{https://doi.org/10.1103/PhysRevD.3.1818}{\emph{Phys. Rev. D} {\bfseries
  3} (1971) 1818}.

\bibitem{Georgi:1974yf}
H.~Georgi, H.~R. Quinn and S.~Weinberg, \emph{{Hierarchy of Interactions in
  Unified Gauge Theories}},
  \href{https://doi.org/10.1103/PhysRevLett.33.451}{\emph{Phys. Rev. Lett.}
  {\bfseries 33} (1974) 451--454}.

\bibitem{Weinberg:1978kz}
S.~Weinberg, \emph{{Phenomenological Lagrangians}},
  \href{https://doi.org/10.1016/0378-4371(79)90223-1}{\emph{Physica A}
  {\bfseries 96} (1979) 327--340}.

\bibitem{Weinberg:1979sa}
S.~Weinberg, \emph{{Baryon and Lepton Nonconserving Processes}},
  \href{https://doi.org/10.1103/PhysRevLett.43.1566}{\emph{Phys. Rev. Lett.}
  {\bfseries 43} (1979) 1566--1570}.

\bibitem{Weinberg:1980wa}
S.~Weinberg, \emph{{Effective Gauge Theories}},
  \href{https://doi.org/10.1016/0370-2693(80)90660-7}{\emph{Phys. Lett. B}
  {\bfseries 91} (1980) 51--55}.

\bibitem{Georgi:1994qn}
H.~Georgi, \emph{{Effective field theory}},
  \href{https://doi.org/10.1146/annurev.ns.43.120193.001233}{\emph{Ann. Rev.
  Nucl. Part. Sci.} {\bfseries 43} (1993) 209--252}.

\bibitem{Manohar:2018aog}
A.~V. Manohar, \emph{{Introduction to Effective Field Theories}},
  \href{https://arxiv.org/abs/1804.05863}{{\ttfamily 1804.05863}}.

\bibitem{Buchmuller:1985jz}
W.~Buchmuller and D.~Wyler, \emph{{Effective Lagrangian Analysis of New
  Interactions and Flavor Conservation}},
  \href{https://doi.org/10.1016/0550-3213(86)90262-2}{\emph{Nucl. Phys. B}
  {\bfseries 268} (1986) 621--653}.

\bibitem{Giudice:2007fh}
G.~F. Giudice, C.~Grojean, A.~Pomarol and R.~Rattazzi, \emph{{The
  Strongly-Interacting Light Higgs}},
  \href{https://doi.org/10.1088/1126-6708/2007/06/045}{\emph{JHEP} {\bfseries
  06} (2007) 045}, [\href{https://arxiv.org/abs/hep-ph/0703164}{{\ttfamily
  hep-ph/0703164}}].

\bibitem{Grzadkowski:2010es}
B.~Grzadkowski, M.~Iskrzynski, M.~Misiak and J.~Rosiek, \emph{{Dimension-Six
  Terms in the Standard Model Lagrangian}},
  \href{https://doi.org/10.1007/JHEP10(2010)085}{\emph{JHEP} {\bfseries 10}
  (2010) 085}, [\href{https://arxiv.org/abs/1008.4884}{{\ttfamily 1008.4884}}].

\bibitem{Alonso:2013hga}
R.~Alonso, E.~E. Jenkins, A.~V. Manohar and M.~Trott, \emph{{Renormalization
  Group Evolution of the Standard Model Dimension Six Operators III: Gauge
  Coupling Dependence and Phenomenology}},
  \href{https://doi.org/10.1007/JHEP04(2014)159}{\emph{JHEP} {\bfseries 04}
  (2014) 159}, [\href{https://arxiv.org/abs/1312.2014}{{\ttfamily 1312.2014}}].

\bibitem{Jenkins:2013wua}
E.~E. Jenkins, A.~V. Manohar and M.~Trott, \emph{{Renormalization Group
  Evolution of the Standard Model Dimension Six Operators II: Yukawa
  Dependence}}, \href{https://doi.org/10.1007/JHEP01(2014)035}{\emph{JHEP}
  {\bfseries 01} (2014) 035},
  [\href{https://arxiv.org/abs/1310.4838}{{\ttfamily 1310.4838}}].

\bibitem{Jenkins:2013zja}
E.~E. Jenkins, A.~V. Manohar and M.~Trott, \emph{{Renormalization Group
  Evolution of the Standard Model Dimension Six Operators I: Formalism and
  lambda Dependence}},
  \href{https://doi.org/10.1007/JHEP10(2013)087}{\emph{JHEP} {\bfseries 10}
  (2013) 087}, [\href{https://arxiv.org/abs/1308.2627}{{\ttfamily 1308.2627}}].

\bibitem{Henning:2014wua}
B.~Henning, X.~Lu and H.~Murayama, \emph{{How to use the Standard Model
  effective field theory}},
  \href{https://doi.org/10.1007/JHEP01(2016)023}{\emph{JHEP} {\bfseries 01}
  (2016) 023}, [\href{https://arxiv.org/abs/1412.1837}{{\ttfamily 1412.1837}}].

\bibitem{Fuentes-Martin:2016uol}
J.~Fuentes-Martin, J.~Portoles and P.~Ruiz-Femenia, \emph{{Integrating out
  heavy particles with functional methods: a simplified framework}},
  \href{https://doi.org/10.1007/JHEP09(2016)156}{\emph{JHEP} {\bfseries 09}
  (2016) 156}, [\href{https://arxiv.org/abs/1607.02142}{{\ttfamily
  1607.02142}}].

\bibitem{Brivio:2017vri}
I.~Brivio and M.~Trott, \emph{{The Standard Model as an Effective Field
  Theory}}, \href{https://doi.org/10.1016/j.physrep.2018.11.002}{\emph{Phys.
  Rept.} {\bfseries 793} (2019) 1--98},
  [\href{https://arxiv.org/abs/1706.08945}{{\ttfamily 1706.08945}}].

\bibitem{Celis:2017hod}
A.~Celis, J.~Fuentes-Martin, A.~Vicente and J.~Virto, \emph{{DsixTools: The
  Standard Model Effective Field Theory Toolkit}},
  \href{https://doi.org/10.1140/epjc/s10052-017-4967-6}{\emph{Eur. Phys. J. C}
  {\bfseries 77} (2017) 405},
  [\href{https://arxiv.org/abs/1704.04504}{{\ttfamily 1704.04504}}].

\bibitem{Wells:2015uba}
J.~D. Wells and Z.~Zhang, \emph{{Effective theories of universal theories}},
  \href{https://doi.org/10.1007/JHEP01(2016)123}{\emph{JHEP} {\bfseries 01}
  (2016) 123}, [\href{https://arxiv.org/abs/1510.08462}{{\ttfamily
  1510.08462}}].

\bibitem{Englert:2019zmt}
C.~Englert, G.~F. Giudice, A.~Greljo and M.~Mccullough, \emph{{The
  $\hat{H}$-Parameter: An Oblique Higgs View}},
  \href{https://doi.org/10.1007/JHEP09(2019)041}{\emph{JHEP} {\bfseries 09}
  (2019) 041}, [\href{https://arxiv.org/abs/1903.07725}{{\ttfamily
  1903.07725}}].

\bibitem{deBlas:2017xtg}
J.~de~Blas, J.~C. Criado, M.~Perez-Victoria and J.~Santiago, \emph{{Effective
  description of general extensions of the Standard Model: the complete
  tree-level dictionary}},
  \href{https://doi.org/10.1007/JHEP03(2018)109}{\emph{JHEP} {\bfseries 03}
  (2018) 109}, [\href{https://arxiv.org/abs/1711.10391}{{\ttfamily
  1711.10391}}].

\bibitem{Fuentes-Martin:2020zaz}
J.~Fuentes-Martin, P.~Ruiz-Femenia, A.~Vicente and J.~Virto, \emph{{DsixTools
  2.0: The Effective Field Theory Toolkit}},
  \href{https://doi.org/10.1140/epjc/s10052-020-08778-y}{\emph{Eur. Phys. J. C}
  {\bfseries 81} (2021) 167},
  [\href{https://arxiv.org/abs/2010.16341}{{\ttfamily 2010.16341}}].

\bibitem{Fuentes-Martin:2020udw}
J.~Fuentes-Martin, M.~K\"onig, J.~Pag\`es, A.~E. Thomsen and F.~Wilsch,
  \emph{{SuperTracer: A Calculator of Functional Supertraces for One-Loop EFT
  Matching}}, \href{https://doi.org/10.1007/JHEP04(2021)281}{\emph{JHEP}
  {\bfseries 04} (2021) 281},
  [\href{https://arxiv.org/abs/2012.08506}{{\ttfamily 2012.08506}}].

\bibitem{Cohen:2020qvb}
T.~Cohen, X.~Lu and Z.~Zhang, \emph{{STrEAMlining EFT Matching}},
  \href{https://doi.org/10.21468/SciPostPhys.10.5.098}{\emph{SciPost Phys.}
  {\bfseries 10} (2021) 098},
  [\href{https://arxiv.org/abs/2012.07851}{{\ttfamily 2012.07851}}].

\bibitem{Carmona:2021xtq}
A.~Carmona, A.~Lazopoulos, P.~Olgoso and J.~Santiago, \emph{{Matchmakereft:
  automated tree-level and one-loop matching}},
  \href{https://arxiv.org/abs/2112.10787}{{\ttfamily 2112.10787}}.

\bibitem{ATLAS:2012yve}
{\scshape ATLAS} collaboration, G.~Aad et~al., \emph{{Observation of a new
  particle in the search for the Standard Model Higgs boson with the ATLAS
  detector at the LHC}},
  \href{https://doi.org/10.1016/j.physletb.2012.08.020}{\emph{Phys. Lett. B}
  {\bfseries 716} (2012) 1--29},
  [\href{https://arxiv.org/abs/1207.7214}{{\ttfamily 1207.7214}}].

\bibitem{CMS:2012qbp}
{\scshape CMS} collaboration, S.~Chatrchyan et~al., \emph{{Observation of a New
  Boson at a Mass of 125 GeV with the CMS Experiment at the LHC}},
  \href{https://doi.org/10.1016/j.physletb.2012.08.021}{\emph{Phys. Lett. B}
  {\bfseries 716} (2012) 30--61},
  [\href{https://arxiv.org/abs/1207.7235}{{\ttfamily 1207.7235}}].

\bibitem{Buckley:2015lku}
A.~Buckley, C.~Englert, J.~Ferrando, D.~J. Miller, L.~Moore, M.~Russell et~al.,
  \emph{{Constraining top quark effective theory in the LHC Run II era}},
  \href{https://doi.org/10.1007/JHEP04(2016)015}{\emph{JHEP} {\bfseries 04}
  (2016) 015}, [\href{https://arxiv.org/abs/1512.03360}{{\ttfamily
  1512.03360}}].

\bibitem{Englert:2016aei}
C.~Englert, L.~Moore, K.~Nordstr\"om and M.~Russell, \emph{{Giving top quark
  effective operators a boost}},
  \href{https://doi.org/10.1016/j.physletb.2016.10.021}{\emph{Phys. Lett. B}
  {\bfseries 763} (2016) 9--15},
  [\href{https://arxiv.org/abs/1607.04304}{{\ttfamily 1607.04304}}].

\bibitem{Hartland:2019bjb}
N.~P. Hartland, F.~Maltoni, E.~R. Nocera, J.~Rojo, E.~Slade, E.~Vryonidou
  et~al., \emph{{A Monte Carlo global analysis of the Standard Model Effective
  Field Theory: the top quark sector}},
  \href{https://doi.org/10.1007/JHEP04(2019)100}{\emph{JHEP} {\bfseries 04}
  (2019) 100}, [\href{https://arxiv.org/abs/1901.05965}{{\ttfamily
  1901.05965}}].

\bibitem{Brivio:2019ius}
I.~Brivio, S.~Bruggisser, F.~Maltoni, R.~Moutafis, T.~Plehn, E.~Vryonidou
  et~al., \emph{{O new physics, where art thou? A global search in the top
  sector}}, \href{https://doi.org/10.1007/JHEP02(2020)131}{\emph{JHEP}
  {\bfseries 02} (2020) 131},
  [\href{https://arxiv.org/abs/1910.03606}{{\ttfamily 1910.03606}}].

\bibitem{Durieux:2018tev}
G.~Durieux, M.~Perell\'o, M.~Vos and C.~Zhang, \emph{{Global and optimal probes
  for the top-quark effective field theory at future lepton colliders}},
  \href{https://doi.org/10.1007/JHEP10(2018)168}{\emph{JHEP} {\bfseries 10}
  (2018) 168}, [\href{https://arxiv.org/abs/1807.02121}{{\ttfamily
  1807.02121}}].

\bibitem{vanBeek:2019evb}
S.~van Beek, E.~R. Nocera, J.~Rojo and E.~Slade, \emph{{Constraining the SMEFT
  with Bayesian reweighting}},
  \href{https://doi.org/10.21468/SciPostPhys.7.5.070}{\emph{SciPost Phys.}
  {\bfseries 7} (2019) 070},
  [\href{https://arxiv.org/abs/1906.05296}{{\ttfamily 1906.05296}}].

\bibitem{Bissmann:2020mfi}
S.~Bi\ss{}mann, C.~Grunwald, G.~Hiller and K.~Kr\"oninger, \emph{{Top and
  Beauty synergies in SMEFT-fits at present and future colliders}},
  \href{https://doi.org/10.1007/JHEP06(2021)010}{\emph{JHEP} {\bfseries 06}
  (2021) 010}, [\href{https://arxiv.org/abs/2012.10456}{{\ttfamily
  2012.10456}}].

\bibitem{Bruggisser:2021duo}
S.~Bruggisser, R.~Sch\"afer, D.~van Dyk and S.~Westhoff, \emph{{The Flavor of
  UV Physics}}, \href{https://doi.org/10.1007/JHEP05(2021)257}{\emph{JHEP}
  {\bfseries 05} (2021) 257},
  [\href{https://arxiv.org/abs/2101.07273}{{\ttfamily 2101.07273}}].

\bibitem{Ethier:2021bye}
{\scshape SMEFiT} collaboration, J.~J. Ethier, G.~Magni, F.~Maltoni,
  L.~Mantani, E.~R. Nocera, J.~Rojo et~al., \emph{{Combined SMEFT
  interpretation of Higgs, diboson, and top quark data from the LHC}},
  \href{https://doi.org/10.1007/JHEP11(2021)089}{\emph{JHEP} {\bfseries 11}
  (2021) 089}, [\href{https://arxiv.org/abs/2105.00006}{{\ttfamily
  2105.00006}}].

\bibitem{Ellis:2020unq}
J.~Ellis, M.~Madigan, K.~Mimasu, V.~Sanz and T.~You, \emph{{Top, Higgs, Diboson
  and Electroweak Fit to the Standard Model Effective Field Theory}},
  \href{https://doi.org/10.1007/JHEP04(2021)279}{\emph{JHEP} {\bfseries 04}
  (2021) 279}, [\href{https://arxiv.org/abs/2012.02779}{{\ttfamily
  2012.02779}}].

\bibitem{Falkowski:2015jaa}
A.~Falkowski, M.~Gonzalez-Alonso, A.~Greljo and D.~Marzocca, \emph{{Global
  constraints on anomalous triple gauge couplings in effective field theory
  approach}}, \href{https://doi.org/10.1103/PhysRevLett.116.011801}{\emph{Phys.
  Rev. Lett.} {\bfseries 116} (2016) 011801},
  [\href{https://arxiv.org/abs/1508.00581}{{\ttfamily 1508.00581}}].

\bibitem{Falkowski:2016cxu}
A.~Falkowski, M.~Gonzalez-Alonso, A.~Greljo, D.~Marzocca and M.~Son,
  \emph{{Anomalous Triple Gauge Couplings in the Effective Field Theory
  Approach at the LHC}},
  \href{https://doi.org/10.1007/JHEP02(2017)115}{\emph{JHEP} {\bfseries 02}
  (2017) 115}, [\href{https://arxiv.org/abs/1609.06312}{{\ttfamily
  1609.06312}}].

\bibitem{Baglio:2017bfe}
J.~Baglio, S.~Dawson and I.~M. Lewis, \emph{{An NLO QCD effective field theory
  analysis of $W^+W^-$ production at the LHC including fermionic operators}},
  \href{https://doi.org/10.1103/PhysRevD.96.073003}{\emph{Phys. Rev. D}
  {\bfseries 96} (2017) 073003},
  [\href{https://arxiv.org/abs/1708.03332}{{\ttfamily 1708.03332}}].

\bibitem{Panico:2017frx}
G.~Panico, F.~Riva and A.~Wulzer, \emph{{Diboson interference resurrection}},
  \href{https://doi.org/10.1016/j.physletb.2017.11.068}{\emph{Phys. Lett. B}
  {\bfseries 776} (2018) 473--480},
  [\href{https://arxiv.org/abs/1708.07823}{{\ttfamily 1708.07823}}].

\bibitem{Grojean:2018dqj}
C.~Grojean, M.~Montull and M.~Riembau, \emph{{Diboson at the LHC vs LEP}},
  \href{https://doi.org/10.1007/JHEP03(2019)020}{\emph{JHEP} {\bfseries 03}
  (2019) 020}, [\href{https://arxiv.org/abs/1810.05149}{{\ttfamily
  1810.05149}}].

\bibitem{Gomez-Ambrosio:2018pnl}
R.~Gomez-Ambrosio, \emph{{Studies of Dimension-Six EFT effects in Vector Boson
  Scattering}},
  \href{https://doi.org/10.1140/epjc/s10052-019-6893-2}{\emph{Eur. Phys. J. C}
  {\bfseries 79} (2019) 389},
  [\href{https://arxiv.org/abs/1809.04189}{{\ttfamily 1809.04189}}].

\bibitem{Dedes:2020xmo}
A.~Dedes, P.~Koz\'ow and M.~Szleper, \emph{{Standard model EFT effects in
  vector-boson scattering at the LHC}},
  \href{https://doi.org/10.1103/PhysRevD.104.013003}{\emph{Phys. Rev. D}
  {\bfseries 104} (2021) 013003},
  [\href{https://arxiv.org/abs/2011.07367}{{\ttfamily 2011.07367}}].

\bibitem{Efrati:2015eaa}
A.~Efrati, A.~Falkowski and Y.~Soreq, \emph{{Electroweak constraints on
  flavorful effective theories}},
  \href{https://doi.org/10.1007/JHEP07(2015)018}{\emph{JHEP} {\bfseries 07}
  (2015) 018}, [\href{https://arxiv.org/abs/1503.07872}{{\ttfamily
  1503.07872}}].

\bibitem{Pomarol:2013zra}
A.~Pomarol and F.~Riva, \emph{{Towards the Ultimate SM Fit to Close in on Higgs
  Physics}}, \href{https://doi.org/10.1007/JHEP01(2014)151}{\emph{JHEP}
  {\bfseries 01} (2014) 151},
  [\href{https://arxiv.org/abs/1308.2803}{{\ttfamily 1308.2803}}].

\bibitem{deBlas:2016ojx}
J.~de~Blas, M.~Ciuchini, E.~Franco, S.~Mishima, M.~Pierini, L.~Reina et~al.,
  \emph{{Electroweak precision observables and Higgs-boson signal strengths in
  the Standard Model and beyond: present and future}},
  \href{https://doi.org/10.1007/JHEP12(2016)135}{\emph{JHEP} {\bfseries 12}
  (2016) 135}, [\href{https://arxiv.org/abs/1608.01509}{{\ttfamily
  1608.01509}}].

\bibitem{deBlas:2017wmn}
J.~de~Blas, M.~Ciuchini, E.~Franco, S.~Mishima, M.~Pierini, L.~Reina et~al.,
  \emph{{The Global Electroweak and Higgs Fits in the LHC era}},
  \href{https://doi.org/10.22323/1.314.0467}{\emph{PoS} {\bfseries EPS-HEP2017}
  (2017) 467}, [\href{https://arxiv.org/abs/1710.05402}{{\ttfamily
  1710.05402}}].

\bibitem{Falkowski:2014tna}
A.~Falkowski and F.~Riva, \emph{{Model-independent precision constraints on
  dimension-6 operators}},
  \href{https://doi.org/10.1007/JHEP02(2015)039}{\emph{JHEP} {\bfseries 02}
  (2015) 039}, [\href{https://arxiv.org/abs/1411.0669}{{\ttfamily 1411.0669}}].

\bibitem{Krauss:2016ely}
F.~Krauss, S.~Kuttimalai and T.~Plehn, \emph{{LHC multijet events as a probe
  for anomalous dimension-six gluon interactions}},
  \href{https://doi.org/10.1103/PhysRevD.95.035024}{\emph{Phys. Rev. D}
  {\bfseries 95} (2017) 035024},
  [\href{https://arxiv.org/abs/1611.00767}{{\ttfamily 1611.00767}}].

\bibitem{Alte:2017pme}
S.~Alte, M.~K\"onig and W.~Shepherd, \emph{{Consistent Searches for SMEFT
  Effects in Non-Resonant Dijet Events}},
  \href{https://doi.org/10.1007/JHEP01(2018)094}{\emph{JHEP} {\bfseries 01}
  (2018) 094}, [\href{https://arxiv.org/abs/1711.07484}{{\ttfamily
  1711.07484}}].

\bibitem{Hirschi:2018etq}
V.~Hirschi, F.~Maltoni, I.~Tsinikos and E.~Vryonidou, \emph{{Constraining
  anomalous gluon self-interactions at the LHC: a reappraisal}},
  \href{https://doi.org/10.1007/JHEP07(2018)093}{\emph{JHEP} {\bfseries 07}
  (2018) 093}, [\href{https://arxiv.org/abs/1806.04696}{{\ttfamily
  1806.04696}}].

\bibitem{Goldouzian:2020wdq}
R.~Goldouzian and M.~D. Hildreth, \emph{{LHC dijet angular distributions as a
  probe for the dimension-six triple gluon vertex}},
  \href{https://doi.org/10.1016/j.physletb.2020.135889}{\emph{Phys. Lett. B}
  {\bfseries 811} (2020) 135889},
  [\href{https://arxiv.org/abs/2001.02736}{{\ttfamily 2001.02736}}].

\bibitem{Cirigliano:2012ab}
V.~Cirigliano, M.~Gonzalez-Alonso and M.~L. Graesser, \emph{{Non-standard
  Charged Current Interactions: beta decays versus the LHC}},
  \href{https://doi.org/10.1007/JHEP02(2013)046}{\emph{JHEP} {\bfseries 02}
  (2013) 046}, [\href{https://arxiv.org/abs/1210.4553}{{\ttfamily 1210.4553}}].

\bibitem{deBlas:2013qqa}
J.~de~Blas, M.~Chala and J.~Santiago, \emph{{Global Constraints on Lepton-Quark
  Contact Interactions}},
  \href{https://doi.org/10.1103/PhysRevD.88.095011}{\emph{Phys. Rev. D}
  {\bfseries 88} (2013) 095011},
  [\href{https://arxiv.org/abs/1307.5068}{{\ttfamily 1307.5068}}].

\bibitem{Gonzalez-Alonso:2016etj}
M.~Gonz\'alez-Alonso and J.~Martin~Camalich, \emph{{Global
  Effective-Field-Theory analysis of New-Physics effects in (semi)leptonic kaon
  decays}}, \href{https://doi.org/10.1007/JHEP12(2016)052}{\emph{JHEP}
  {\bfseries 12} (2016) 052},
  [\href{https://arxiv.org/abs/1605.07114}{{\ttfamily 1605.07114}}].

\bibitem{Faroughy:2016osc}
D.~A. Faroughy, A.~Greljo and J.~F. Kamenik, \emph{{Confronting lepton flavor
  universality violation in B decays with high-$p_T$ tau lepton searches at
  LHC}}, \href{https://doi.org/10.1016/j.physletb.2016.11.011}{\emph{Phys.
  Lett. B} {\bfseries 764} (2017) 126--134},
  [\href{https://arxiv.org/abs/1609.07138}{{\ttfamily 1609.07138}}].

\bibitem{Greljo:2017vvb}
A.~Greljo and D.~Marzocca, \emph{{High-$p_T$ dilepton tails and flavor
  physics}}, \href{https://doi.org/10.1140/epjc/s10052-017-5119-8}{\emph{Eur.
  Phys. J. C} {\bfseries 77} (2017) 548},
  [\href{https://arxiv.org/abs/1704.09015}{{\ttfamily 1704.09015}}].

\bibitem{Cirigliano:2018dyk}
V.~Cirigliano, A.~Falkowski, M.~Gonz\'alez-Alonso and
  A.~Rodr\'\i{}guez-S\'anchez, \emph{{Hadronic \ensuremath{\tau} Decays as New
  Physics Probes in the LHC Era}},
  \href{https://doi.org/10.1103/PhysRevLett.122.221801}{\emph{Phys. Rev. Lett.}
  {\bfseries 122} (2019) 221801},
  [\href{https://arxiv.org/abs/1809.01161}{{\ttfamily 1809.01161}}].

\bibitem{Greljo:2018tzh}
A.~Greljo, J.~Martin~Camalich and J.~D. Ruiz-\'Alvarez, \emph{{Mono-$\tau$
  Signatures at the LHC Constrain Explanations of $B$-decay Anomalies}},
  \href{https://doi.org/10.1103/PhysRevLett.122.131803}{\emph{Phys. Rev. Lett.}
  {\bfseries 122} (2019) 131803},
  [\href{https://arxiv.org/abs/1811.07920}{{\ttfamily 1811.07920}}].

\bibitem{Bansal:2018eha}
S.~Bansal, R.~M. Capdevilla, A.~Delgado, C.~Kolda, A.~Martin and N.~Raj,
  \emph{{Hunting leptoquarks in monolepton searches}},
  \href{https://doi.org/10.1103/PhysRevD.98.015037}{\emph{Phys. Rev. D}
  {\bfseries 98} (2018) 015037},
  [\href{https://arxiv.org/abs/1806.02370}{{\ttfamily 1806.02370}}].

\bibitem{Angelescu:2020uug}
A.~Angelescu, D.~A. Faroughy and O.~Sumensari, \emph{{Lepton Flavor Violation
  and Dilepton Tails at the LHC}},
  \href{https://doi.org/10.1140/epjc/s10052-020-8210-5}{\emph{Eur. Phys. J. C}
  {\bfseries 80} (2020) 641},
  [\href{https://arxiv.org/abs/2002.05684}{{\ttfamily 2002.05684}}].

\bibitem{Farina:2016rws}
M.~Farina, G.~Panico, D.~Pappadopulo, J.~T. Ruderman, R.~Torre and A.~Wulzer,
  \emph{{Energy helps accuracy: electroweak precision tests at hadron
  colliders}},
  \href{https://doi.org/10.1016/j.physletb.2017.06.043}{\emph{Phys. Lett. B}
  {\bfseries 772} (2017) 210--215},
  [\href{https://arxiv.org/abs/1609.08157}{{\ttfamily 1609.08157}}].

\bibitem{Alioli:2017nzr}
S.~Alioli, M.~Farina, D.~Pappadopulo and J.~T. Ruderman, \emph{{Catching a New
  Force by the Tail}},
  \href{https://doi.org/10.1103/PhysRevLett.120.101801}{\emph{Phys. Rev. Lett.}
  {\bfseries 120} (2018) 101801},
  [\href{https://arxiv.org/abs/1712.02347}{{\ttfamily 1712.02347}}].

\bibitem{Raj:2016aky}
N.~Raj, \emph{{Anticipating nonresonant new physics in dilepton angular spectra
  at the LHC}}, \href{https://doi.org/10.1103/PhysRevD.95.015011}{\emph{Phys.
  Rev. D} {\bfseries 95} (2017) 015011},
  [\href{https://arxiv.org/abs/1610.03795}{{\ttfamily 1610.03795}}].

\bibitem{Schmaltz:2018nls}
M.~Schmaltz and Y.-M. Zhong, \emph{{The leptoquark Hunter\textquoteright{}s
  guide: large coupling}},
  \href{https://doi.org/10.1007/JHEP01(2019)132}{\emph{JHEP} {\bfseries 01}
  (2019) 132}, [\href{https://arxiv.org/abs/1810.10017}{{\ttfamily
  1810.10017}}].

\bibitem{Brooijmans:2020yij}
G.~Brooijmans et~al., \emph{{Les Houches 2019 Physics at TeV Colliders: New
  Physics Working Group Report}},  in \emph{{11th Les Houches Workshop on
  Physics at TeV Colliders}: {PhysTeV Les Houches}}, 2, 2020,
  \href{https://arxiv.org/abs/2002.12220}{{\ttfamily 2002.12220}}.

\bibitem{Ricci:2020xre}
R.~Torre, L.~Ricci and A.~Wulzer, \emph{{On the W\&Y interpretation of
  high-energy Drell-Yan measurements}},
  \href{https://doi.org/10.1007/JHEP02(2021)144}{\emph{JHEP} {\bfseries 02}
  (2021) 144}, [\href{https://arxiv.org/abs/2008.12978}{{\ttfamily
  2008.12978}}].

\bibitem{Fuentes-Martin:2020lea}
J.~Fuentes-Martin, A.~Greljo, J.~Martin~Camalich and J.~D. Ruiz-Alvarez,
  \emph{{Charm physics confronts high-p$_{T}$ lepton tails}},
  \href{https://doi.org/10.1007/JHEP11(2020)080}{\emph{JHEP} {\bfseries 11}
  (2020) 080}, [\href{https://arxiv.org/abs/2003.12421}{{\ttfamily
  2003.12421}}].

\bibitem{Alioli:2017ces}
S.~Alioli, V.~Cirigliano, W.~Dekens, J.~de~Vries and E.~Mereghetti,
  \emph{{Right-handed charged currents in the era of the Large Hadron
  Collider}}, \href{https://doi.org/10.1007/JHEP05(2017)086}{\emph{JHEP}
  {\bfseries 05} (2017) 086},
  [\href{https://arxiv.org/abs/1703.04751}{{\ttfamily 1703.04751}}].

\bibitem{Alioli:2017jdo}
S.~Alioli, M.~Farina, D.~Pappadopulo and J.~T. Ruderman, \emph{{Precision
  Probes of QCD at High Energies}},
  \href{https://doi.org/10.1007/JHEP07(2017)097}{\emph{JHEP} {\bfseries 07}
  (2017) 097}, [\href{https://arxiv.org/abs/1706.03068}{{\ttfamily
  1706.03068}}].

\bibitem{Alioli:2018ljm}
S.~Alioli, W.~Dekens, M.~Girard and E.~Mereghetti, \emph{{NLO QCD corrections
  to SM-EFT dilepton and electroweak Higgs boson production, matched to parton
  shower in POWHEG}},
  \href{https://doi.org/10.1007/JHEP08(2018)205}{\emph{JHEP} {\bfseries 08}
  (2018) 205}, [\href{https://arxiv.org/abs/1804.07407}{{\ttfamily
  1804.07407}}].

\bibitem{Alioli:2020kez}
S.~Alioli, R.~Boughezal, E.~Mereghetti and F.~Petriello, \emph{{Novel angular
  dependence in Drell-Yan lepton production via dimension-8 operators}},
  \href{https://doi.org/10.1016/j.physletb.2020.135703}{\emph{Phys. Lett. B}
  {\bfseries 809} (2020) 135703},
  [\href{https://arxiv.org/abs/2003.11615}{{\ttfamily 2003.11615}}].

\bibitem{Panico:2021vav}
G.~Panico, L.~Ricci and A.~Wulzer, \emph{{High-energy EFT probes with fully
  differential Drell-Yan measurements}},
  \href{https://doi.org/10.1007/JHEP07(2021)086}{\emph{JHEP} {\bfseries 07}
  (2021) 086}, [\href{https://arxiv.org/abs/2103.10532}{{\ttfamily
  2103.10532}}].

\bibitem{Sirunyan:2021khd}
{\scshape CMS} collaboration, A.~M. Sirunyan et~al., \emph{{Search for resonant
  and nonresonant new phenomena in high-mass dilepton final states at $
  \sqrt{s} $ = 13 TeV}},
  \href{https://doi.org/10.1007/JHEP07(2021)208}{\emph{JHEP} {\bfseries 07}
  (2021) 208}, [\href{https://arxiv.org/abs/2103.02708}{{\ttfamily
  2103.02708}}].

\bibitem{ATLAS:2021pvh}
{\scshape ATLAS} collaboration, \emph{{Search for new phenomena in final states
  with two leptons and one or no $b$-tagged jets at $\sqrt{s} = 13$ TeV using
  the ATLAS detector}}, .

\bibitem{Marzocca:2020ueu}
D.~Marzocca, U.~Min and M.~Son, \emph{{Bottom-Flavored Mono-Tau Tails at the
  LHC}}, \href{https://doi.org/10.1007/JHEP12(2020)035}{\emph{JHEP} {\bfseries
  12} (2020) 035}, [\href{https://arxiv.org/abs/2008.07541}{{\ttfamily
  2008.07541}}].

\bibitem{Afik:2019htr}
Y.~Afik, S.~Bar-Shalom, J.~Cohen and Y.~Rozen, \emph{{Searching for New Physics
  with $b\bar{b} \ell^+ \ell^-$ contact interactions}},
  \href{https://doi.org/10.1016/j.physletb.2020.135541}{\emph{Phys. Lett. B}
  {\bfseries 807} (2020) 135541},
  [\href{https://arxiv.org/abs/1912.00425}{{\ttfamily 1912.00425}}].

\bibitem{Alves:2018krf}
A.~Alves, O.~J. P.~t. Eboli, G.~Grilli Di~Cortona and R.~R. Moreira,
  \emph{{Indirect and monojet constraints on scalar leptoquarks}},
  \href{https://doi.org/10.1103/PhysRevD.99.095005}{\emph{Phys. Rev. D}
  {\bfseries 99} (2019) 095005},
  [\href{https://arxiv.org/abs/1812.08632}{{\ttfamily 1812.08632}}].

\bibitem{Greljo:2021kvv}
A.~Greljo, S.~Iranipour, Z.~Kassabov, M.~Madigan, J.~Moore, J.~Rojo et~al.,
  \emph{{Parton distributions in the SMEFT from high-energy Drell-Yan tails}},
  \href{https://doi.org/10.1007/JHEP07(2021)122}{\emph{JHEP} {\bfseries 07}
  (2021) 122}, [\href{https://arxiv.org/abs/2104.02723}{{\ttfamily
  2104.02723}}].

\bibitem{Bonnefoy:2021tbt}
Q.~Bonnefoy, E.~Gendy, C.~Grojean and J.~T. Ruderman, \emph{{Beyond Jarlskog:
  699 invariants for CP violation in SMEFT}},
  \href{https://arxiv.org/abs/2112.03889}{{\ttfamily 2112.03889}}.

\bibitem{Yu:2022nxj}
B.~Yu and S.~Zhou, \emph{{Spelling Out Leptonic CP Violation in the Language of
  Invariant Theory}},  \href{https://arxiv.org/abs/2203.00574}{{\ttfamily
  2203.00574}}.

\bibitem{Yu:2022ttm}
B.~Yu and S.~Zhou, \emph{{CP violation and flavor invariants in the seesaw
  effective field theory}},  \href{https://arxiv.org/abs/2203.10121}{{\ttfamily
  2203.10121}}.

\bibitem{Isidori:2010kg}
G.~Isidori, Y.~Nir and G.~Perez, \emph{{Flavor Physics Constraints for Physics
  Beyond the Standard Model}},
  \href{https://doi.org/10.1146/annurev.nucl.012809.104534}{\emph{Ann. Rev.
  Nucl. Part. Sci.} {\bfseries 60} (2010) 355},
  [\href{https://arxiv.org/abs/1002.0900}{{\ttfamily 1002.0900}}].

\bibitem{EuropeanStrategyforParticlePhysicsPreparatoryGroup:2019qin}
R.~K. Ellis et~al., \emph{{Physics Briefing Book}: {Input for the European
  Strategy for Particle Physics Update 2020}},
  \href{https://arxiv.org/abs/1910.11775}{{\ttfamily 1910.11775}}.

\bibitem{Aebischer:2020dsw}
J.~Aebischer, C.~Bobeth, A.~J. Buras and J.~Kumar, \emph{{SMEFT ATLAS of
  $\Delta$F = 2 transitions}},
  \href{https://doi.org/10.1007/JHEP12(2020)187}{\emph{JHEP} {\bfseries 12}
  (2020) 187}, [\href{https://arxiv.org/abs/2009.07276}{{\ttfamily
  2009.07276}}].

\bibitem{Silvestrini:2018dos}
L.~Silvestrini and M.~Valli, \emph{{Model-independent Bounds on the Standard
  Model Effective Theory from Flavour Physics}},
  \href{https://doi.org/10.1016/j.physletb.2019.135062}{\emph{Phys. Lett. B}
  {\bfseries 799} (2019) 135062},
  [\href{https://arxiv.org/abs/1812.10913}{{\ttfamily 1812.10913}}].

\bibitem{Pruna:2014asa}
G.~M. Pruna and A.~Signer, \emph{{The $\mu\to e\gamma$ decay in a systematic
  effective field theory approach with dimension 6 operators}},
  \href{https://doi.org/10.1007/JHEP10(2014)014}{\emph{JHEP} {\bfseries 10}
  (2014) 014}, [\href{https://arxiv.org/abs/1408.3565}{{\ttfamily 1408.3565}}].

\bibitem{Feruglio:2015yua}
F.~Feruglio, \emph{{Theoretical Aspects of Flavour and CP Violation in the
  Lepton Sector}},  in \emph{{27th Rencontres de Blois on Particle Physics and
  Cosmology}}, 9, 2015, \href{https://arxiv.org/abs/1509.08428}{{\ttfamily
  1509.08428}}.

\bibitem{Hiller:2003js}
G.~Hiller and F.~Kruger, \emph{{More model-independent analysis of $b \to s$
  processes}}, \href{https://doi.org/10.1103/PhysRevD.69.074020}{\emph{Phys.
  Rev. D} {\bfseries 69} (2004) 074020},
  [\href{https://arxiv.org/abs/hep-ph/0310219}{{\ttfamily hep-ph/0310219}}].

\bibitem{LHCb:2017avl}
{\scshape LHCb} collaboration, R.~Aaij et~al., \emph{{Test of lepton
  universality with $B^{0} \rightarrow K^{*0}\ell^{+}\ell^{-}$ decays}},
  \href{https://doi.org/10.1007/JHEP08(2017)055}{\emph{JHEP} {\bfseries 08}
  (2017) 055}, [\href{https://arxiv.org/abs/1705.05802}{{\ttfamily
  1705.05802}}].

\bibitem{LHCb:2021trn}
{\scshape LHCb} collaboration, R.~Aaij et~al., \emph{{Test of lepton
  universality in beauty-quark decays}},
  \href{https://arxiv.org/abs/2103.11769}{{\ttfamily 2103.11769}}.

\bibitem{LHCb:2020lmf}
{\scshape LHCb} collaboration, R.~Aaij et~al., \emph{{Measurement of
  $CP$-Averaged Observables in the $B^{0}\rightarrow K^{*0}\mu^{+}\mu^{-}$
  Decay}}, \href{https://doi.org/10.1103/PhysRevLett.125.011802}{\emph{Phys.
  Rev. Lett.} {\bfseries 125} (2020) 011802},
  [\href{https://arxiv.org/abs/2003.04831}{{\ttfamily 2003.04831}}].

\bibitem{LHCb:2020gog}
{\scshape LHCb} collaboration, R.~Aaij et~al., \emph{{Angular Analysis of the
  $B^{+}\rightarrow K^{\ast+}\mu^{+}\mu^{-}$ Decay}},
  \href{https://doi.org/10.1103/PhysRevLett.126.161802}{\emph{Phys. Rev. Lett.}
  {\bfseries 126} (2021) 161802},
  [\href{https://arxiv.org/abs/2012.13241}{{\ttfamily 2012.13241}}].

\bibitem{LHCb:2020zud}
{\scshape LHCb, ATLAS, CMS} collaboration, \emph{{Combination of the ATLAS, CMS
  and LHCb results on the $B^0_{(s)} \to \mu^+ \mu^-$ decays}}, .

\bibitem{LHCb:2021awg}
{\scshape LHCb} collaboration, R.~Aaij et~al., \emph{{Measurement of the
  $B^0_s\to\mu^+\mu^-$ decay properties and search for the $B^0\to\mu^+\mu^-$
  and $B^0_s\to\mu^+\mu^-\gamma$ decays}},
  \href{https://doi.org/10.1103/PhysRevD.105.012010}{\emph{Phys. Rev. D}
  {\bfseries 105} (2022) 012010},
  [\href{https://arxiv.org/abs/2108.09283}{{\ttfamily 2108.09283}}].

\bibitem{LHCb:2021vsc}
{\scshape LHCb} collaboration, R.~Aaij et~al., \emph{{Analysis of Neutral
  B-Meson Decays into Two Muons}},
  \href{https://doi.org/10.1103/PhysRevLett.128.041801}{\emph{Phys. Rev. Lett.}
  {\bfseries 128} (2022) 041801},
  [\href{https://arxiv.org/abs/2108.09284}{{\ttfamily 2108.09284}}].

\bibitem{LHCb:2014cxe}
{\scshape LHCb} collaboration, R.~Aaij et~al., \emph{{Differential branching
  fractions and isospin asymmetries of $B \to K^{(*)} \mu^+ \mu^-$ decays}},
  \href{https://doi.org/10.1007/JHEP06(2014)133}{\emph{JHEP} {\bfseries 06}
  (2014) 133}, [\href{https://arxiv.org/abs/1403.8044}{{\ttfamily 1403.8044}}].

\bibitem{LHCb:2015wdu}
{\scshape LHCb} collaboration, R.~Aaij et~al., \emph{{Angular analysis and
  differential branching fraction of the decay $B^0_s\to\phi\mu^+\mu^-$}},
  \href{https://doi.org/10.1007/JHEP09(2015)179}{\emph{JHEP} {\bfseries 09}
  (2015) 179}, [\href{https://arxiv.org/abs/1506.08777}{{\ttfamily
  1506.08777}}].

\bibitem{LHCb:2016ykl}
{\scshape LHCb} collaboration, R.~Aaij et~al., \emph{{Measurements of the
  S-wave fraction in $B^{0}\rightarrow K^{+}\pi^{-}\mu^{+}\mu^{-}$ decays and
  the $B^{0}\rightarrow K^{\ast}(892)^{0}\mu^{+}\mu^{-}$ differential branching
  fraction}}, \href{https://doi.org/10.1007/JHEP11(2016)047}{\emph{JHEP}
  {\bfseries 11} (2016) 047},
  [\href{https://arxiv.org/abs/1606.04731}{{\ttfamily 1606.04731}}].

\bibitem{LHCb:2021zwz}
{\scshape LHCb} collaboration, R.~Aaij et~al., \emph{{Branching Fraction
  Measurements of the Rare $B^0_s\rightarrow\phi\mu^+\mu^-$ and
  $B^0_s\rightarrow f_2^\prime(1525)\mu^+\mu^-$- Decays}},
  \href{https://doi.org/10.1103/PhysRevLett.127.151801}{\emph{Phys. Rev. Lett.}
  {\bfseries 127} (2021) 151801},
  [\href{https://arxiv.org/abs/2105.14007}{{\ttfamily 2105.14007}}].

\bibitem{Isidori:2021vtc}
G.~Isidori, D.~Lancierini, P.~Owen and N.~Serra, \emph{{On the significance of
  new physics in $ b \to s \ell^+ \ell^- $ decays}},
  \href{https://doi.org/10.1016/j.physletb.2021.136644}{\emph{Phys. Lett. B}
  {\bfseries 822} (2021) 136644},
  [\href{https://arxiv.org/abs/2104.05631}{{\ttfamily 2104.05631}}].

\bibitem{Muong-2:2006rrc}
{\scshape Muon g-2} collaboration, G.~W. Bennett et~al., \emph{{Final Report of
  the Muon E821 Anomalous Magnetic Moment Measurement at BNL}},
  \href{https://doi.org/10.1103/PhysRevD.73.072003}{\emph{Phys. Rev. D}
  {\bfseries 73} (2006) 072003},
  [\href{https://arxiv.org/abs/hep-ex/0602035}{{\ttfamily hep-ex/0602035}}].

\bibitem{Muong-2:2021ojo}
{\scshape Muon g-2} collaboration, B.~Abi et~al., \emph{{Measurement of the
  Positive Muon Anomalous Magnetic Moment to 0.46 ppm}},
  \href{https://doi.org/10.1103/PhysRevLett.126.141801}{\emph{Phys. Rev. Lett.}
  {\bfseries 126} (2021) 141801},
  [\href{https://arxiv.org/abs/2104.03281}{{\ttfamily 2104.03281}}].

\bibitem{DAmbrosio:2002vsn}
G.~D'Ambrosio, G.~F. Giudice, G.~Isidori and A.~Strumia, \emph{{Minimal flavor
  violation: An Effective field theory approach}},
  \href{https://doi.org/10.1016/S0550-3213(02)00836-2}{\emph{Nucl. Phys. B}
  {\bfseries 645} (2002) 155--187},
  [\href{https://arxiv.org/abs/hep-ph/0207036}{{\ttfamily hep-ph/0207036}}].

\bibitem{Buras:2003td}
A.~J. Buras, \emph{{Relations between $\Delta$ M($s$, $d^{)}$ and B($s$, $d^{)}
  \to \mu \bar{\mu}$ in models with minimal flavor violation}},
  \href{https://doi.org/10.1016/S0370-2693(03)00561-6}{\emph{Phys. Lett. B}
  {\bfseries 566} (2003) 115--119},
  [\href{https://arxiv.org/abs/hep-ph/0303060}{{\ttfamily hep-ph/0303060}}].

\bibitem{Cirigliano:2005ck}
V.~Cirigliano, B.~Grinstein, G.~Isidori and M.~B. Wise, \emph{{Minimal flavor
  violation in the lepton sector}},
  \href{https://doi.org/10.1016/j.nuclphysb.2005.08.037}{\emph{Nucl. Phys. B}
  {\bfseries 728} (2005) 121--134},
  [\href{https://arxiv.org/abs/hep-ph/0507001}{{\ttfamily hep-ph/0507001}}].

\bibitem{Blanke:2006ig}
M.~Blanke, A.~J. Buras, D.~Guadagnoli and C.~Tarantino, \emph{{Minimal Flavour
  Violation Waiting for Precise Measurements of $\Delta M_s$, $S_{\psi\phi}$,
  $A_{SL}^s$, $|V_{ub}|$, $\gamma$ and $B^0_{s,d}\to\mu^+ \mu^-$}},
  \href{https://doi.org/10.1088/1126-6708/2006/10/003}{\emph{JHEP} {\bfseries
  10} (2006) 003}, [\href{https://arxiv.org/abs/hep-ph/0604057}{{\ttfamily
  hep-ph/0604057}}].

\bibitem{UTfit:2005lis}
{\scshape UTfit} collaboration, M.~Bona et~al., \emph{{The UTfit collaboration
  report on the status of the unitarity triangle beyond the standard model. I.
  Model-independent analysis and minimal flavor violation}},
  \href{https://doi.org/10.1088/1126-6708/2006/03/080}{\emph{JHEP} {\bfseries
  03} (2006) 080}, [\href{https://arxiv.org/abs/hep-ph/0509219}{{\ttfamily
  hep-ph/0509219}}].

\bibitem{Csaki:2011ge}
C.~Csaki, Y.~Grossman and B.~Heidenreich, \emph{{MFV SUSY: A Natural Theory for
  R-Parity Violation}},
  \href{https://doi.org/10.1103/PhysRevD.85.095009}{\emph{Phys. Rev. D}
  {\bfseries 85} (2012) 095009},
  [\href{https://arxiv.org/abs/1111.1239}{{\ttfamily 1111.1239}}].

\bibitem{Fitzpatrick:2007sa}
A.~L. Fitzpatrick, G.~Perez and L.~Randall, \emph{{Flavor anarchy in a
  Randall-Sundrum model with 5D minimal flavor violation and a low Kaluza-Klein
  scale}}, \href{https://doi.org/10.1103/PhysRevLett.100.171604}{\emph{Phys.
  Rev. Lett.} {\bfseries 100} (2008) 171604},
  [\href{https://arxiv.org/abs/0710.1869}{{\ttfamily 0710.1869}}].

\bibitem{Davidson:2006bd}
S.~Davidson and F.~Palorini, \emph{{Various definitions of Minimal Flavour
  Violation for Leptons}},
  \href{https://doi.org/10.1016/j.physletb.2006.09.016}{\emph{Phys. Lett. B}
  {\bfseries 642} (2006) 72--80},
  [\href{https://arxiv.org/abs/hep-ph/0607329}{{\ttfamily hep-ph/0607329}}].

\bibitem{Buras:2010wr}
A.~J. Buras, \emph{{Minimal flavour violation and beyond: Towards a flavour
  code for short distance dynamics}}, {\emph{Acta Phys. Polon. B} {\bfseries
  41} (2010) 2487--2561}, [\href{https://arxiv.org/abs/1012.1447}{{\ttfamily
  1012.1447}}].

\bibitem{Isidori:2012ts}
G.~Isidori and D.~M. Straub, \emph{{Minimal Flavour Violation and Beyond}},
  \href{https://doi.org/10.1140/epjc/s10052-012-2103-1}{\emph{Eur. Phys. J. C}
  {\bfseries 72} (2012) 2103},
  [\href{https://arxiv.org/abs/1202.0464}{{\ttfamily 1202.0464}}].

\bibitem{Hurth:2008jc}
T.~Hurth, G.~Isidori, J.~F. Kamenik and F.~Mescia, \emph{{Constraints on New
  Physics in MFV models: A Model-independent analysis of $\Delta$ F = 1
  processes}},
  \href{https://doi.org/10.1016/j.nuclphysb.2008.09.040}{\emph{Nucl. Phys. B}
  {\bfseries 808} (2009) 326--346},
  [\href{https://arxiv.org/abs/0807.5039}{{\ttfamily 0807.5039}}].

\bibitem{Giudice:1998xp}
G.~F. Giudice, M.~A. Luty, H.~Murayama and R.~Rattazzi, \emph{{Gaugino mass
  without singlets}},
  \href{https://doi.org/10.1088/1126-6708/1998/12/027}{\emph{JHEP} {\bfseries
  12} (1998) 027}, [\href{https://arxiv.org/abs/hep-ph/9810442}{{\ttfamily
  hep-ph/9810442}}].

\bibitem{Dine:1994vc}
M.~Dine, A.~E. Nelson and Y.~Shirman, \emph{{Low-energy dynamical supersymmetry
  breaking simplified}},
  \href{https://doi.org/10.1103/PhysRevD.51.1362}{\emph{Phys. Rev. D}
  {\bfseries 51} (1995) 1362--1370},
  [\href{https://arxiv.org/abs/hep-ph/9408384}{{\ttfamily hep-ph/9408384}}].

\bibitem{Barbieri:2011ci}
R.~Barbieri, G.~Isidori, J.~Jones-Perez, P.~Lodone and D.~M. Straub,
  \emph{{$U(2)$ and Minimal Flavour Violation in Supersymmetry}},
  \href{https://doi.org/10.1140/epjc/s10052-011-1725-z}{\emph{Eur. Phys. J. C}
  {\bfseries 71} (2011) 1725},
  [\href{https://arxiv.org/abs/1105.2296}{{\ttfamily 1105.2296}}].

\bibitem{Kagan:2009bn}
A.~L. Kagan, G.~Perez, T.~Volansky and J.~Zupan, \emph{{General Minimal Flavor
  Violation}}, \href{https://doi.org/10.1103/PhysRevD.80.076002}{\emph{Phys.
  Rev. D} {\bfseries 80} (2009) 076002},
  [\href{https://arxiv.org/abs/0903.1794}{{\ttfamily 0903.1794}}].

\bibitem{Barbieri:2011fc}
R.~Barbieri, P.~Campli, G.~Isidori, F.~Sala and D.~M. Straub, \emph{{$B$-decay
  CP-asymmetries in SUSY with a $U(2)^3$ flavour symmetry}},
  \href{https://doi.org/10.1140/epjc/s10052-011-1812-1}{\emph{Eur. Phys. J. C}
  {\bfseries 71} (2011) 1812},
  [\href{https://arxiv.org/abs/1108.5125}{{\ttfamily 1108.5125}}].

\bibitem{Blankenburg:2012nx}
G.~Blankenburg, G.~Isidori and J.~Jones-Perez, \emph{{Neutrino Masses and LFV
  from Minimal Breaking of $U(3)^5$ and $U(2)^5$ flavor Symmetries}},
  \href{https://doi.org/10.1140/epjc/s10052-012-2126-7}{\emph{Eur. Phys. J. C}
  {\bfseries 72} (2012) 2126},
  [\href{https://arxiv.org/abs/1204.0688}{{\ttfamily 1204.0688}}].

\bibitem{Barbieri:2012uh}
R.~Barbieri, D.~Buttazzo, F.~Sala and D.~M. Straub, \emph{{Flavour physics from
  an approximate $U(2)^3$ symmetry}},
  \href{https://doi.org/10.1007/JHEP07(2012)181}{\emph{JHEP} {\bfseries 07}
  (2012) 181}, [\href{https://arxiv.org/abs/1203.4218}{{\ttfamily 1203.4218}}].

\bibitem{Barbieri:2012bh}
R.~Barbieri, D.~Buttazzo, F.~Sala and D.~M. Straub, \emph{{Less Minimal Flavour
  Violation}}, \href{https://doi.org/10.1007/JHEP10(2012)040}{\emph{JHEP}
  {\bfseries 10} (2012) 040},
  [\href{https://arxiv.org/abs/1206.1327}{{\ttfamily 1206.1327}}].

\bibitem{Kley:2021yhn}
J.~Kley, T.~Theil, E.~Venturini and A.~Weiler, \emph{{Electric dipole moments
  at one-loop in the dimension-6 SMEFT}},
  \href{https://arxiv.org/abs/2109.15085}{{\ttfamily 2109.15085}}.

\bibitem{Greljo:2015mma}
A.~Greljo, G.~Isidori and D.~Marzocca, \emph{{On the breaking of Lepton Flavor
  Universality in B decays}},
  \href{https://doi.org/10.1007/JHEP07(2015)142}{\emph{JHEP} {\bfseries 07}
  (2015) 142}, [\href{https://arxiv.org/abs/1506.01705}{{\ttfamily
  1506.01705}}].

\bibitem{Barbieri:2015yvd}
R.~Barbieri, G.~Isidori, A.~Pattori and F.~Senia, \emph{{Anomalies in
  $B$-decays and $U(2)$ flavour symmetry}},
  \href{https://doi.org/10.1140/epjc/s10052-016-3905-3}{\emph{Eur. Phys. J. C}
  {\bfseries 76} (2016) 67},
  [\href{https://arxiv.org/abs/1512.01560}{{\ttfamily 1512.01560}}].

\bibitem{Buttazzo:2017ixm}
D.~Buttazzo, A.~Greljo, G.~Isidori and D.~Marzocca, \emph{{B-physics anomalies:
  a guide to combined explanations}},
  \href{https://doi.org/10.1007/JHEP11(2017)044}{\emph{JHEP} {\bfseries 11}
  (2017) 044}, [\href{https://arxiv.org/abs/1706.07808}{{\ttfamily
  1706.07808}}].

\bibitem{Cornella:2021sby}
C.~Cornella, D.~A. Faroughy, J.~Fuentes-Martin, G.~Isidori and M.~Neubert,
  \emph{{Reading the footprints of the B-meson flavor anomalies}},
  \href{https://doi.org/10.1007/JHEP08(2021)050}{\emph{JHEP} {\bfseries 08}
  (2021) 050}, [\href{https://arxiv.org/abs/2103.16558}{{\ttfamily
  2103.16558}}].

\bibitem{Fuentes-Martin:2019mun}
J.~Fuentes-Mart\'\i{}n, G.~Isidori, J.~Pag\`es and K.~Yamamoto, \emph{{With or
  without U(2)? Probing non-standard flavor and helicity structures in
  semileptonic B decays}},
  \href{https://doi.org/10.1016/j.physletb.2019.135080}{\emph{Phys. Lett. B}
  {\bfseries 800} (2020) 135080},
  [\href{https://arxiv.org/abs/1909.02519}{{\ttfamily 1909.02519}}].

\bibitem{Marzocca:2021miv}
D.~Marzocca, S.~Trifinopoulos and E.~Venturini, \emph{{From B-meson anomalies
  to Kaon physics with scalar leptoquarks}},
  \href{https://arxiv.org/abs/2106.15630}{{\ttfamily 2106.15630}}.

\bibitem{Bordone:2017anc}
M.~Bordone, G.~Isidori and S.~Trifinopoulos, \emph{{Semileptonic $B$-physics
  anomalies: A general EFT analysis within $U(2)^n$ flavor symmetry}},
  \href{https://doi.org/10.1103/PhysRevD.96.015038}{\emph{Phys. Rev. D}
  {\bfseries 96} (2017) 015038},
  [\href{https://arxiv.org/abs/1702.07238}{{\ttfamily 1702.07238}}].

\bibitem{Bordone:2019uzc}
M.~Bordone, O.~Cat\`a and T.~Feldmann, \emph{{Effective Theory Approach to New
  Physics with Flavour: General Framework and a Leptoquark Example}},
  \href{https://doi.org/10.1007/JHEP01(2020)067}{\emph{JHEP} {\bfseries 01}
  (2020) 067}, [\href{https://arxiv.org/abs/1910.02641}{{\ttfamily
  1910.02641}}].

\bibitem{Kobayashi:2021pav}
T.~Kobayashi, H.~Otsuka, M.~Tanimoto and K.~Yamamoto, \emph{{Modular symmetry
  in the SMEFT}},  \href{https://arxiv.org/abs/2112.00493}{{\ttfamily
  2112.00493}}.

\bibitem{Greljo:2021xmg}
A.~Greljo, P.~Stangl and A.~E. Thomsen, \emph{{A model of muon anomalies}},
  \href{https://doi.org/10.1016/j.physletb.2021.136554}{\emph{Phys. Lett. B}
  {\bfseries 820} (2021) 136554},
  [\href{https://arxiv.org/abs/2103.13991}{{\ttfamily 2103.13991}}].

\bibitem{Greljo:2021npi}
A.~Greljo, Y.~Soreq, P.~Stangl, A.~E. Thomsen and J.~Zupan, \emph{{Muonic Force
  Behind Flavor Anomalies}},
  \href{https://arxiv.org/abs/2107.07518}{{\ttfamily 2107.07518}}.

\bibitem{Davighi:2022qgb}
J.~Davighi, A.~Greljo and A.~E. Thomsen, \emph{{Leptoquarks with Exactly Stable
  Protons}},  \href{https://arxiv.org/abs/2202.05275}{{\ttfamily 2202.05275}}.

\bibitem{Isidori:2021gqe}
G.~Isidori, J.~Pag\`es and F.~Wilsch, \emph{{Flavour alignment of New Physics
  in light of the $(g-2)_\mu$ anomaly}},
  \href{https://arxiv.org/abs/2111.13724}{{\ttfamily 2111.13724}}.

\bibitem{Aoude:2020dwv}
R.~Aoude, T.~Hurth, S.~Renner and W.~Shepherd, \emph{{The impact of flavour
  data on global fits of the MFV SMEFT}},
  \href{https://doi.org/10.1007/JHEP12(2020)113}{\emph{JHEP} {\bfseries 12}
  (2020) 113}, [\href{https://arxiv.org/abs/2003.05432}{{\ttfamily
  2003.05432}}].

\bibitem{Brod:2014hsa}
J.~Brod, A.~Greljo, E.~Stamou and P.~Uttayarat, \emph{{Probing anomalous $
  t\overline{t}Z $ interactions with rare meson decays}},
  \href{https://doi.org/10.1007/JHEP02(2015)141}{\emph{JHEP} {\bfseries 02}
  (2015) 141}, [\href{https://arxiv.org/abs/1408.0792}{{\ttfamily 1408.0792}}].

\bibitem{Bobeth:2015zqa}
C.~Bobeth and U.~Haisch, \emph{{Anomalous triple gauge couplings from $B$-meson
  and kaon observables}},
  \href{https://doi.org/10.1007/JHEP09(2015)018}{\emph{JHEP} {\bfseries 09}
  (2015) 018}, [\href{https://arxiv.org/abs/1503.04829}{{\ttfamily
  1503.04829}}].

\bibitem{ParticleDataGroup:2020ssz}
{\scshape Particle Data Group} collaboration, P.~A. Zyla et~al., \emph{{Review
  of Particle Physics}},
  \href{https://doi.org/10.1093/ptep/ptaa104}{\emph{PTEP} {\bfseries 2020}
  (2020) 083C01}.

\bibitem{Martin:2019lqd}
S.~P. Martin and D.~G. Robertson, \emph{{Standard model parameters in the
  tadpole-free pure $\overline{\rm{MS}}$ scheme}},
  \href{https://doi.org/10.1103/PhysRevD.100.073004}{\emph{Phys. Rev. D}
  {\bfseries 100} (2019) 073004},
  [\href{https://arxiv.org/abs/1907.02500}{{\ttfamily 1907.02500}}].

\end{thebibliography}\endgroup

\end{document}